\documentclass{article}
\usepackage[centertags]{amsmath}
\usepackage{hyperref}
\usepackage{color}
\usepackage{amsfonts}
\usepackage{amssymb}
\usepackage{amsthm}
\usepackage{cite}
\usepackage{fullpage}
\usepackage{graphicx}
\usepackage{pgf,pgfarrows,pgfnodes}
\usepackage{graphics}
\usepackage{tikz}
\usepackage{rotating}
\usepackage{bm}\usepackage{authblk}
\usetikzlibrary{decorations.pathreplacing,calligraphy,backgrounds}
\usetikzlibrary{decorations.shapes}
\usetikzlibrary{patterns}

\numberwithin{equation}{section}

\definecolor{colour1}{rgb}{1,0.6,0.6}
\definecolor{colour2}{rgb}{0,0.9,0}

\newcommand{\drawYT}[6]{
\foreach \l [count=\i] in {#6}
    \foreach \j in {1,...,\l}
    {
      \filldraw[#1] (#2+#4*\j,#3+#5*\i) -- ++(-#4,0) -- ++(0,-#5) -- ++(#4,0) -- cycle;
    }    
}

\title{Recurrence relation for instanton partition function in SU(N) gauge theory}

\date{\vspace{-5ex}}
\author{}

\begin{document}

\maketitle

\vspace{6pt}
\begin{center}	
	{\textsl
	Ekaterina Sysoeva$^{\dagger}$\footnote{\scriptsize \tt esysoeva@sissa.it}, Aleksei Bykov$^{\ddagger}$\footnote{\scriptsize \tt spirats@yandex.ru}} \\
\vspace{1cm}
$\dagger$\textit{\small SISSA, Via Bonomea 265, 34136 Trieste, Italy\\  I.N.F.N, Sezione di Trieste} \\ $\ddagger$\textit{\small 34141 Trieste, Italy}
\vspace{6pt}
\end{center}

\begin{center}
\textbf{Abstract}
\end{center}

\vspace{4pt} {\small

\noindent We derive a residue formula and as a consequence a recurrence relation for the instanton partition function in $\mathcal{N}=2$ supersymmetric theory on $\mathbb{C}^2$ with $SU(N)$ gauge group. The particular cases of $SU(2)$ and $SU(3)$ gauge groups were considered in the literature before. The recurrence relation with $SU(2)$ gauge group is long well known and was found as the Alday-Gaiotto-Tachikawa (AGT) counterpart of the Zamolodchikov relation for the Virasoro conformal blocks. In the $SU(3)$ case a residue formula for the term with the minimal number of instantons was found and basing on it a recurrence relation was conjectured.
\par We give a complete proof of the residue formula in all instanton orders in presence of any number of matter hypermultiplets in the adjoint and fundamental representations. The recurrence relation however describes only theories with not too much matter hypermultiplets so that the behaviour at infinity is moderate. The guideline of the proof is an algebro-geometric interpretation of the $\mathcal{N}=2$ supersymmetric gauge theory partition function in terms of the framed torsion-free sheaves. Lead by this interpretation we formulate a refined version of the residue formula and prove it by direct algebraic manipulations.

}
\newpage

\tableofcontents

\section{Introduction}

Instanton non-perturbative corrections make an essential contribution to the dynamic of supersymmetric gauge theories. In \cite{Nekrasov} Nekrasov proposed a very convenient way to compute the instanton part of the partition function of $\mathcal{N}=2$ SYM theory with $SU(N)$ gauge group. The instanton partition function can be viewed as a generating function of the contributions of the $k$-instanton sectors
\begin{equation}
  Z=\sum_{k=1}^\infty q^k Z_k
\end{equation}
and a single term $Z_k$ Nekrasov found in the integral form by equivariantisation of the theory and applying the localisation technique. The poles defining the value of this integral are parameterised by the $N$-tuples of Young diagrams with the total number of $k$ boxes.
\par Although this approach provides a direct way to compute the instanton contribution to the partition  function, the difficulty of calculations increases when the number of instantons $k$ grows, and increases the faster the higher the rank $N$ of the theory is.
\par A big step had been made when Poghossian in \cite{Poghossian2} noticed that a recurrence relation found by Zamolodchikov in \cite{Zamolodchikov} for the conformal blocks in $2d$ CFT theory with $SU(2)$ gauge group can be translated to the language of the Nekrasov partition function. By this the eminent Zamolodchikov recurrence relation for the instanton partition function in $SU(2)$ gauge theory appeared
\begin{eqnarray} \label{theveryrelation}
    Z(a)= 1+\sum_{m,\, n=1}^{\infty} \frac{q^{mn} Z(\epsilon_{m,-n})}{(-a+\epsilon_{m,n})(a+\epsilon_{m,n})}\frac{2 \epsilon_{m,n}}{\underset{(i,j)\neq(0,0)}{\prod_{i=-m+1}^m\prod_{j=-n+1}^n}\epsilon_{i,j}}.
\end{eqnarray}
Here $\epsilon_{m,n}=m \epsilon_1+n\epsilon_2$, $a$ is the difference $a=a_1-a_2$, $a_u$ are vacuum expectation values of eigenvalues of the scalar field $\phi$ of the vector multiplet and $\epsilon_1$, $\epsilon_2$ are the equivalent parameters.
\par Not only this relation allows us to calculate the Nekrasov partition function recurrently in terms of the parameter $q$, but it also grants us a clear understanding of the positions and of the orders of the poles of the partition function with respect to the variable $a$, which are not obvious from the integral form. The relation (\ref{theveryrelation}) proved to be quite useful in the computations related to $\mathcal{N}=2$ SYM $SU(2)$ gauge theory \cite{Bershtein, Bonelli}.
\par Later in \cite{Poghossian} Poghossian basing on the analysis of the instantonic partition function suggested a similar recurrence relation for $\mathcal{N}=2$ SYM $SU(3)$ theory, and, by translating it to the AGT-dual conformal theory language, a generalisation of the Zamolodchikov conformal block recurrence relation. However, rigorous proof for the case of $SU(3)$ gauge group was lacking.
\par In \cite{Bonelli} with the help of the Zamolodchikov recurrence relation an interesting relation for the full partition function, consisting of the instanton, classical and one-loop parts, was proved
\begin{equation} \label{flip}
    \lim_{\alpha \rightarrow 0}\frac{\mathcal{Z}(\alpha+\epsilon_{m,n})}{\mathcal{Z}(\alpha+\epsilon_{m,-n})}=-{\rm Sign}(\epsilon_1).
\end{equation}
A similar relation for a higher rank theory was suggested, but, firstly, the conjecture was not strong enough to recover a recurrence relation for the instanton partition function, and, secondly, there was no proof.
\par In the present paper we fix these flaws. Directly from the Nekrasov's integral representation of the $k$-instanton term $Z_k$ we determine the positions and orders of its poles and express its residue via the contribution of a smaller number of instantons. In terms of the full partition function this relation can be written in a nice form generalising (\ref{flip})
\begin{equation} \label{PartialWeylI}\lim_{a_{uv} \rightarrow \epsilon_{m,n}}\frac{\mathcal{Z}({\bf a})}{\mathcal{Z}({\bf \hat{a}}^{(uv)})}=-{\rm Sign}(\epsilon_1),
\end{equation}
where $u, \, v \in \{1,\ldots,N\}$, $a_{uv}=a_u-a_v$ and the $N$-dimensional vectors ${\bf a}$ and ${\bf \hat{a}}$ are related by the partial Weyl permutation between the $\epsilon_2$ components of $a_u$ and $a_v$ as defined in the Section \ref{ResidueFormula}. Unlike the formula proposed in \cite{Bonelli} for the $SU(N)$ case, which was written in the leading order with respect to all $N-1$ independent arguments of $\mathcal{Z}$, the relation (\ref{PartialWeylI}) is exact with respect to all variables except $a_{uv}$.
\par We prove the residue formula for the pure gauge theory, the theory with adjoint hypermultiplet and a theory with any number of fundamental and anti-fundamental hypermultiplets. 
\par Basing on the residue formula we write recurrence relation for the partition function in two different ways. In terms of the variables $a_{uv}$ we present it only for the pure theory, while in terms of the Weyl-symmetric variables we write it for all the listed above theories except the case of total number of the fundamental and anti-fundamental hypermultiplets greater than critical ($N_f+N_a) > 2(N-1)$.
\par Proving the wanted relation between two different instanton sectors looks like a rather sophisticated problem at the first glance. We approach it by establishing a refined duality between the terms contributing to the partition function. Namely, instead of treating all the Young diagrams with the total number of $k$ boxes together, we group them in smaller families of Young diagrams and we prove the residue formula for sums running over these families. This refinement is based on the interpretation of the partition function in the language of the framed torsion-free sheaves on $\mathbb{CP}^2$ and twisting of the symmetric group.
\par It would be interesting to find the AGT-dual relation on the CFT side, but we do not consider this problem in the present paper.

\par The paper is organised as follows:
\begin{itemize}
    \item In Section \ref{InstPartFunc} we define the main objects which we use throughout the computations.
    \item Section \ref{ResidueFormula} is the central part of the paper containing the formulation of the residue formula, its refined version, its interpretation in terms of the framed torsion-free sheaves, and finally the rigorous proof of the residue formula.
    \item In Section \ref{ZamRecRelSUN} we provide the recurrence relation in terms of two different sets of variables.
    \item In Section \ref{summary} we collect the  main results of this paper.
\end{itemize}

\paragraph{Acknowledgements. } This work arose from an insightful note of Jose F. Morales made during the work on \cite{Bonelli} that the leading orders of the partition function at the points related by the partial Weyl permutation coincide up to a sign.
\par E.S. would like to thank Alessandro Tanzini and Giulio Bonelli for encouragement and many discussions which helped significantly improve the paper and specifically for pointing out the paper \cite{Poghossian}.
\par E.S. also thanks Egor Zenkevich for a short yet fruitful discussion.
\par The research of E.S. is partly supported by the INFN Iniziativa Specifica GAST.

\section{Instanton partition function} \label{InstPartFunc}

We consider the $\mathcal{N}=2$ topologically twisted gauge theory with gauge group $SU(N)$ on $\mathbb{R}^4$. We identify the space $\mathbb{R}^4$ with $\mathbb{C}^2$ with coordinates $x$ and $y$, and endow it with action of $U(1)^2\subset SO(2)$ defined as
\begin{equation}\label{epsilonEq}
(\epsilon_1,\epsilon_2): (x,y)\mapsto (e^{i\epsilon_1}x, e^{i\epsilon_2}y)
\end{equation}
for $\epsilon=(\epsilon_1,\epsilon_2)\in\mathfrak{u}(1)\oplus \mathfrak{u}(1)$.
\par The main object of our interest is the instanton partition function of this theory derived in \cite{Nekrasov}. The instanton partition function is constructed by integration in the equivariant cohomology and as a result depends on the formal parameters $\epsilon_1$ and $\epsilon_2$. Physically these parameters characterise the non-trivial geometry of $\Omega$-background (see \cite{NekrasovOkunkov}). It also depends on a vector ${\bf a}=(a_1,\ldots,a_N)\in\mathbb{C}^N$ with $\sum_{u=1}^{N}a_u=0$ which again has two equivalent interpretation. In the language of the equivariant cohomology these are the coordinates on the complexified Lie algebra of the maximal torus of the gauge group $SU(N)$. Physically they are the vacuum expectation values of the Higgs field.
\par Partition function is presented as a sum over a number of instantons
\begin{equation}  \label{zinstsum}
    Z^{(\rm R)}=\sum_{k=0}^\infty q^k Z^{(\rm R)}_k({\bf a}),
\end{equation}
where $\rm R$ stands for a representation of the matter hypermultiplet, and $R=0$ corresponds to the pure theory.
\begin{equation} \label{Zk}
    Z^{(0)}_k({\bf a})=\frac{\epsilon^k}{(2 \pi {\rm i} \epsilon_1 \epsilon_2 )^k} \oint \prod_{i=1}^k \frac{{\rm d} \phi_i }{\prod_{u=1}^N \left[ (\phi_i-a_u)(a_u-\phi_i+\epsilon)\right]} \prod_{ j <i}\frac{\phi_{ij}^2(\phi_{ij}^2-\epsilon^2)}{(\phi_{ij}^2-\epsilon_1^2)(\phi_{ij}^2-\epsilon_2^2)}
\end{equation}
The poles of the integrand in (\ref{Zk}) located inside the integration contour are parametrized by $N$ Young diagrams $\vec{Y}=(Y_1, \, \ldots, \, Y_N)$ with the total number of boxes equal to the number of instantons $|\vec{Y}|= k$. The poles of integral corresponding to $\vec{Y}$ are at the points $\Phi_I$
\begin{equation} \label{statpoints}
    \Phi_I=a_I-\epsilon_1(\alpha_I-1)-\epsilon_2(\beta_I-1),
\end{equation}
where $I$ is labelling a box belonging to one of the Young diagrams $Y_u \in \vec{Y}$, $(\alpha_I, \, \beta_I)$ are coordinates of the box $I$ in $Y_u$ and $a_I=a_u$.
\par In the case of gauge theory with a matter field in the adjoint representation the 
contribution to the $k$-th instanton sector can be written as
\begin{eqnarray}
 \label{Zkadj}
    Z_k^{({\rm adj})}({\bf a})=\left(\frac{\epsilon (\epsilon_1+M)( \epsilon_2+M)}{(2 \pi {\rm i}) \epsilon_1 \epsilon_2 M (\epsilon+M)}\right)^k \oint \prod_{i=1}^k{\rm d} \phi_i  \frac{\prod_{u=1}^N \left[ (\phi_i-a_u+M)(a_u-\phi_i+\epsilon+M)\right]}{\prod_{u=1}^N \left[ (\phi_i-a_u)(a_u-\phi_i+\epsilon) \right]} \cdot \\
    \prod_{ j <i}\frac{\phi_{ij}^2(\phi_{ij}^2-\epsilon^2)(\phi_{ij}^2-(\epsilon_1+M)^2)(\phi_{ij}^2-(\epsilon_2+M)^2)}{(\phi_{ij}^2-\epsilon_1^2)(\phi_{ij}^2-\epsilon_2^2)(\phi_{ij}^2-M^2)(\phi_{ij}^2-(\epsilon+M)^2)}. \nonumber
\end{eqnarray}
The contours of integration are chosen in such a way that there are no new poles inside the contours compared to (\ref{Zk}).
\par In the case of presence of $N_f$ fundamental hypermultiplets and $N_{a}$ anti-fundamental the contribution of the $k$-sector is
\begin{equation} \label{Zkfund}
        Z_k^{(\rm fund)}({\bf a})=\frac{\epsilon^k}{(2 \pi {\rm i} \epsilon_1 \epsilon_2 )^k} \oint \prod_{i=1}^k {\rm d} \phi_i\frac{ \prod_{t=1}^{N_f} (\phi_i-m_t) \prod_{t=1}^{N_{a}} (-\phi_i+\epsilon+m_{t}) }{\prod_{u=1}^N \left[ (\phi_i-a_u)(a_u-\phi_i+\epsilon)\right]} \prod_{ j <i}\frac{\phi_{ij}^2(\phi_{ij}^2-\epsilon^2)}{(\phi_{ij}^2-\epsilon_1^2)(\phi_{ij}^2-\epsilon_2^2)}.
\end{equation}
One may notice that the signs in the (\ref{Zk}-\ref{Zkfund}) differ from \cite{Nekrasov}, but this choice of signs is in agreement with \cite{Bruzzo} and \cite{NakajimaYoshiokaZ}. To be completely clear with our sign convention let us write the same partition functions evaluated in the manner of \cite{NakajimaYoshiokaZ}.
\begin{equation} \label{ZkNak}
    Z_k^{(0)}({\bf a})=\sum_{\underset{|\vec{Y}|=k}{\vec{Y}}}\frac{1}{\prod_{u,v=1}^{N}Z_{Y_u,Y_v}(a_u,a_v)},
\end{equation}
\begin{equation} \label{ZkadjNak}
     Z_k^{({\rm adj})}({\bf a})=\sum_{\underset{|\vec{Y}|=k}{\vec{Y}}}\prod_{u,v=1}^{N}\frac{Z_{Y_u,Y_v}(a_u,a_v+M)}{Z_{Y_u,Y_v}(a_u,a_v)},
\end{equation}
\begin{equation} \label{ZkfundNak}
     Z_k^{({\rm fund})}({\bf a})=\sum_{\underset{|\vec{Y}|=k}{\vec{Y}}}\frac{\prod_{u,v=1}^{N_f}Z_{\o,Y_v}(m_u,a_v) \prod_{u,v=1}^{N_a} Z_{Y_u,\o}(a_u,m_v)}{\prod_{u,v=1}^{N}Z_{Y_u,Y_v}(a_u,a_v)},
\end{equation}
where
\begin{eqnarray}
    Z_{Y_u,Y_v}(a_u,a_v)= \prod_{(i,j)\in Y_u}(a_v-a_u+\epsilon_1(i-\tilde{l}_{Y_v,j})-\epsilon_2(j-1-l_{Y_u,i})) \nonumber \\
     \prod_{(i,j)\in Y_v}(a_v-a_u-\epsilon_1(i-1-\tilde{l}_{Y_u,j})+\epsilon_2(j-l_{Y_v,i}))
\end{eqnarray}
and $l_{Y,i}$ is the length of the $i$-th row of tableau $Y$, $\tilde{l}_{Y,i}$ is the length of the $i$-th column of tableau $Y$.

\section{Residue formula} \label{ResidueFormula}

\subsection{Dual points and the residue formula}

The first step to establish the recurrence relation for the instanton partition functions is to connect its residue with its value at some other point, which we will call the dual point.
\par As we can see from (\ref{ZkNak}-\ref{ZkfundNak}), $Z^{({\rm R})}({\bf a})$ has poles only with respect to the differences $a_u-a_v \triangleq a_{uv}$ and only at the integer lattice points $a_{uv}=m \epsilon_1+n\epsilon_2 \triangleq \epsilon_{m,n}$ with $m, n \in \mathbb{Z}$.
\par In order to find the point dual to the pole at $a_{uv}=\epsilon_{m,n}$ let us introduce the partial Weyl permutation.
\par We assume that there is exactly one pair of indices $u, \, v\in \{1,\ldots,N\}$ such that their difference is exactly in an integer lattice point.
\begin{eqnarray}\label{unpermutation}
  a_u&=&\alpha+m_u \epsilon_1 + n_u \epsilon_2 \nonumber \\
  a_v&=&\alpha+m_v \epsilon_1 + n_v \epsilon_2 \nonumber \\
  a_{uv}&=&m_{uv}\epsilon_1 + n_{uv} \epsilon_2 \triangleq  m\epsilon_1 + n \epsilon_2\nonumber
\end{eqnarray}
 By the partial Weyl permutation we understand a permutation of either $\epsilon_1$-components $m_u$ and $m_v$ or $\epsilon_2$-components $n_u$ and $n_v$. Both choices are equivalent here due to the symmetry under complete Weyl permutation. For definiteness we consider
\begin{eqnarray} \label{permutation}
  \hat{a}_u^{(uv)}=\alpha+m_u \epsilon_1 + n_v \epsilon_2  \\
  \hat{a}_v^{(uv)}=\alpha+m_v \epsilon_1 + n_u \epsilon_2 \nonumber
\end{eqnarray}
We will denote the set of vacuum expectation values with the partial Weyl permutation performed between $a_u$, $a_v$ by $\hat{\bf a}^{(uv)}$.
\par Our claim is that the instanton partition function has poles only at $a_{uv}=\epsilon_{m,n}$ with $m \cdot n>0$, the poles are simple and a residue in the case of $m>0$, $n>0$ is the following
\begin{equation} \label{zinstrelation}
    {\rm Res}_{a_{uv}=\epsilon_{m,n}} Z^{({\rm R})}( {\bf a})=q^{m n}\frac{\mathcal{P}^{(uv)}_{N,{\rm R}}(m,n|{\bf a})}{\mathcal{P}^{(uv)}_{N}(m,n|{\bf a}) }Z^{({\rm R})}(\hat{{\bf a}}^{(uv)}),
\end{equation}
where
\begin{eqnarray} \label{polynomialP}
   \mathcal{P}^{(uv)}_{N}(m,n|{\bf a})=
   \prod_{i=-m}^{m-1}\sideset{}{'} \prod_{j=-n}^{n-1} \epsilon_{i,j} \cdot\underset{w \neq u, \, v}{\prod_{w=1}^{N}} \prod_{i=1}^m\prod_{j=1}^n\left[  (a_{vw}+\epsilon_{i,j})(-a_{uw}+\epsilon_{i,j}) \right]
\end{eqnarray}
The prime in the first product means that the factor with $(i,j)=(0,0)$ is not included.
\begin{equation}
    \mathcal{P}^{(uv)}_{N,{\rm 0}}=1,
\end{equation}
\begin{eqnarray} \label{polynomialPAdjmass}
 \mathcal{P}^{(uv)}_{N,{\rm adj}}(m,n|{\bf a})=    \prod_{i=-m}^{m-1}\prod_{j=-n}^{n-1}  \left(\epsilon_{i,j}-M \right)  \cdot\underset{w \neq u, \, v}{\prod_{w=1}^{N}} \prod_{i=1}^m\prod_{j=1}^n\left[  (a_{vw}+\epsilon_{i,j}+M)(-a_{uw}+\epsilon_{i,j}+M) \right],
\end{eqnarray}
\begin{eqnarray} \label{polynomialPFundmass}
 \mathcal{P}^{(uv)}_{N,{\rm fund}}(m,n|{\bf a})=\prod_{i=1}^{m}\prod_{j=1}^{n} \Bigg[ \prod_{t=1}^{N_f}   \left(-\frac{1}{2}\epsilon_{m,n}+\epsilon_{i,j}-m_t-\frac{1}{2}\sum_{\underset{w \neq u,v}{w=1}}^N a_w \right) \cdot \\
 \prod_{t=1}^{N_a}  \left(-\frac{1}{2}\epsilon_{m,n}+\epsilon_{i,j}+m_t+\frac{1}{2}\sum_{\underset{w \neq u,v}{w=1}}^N a_w \right) \Bigg]. \nonumber
\end{eqnarray}
\par The relation (\ref{zinstrelation}) can be written more elegantly if one adds in the consideration the classical and the one-loop parts of the full partition function of the gauge theory.
\par The classical part defined as
   \begin{equation} \label{Zclass}
  Z_{{\rm class}}  = q^{- { \sum_{u}  \frac{a_{u}^2}{2   \epsilon_1 \epsilon_2}} }
  = q^{- {  \sum_{u,v} \frac{ a_{uv}^2 }{ 4N   \epsilon_1 \epsilon_2}} }
 \end{equation}
transforms under the partial Weyl permutation (\ref{permutation}) as
\begin{equation} \label{zclass}
    Z_{{\rm class}}({\bf a})=q^{-mn}Z_{{\rm class}}(\hat{\bf a}^{(uv)}).
\end{equation}
The one-loop part depends on the representation of the matter hypermultiplet and can be conveniently written in terms of the character \cite{NekrasovOkunkov}
\begin{equation}\label{Z1loop}
    Z_{{\rm 1-loop}}^{({\rm R})}({\bf a})={\rm exp}\left( -\frac{\rm d}{{\rm d}s}\left[ \frac{\Lambda^s}{\Gamma(s)} \int_0^\infty \frac{{\rm d}t}{t}t^s (\chi(y,t_1,t_2)-\chi^{({\rm R})}(y,t_1,t_2))\right] \bigg|_{s=0}\right),
\end{equation}
where the common part of the character $\chi(y,t_1,t_2)$ is
\begin{equation}
    \chi(y,t_1,t_2)=\frac{\sum_{u<v}(y_{uv}+y_{uv}^{-1})}{(1-t_1)(1-t_2)},
\end{equation}
the representation-depending parts of the character are
\begin{eqnarray}
    \chi^{({\rm 0})}(y,t_1,t_2)&=&0 \nonumber \\
    \chi^{({\rm adj})}(y,t_1,t_2)&=&\frac{\sum_{u<v}(y_{uv}e^{-t M}+y_{uv}^{-1}e^{-t M})}{(1-t_1)(1-t_2)} \\
    \chi^{({\rm fund})}(y,t_1,t_2)&=&\frac{\sum_{u}(\sum_{f=1}^{N_f}y_{u}e^{t m_f}+\sum_{f=1}^{N_a}y_{u}^{-1}e^{-t m_f})}{(1-t_1)(1-t_2)}
\end{eqnarray}
and the arguments of the characters are
\begin{equation}
    y_{uv}=e^{-t a_{uv}}, \quad y_{u}=e^{-t a_{u}}, \quad t_1=e^{-t \epsilon_1}, \quad t_2=e^{-t \epsilon_2}.
\end{equation}
One can easily derive how the characters change under the partial Weyl permutation (\ref{permutation}). For example, for the common part of the character one gets
\begin{eqnarray} \label{charactertransform}
    \chi(y,t_1,t_2)-\chi(\hat{y},t_1,t_2)=\frac{1}{(1-t_1)(1-t_2)}\big[(t_1^{m}-t_1^{-m})(t_2^{n}-t_2^{-n})+(\eta^{-{\rm Sign}(\epsilon_1)}-1)  \nonumber \\
    +\sum_{\underset{w\neq u,v}{w}}y_{uw}(1-t_1^{-m})(1-t_2^{-n})+\sum_{\underset{w\neq u,v}{w}}y^{-1}_{uw}(1-t_1^{m})(1-t_2^{n})\big],
\end{eqnarray}
where $\eta=e^{-t(a_{uv}-\epsilon_{m,n})} \rightarrow 1$ and it immediately gives in the case of pure theory ${\rm R}=0$
\begin{equation} \label{Z1looptransform}
     \underset{a_{uv}\rightarrow \epsilon_{m,n}}{\rm Lead}\frac{Z_{\rm 1-loop}^{({\rm R})}(\hat{\bf a}^{(uv)})}{Z_{\rm 1-loop}^{({\rm R})}({\bf a}^{(uv)})} = - \frac{{\rm Sign}(\epsilon_1)}{\alpha_{uv}}\frac{\mathcal{P}_{N, {\rm R}}^{(uv)}(m,n|{\bf a})}{\mathcal{P}_{N}^{(uv)}(m,n|{\bf a})}.
\end{equation}
Treating the representation-dependent part of the character in the presence of the adjoint hypermultiplet exactly in the same way as the common part (\ref{charactertransform}) we see that (\ref{Z1looptransform}) holds also for this theory. To show that (\ref{Z1looptransform}) works also in the case of a theory with the fundamental and anti-fundamental hypermultiplets we have to explicitly use that $\sum_u a_u =0$.
\par If we look at the point $a_{uv}=\epsilon_{-m,-n}$, then  (\ref{zinstrelation}), (\ref{Z1looptransform}) gain an additional minus sign.
\par Combining together (\ref{zinstrelation}), (\ref{zclass}) and (\ref{Z1looptransform}) for all types of theories we see that the full partition function consisting of the classical, one-loop and instanton contributions
\begin{equation}
    \mathcal{Z}^{({\rm R})}=Z_{\rm class}Z_{\rm 1-loop}^{({\rm R})}Z^{({\rm R})}
\end{equation}
transforms very simply under the partial Weyl permutation of $\epsilon_2$ components as
\begin{equation} \label{PartialWeyl}
    \lim_{a_{uv} \rightarrow \epsilon_{m,n}}\frac{\mathcal{Z}^{({\rm R})}({\bf a})}{\mathcal{Z}^{({\rm R})}({\bf \hat{a}}^{(uv)})}=-{\rm Sign}(\epsilon_1), \quad m,n \in \mathbb{Z}\setminus \{0\}
\end{equation}
and accordingly under the partial Weyl permutation of $\epsilon_1$ components the obtained factor is $-{\rm Sign}(\epsilon_2)$. If $m=0$ or $n=0$ the points ${\bf a}$ and ${\bf\hat{a}}^{(uv)}$ coincide or differ by a complete Weyl permutation and hence the partition function at these points is the same.

\subsection{Refined formula and geometric motivation}\label{Refined}
Proving (\ref{zinstrelation}) is complicated by the fact that there are a lot of terms in the sums on the both sides of the equality. Indeed, the instanton partition function can be written as
\begin{equation} \label{sums}
    Z^{({\rm R})}( {\bf a})=\sum_{\bf Y} Z_{\bf Y}^{({\rm R})}( {\bf a}),
\end{equation}
where the sum runs over all possible $N$-tuples of the Young diagrams $\bf Y$ and $Z_{\bf Y}^{({\rm R})}( {\bf a})$ is a contribution to the integral from a pole parameterised by $\bf Y$, so the relation (\ref{zinstrelation}) connects two sums of the type (\ref{sums}).
\par Of course one can reduce the number of terms in the sums by considering the different instanton sectors separately,
\begin{equation} \label{sumoversect}
    {\rm Res}_{a_{uv}=\epsilon_{m,n}} \sum_{\underset{|\bf Y|=k}{\bf Y}} Z_{\bf Y}^{({\rm R})}( {\bf a})=q^{mn}\frac{\mathcal{P}^{(uv)}_{N,{\rm R}}(m,n|{\bf a})}{\mathcal{P}^{(uv)}_{N}(m,n|{\bf a}) }\sum_{\underset{|\bf Y|=k-mn}{\bf Y}} Z_{\bf Y}^{({\rm R})}(\hat{{\bf a}}^{(uv)}),
\end{equation}
but this relation still has many terms on the both sides and is difficult to prove.
\par We are about to show that (\ref{sumoversect}) can be refined even more, \textit{i.e.} that the sums on the both sides of (\ref{sumoversect}) can be divided in smaller subsums and that the equality holds between these subsums independently.
\begin{equation} \label{veryrefined}
    {\rm Res}_{a_{uv}=\epsilon_{m,n}} \sum_{\underset{|\bf Y|=k}{\bf Y \in \mathcal{F}}} Z_{\bf Y}^{({\rm R})}( {\bf a})=q^{mn}\frac{\mathcal{P}^{(uv)}_{N,{\rm R}}(m,n|{\bf a})}{\mathcal{P}^{(uv)}_{N}(m,n|{\bf a}) }\sum_{\underset{|\bf Y|=k-mn}{\bf Y \in \tilde{\mathcal{F}}}} Z_{\bf Y}^{({\rm R})}(\hat{{\bf a}}^{(uv)}),
\end{equation}
where $\mathcal{F}$ (or $\tilde{\mathcal{F}}$) is a subset of the set of all the $N$-tuples of Young diagrams with $k$ cells (or $k-mn$ cells), which we will call a family of Young diagrams (or a dual family). Dividing all the Young diagrams into smaller families is the crucial point of the proof.
\par The next subsection contains a rigorous proof of (\ref{zinstrelation}) and a precise recipe of combining Young diagrams into the families, however it lacks an explanation why the recipe is exactly as it is given. We discuss the algebro-geometric picture behind this refinement and explain how the families appear in the first place in the current subsection. A reader not interested in this side of the problem can safely skip it and go directly to Subsection \ref{rigorousproof}, since the proof provided there is self-consistent.
\par For simplicity in this subsection we consider only the pure theory, although the resulting relation holds for all cases. Instead of (\ref{zinstrelation}) we deal here with its equivalent form (\ref{PartialWeyl}).
\paragraph{Algebro-geometric interpretation of the partition function.}
The partition function on $\mathbb{C}^2$ can be interpreted in terms of the framed torsion-free sheaves on $\mathbb{CP}^2$. \cite{GNY, NakajimaYoshioka,NakajimaYoshiokaZ,Nakajima}
\par 
From this point of view the functions $Z_k$ are considered as integrals over the moduli space of framed rank-$N$ torsion-free sheaves on $\mathbb{CP}^2$ with the second Chern class $k$. This space is equipped with the natural action of $T={\mathbb{C}^{*}}^2\times {\mathbb{C}^{*}}^N$, where the first factor is the complexification of the geometric rotations $U(1)^2$ and the second factor is the maximal torus of $GL(N)$ acting on the framings. On the physical side the latter can be interpreted as a complexification of the gauge group $U(N)$\footnote
{We can instead deal with $C^{*(N-1)}$, the maximal torus of $SL(N)$, which is a complexification of $SU(N)$, but it would introduce unnecessary technicalities.}. The integrals can be computed by means of the equivariant localisation. 
\par The fixed points turn out to be direct sums of $N$ rank-1 equivariant ideal sheaves with trivial framing
$$
\mathcal{E}=\bigoplus_{u=1}^N \mathcal{I}_{u}.
$$
In the sequel we call such sheaves the fixed-point sheaves.
\par
To describe each $\mathcal{I}_{u}$ it is enough to define the space $I_u$ of its sections on $\mathbb{C}^2$ as an ideal of the coordinate ring $\mathbb{C}[x,y]$. 
 Each ideal $I_u$ in its turn is described by a set of monomials which do not belong to it 
\begin{equation} \label{ideal}
 Y_u=\{(i,j)|x^{i-1} y^{j-1}\notin I_u\}.
\end{equation}
Since $I_u$ is an ideal of $\mathbb{C}[x,y]$, if a monomial $x^i y^j$ belongs to it, then so do $x^{i+1} y^j$ and $x^i y^{j+1}$. Therefore the set $Y_u$ always has the shape of a Young diagram.
\par The space of sections of the whole sheaf $\mathcal{E}$ is then
\begin{equation}
E=\bigoplus_{u=1}^N I_u. 
\end{equation}
As it was in the gauge theory picture, we see that the fixed points are characterised by the $N$-tuples of Young diagrams.
\par 
By the equivariant localisation the partition function on $\mathbb{C}^2$ is given by a sum over the $N$-tuples of the Young diagrams
\begin{equation} \label{ZfullviaY}
    \mathcal{Z}=\sum_{\bf Y} \mathcal{Z}_{\bf Y}
\end{equation}
The contribution of a fixed point $\mathcal{Z}_{\bf Y}$ is exactly the contribution of the pole of (\ref{Zk}) parametrized by the $N$-tuple of the Young diagrams $\mathbf{Y}$ in the sense of (\ref{statpoints})\cite{Nekrasov}.
\par 
From the equivariant localization formula \cite{AtiyhaBott} we expect that the integral over the moduli space of framed torsion-free sheaves can be given in terms of the weights of the representation of group $T$ acting on its tangent space at the fixed point. It can be shown that the latter is determined by the representation of this group acting on the space of sections $E$, so let us describe it.
 
\paragraph{Twisted equivariant structure.} \par Sections of a fixed-point sheaf transform into the sections of the same sheaf under the action of $T$. We will mark  a section $p(x,y) \in I_u\subset \mathbb{C}[x,y]$ by a subindex $u$ as $(p)_u$ to indicate that it is considered as an element of the $u$-th summand in $E$ and to distinguish it from identical polynomials which may appear in $I_v$, $v\neq u$.
\par 
The equivariant structure is given by the action of $(e^{i\epsilon_1},e^{i\epsilon_2},e^{ia_1},\ldots,e^{ia_n})\in T$
\begin{equation}
I_u\ni (p(x,y))_u \mapsto (e^{ia_u}p(x e^{-i\epsilon_1},ye^{-i\epsilon_2}))_u \in I_u.
\label{TActsOnIu}    
\end{equation}
In particular, each monomial $(p_{i,j}(x,y))_u=(x^{i-1}y^{j-1})_u\in I_u$ spans a space carrying an irreducible representation of $T$ of weight $\chi_{u,i,j}(\epsilon_1,\epsilon_2,\bm{a})=a_u-(i-1)\epsilon_1-(j-1)\epsilon_2$. 
 
\par We interpret the shifted argument of the partition function in the duality (\ref{PartialWeyl}) as a twist of the group $T$, which makes the geometric group to act on the framing.  Physically it corresponds to a mixing of the geometric and the global gauge groups.
\par To 
do so we define new coordinates $(\epsilon_1,\epsilon_2,\alpha_1,\ldots,\alpha_N)$ on $T$ instead of the old ones $(\epsilon_1,\epsilon_2, a_1,\ldots,a_N)$ by setting 
\begin{equation}
a_u=\alpha_u+m_u \epsilon_1+n_u \epsilon_2,\qquad u=1,\ldots, N,
\label{twist}    
\end{equation}
where $m_u$, $n_u$ are arbitrary integers.
\par Then the weight of a monomial $(p_{i,j}(x,y))_{u}$ becomes
\begin{equation} \label{chiAlpha}
    \alpha_u+(m_u-i+1)\epsilon_1+(n_u-j+1)\epsilon_2.
\end{equation}
We understand now $\alpha_u$ as a weight of a representation of $\mathbb{C}^{*N}$ and $(m_u-i+1)\epsilon_1+(n_u-j+1)\epsilon_2$ as a weight of a representation of $\mathbb{C}^{*2}$. The latter is not trivial even for a constant section $(p_{1,1})_u=(x^0y^0)_u$, which reflects that the groups were twisted.
\par 
In (\ref{PartialWeylI}) we are interested in the limit $\alpha_{uv}\rightarrow 0$ for some fixed pair $u,v\in\{1,\ldots, N\}$. It is equivalent to breaking the symmetry group down to $T^{(uv)}=C^{*2}\times C^{*(N-1)}\subset T$, where the subgroup is fixed by the equation $\alpha_u=\alpha_v$.
From the discussion above, we conclude that the behavior of $\mathcal{Z}$ in this limit is essentially determined by the representation of $T^{(uv)}$
on $E$. For this reason, below we look for such a description of a fixed point sheaf $\mathcal{E}$ that the representation of $T^{(uv)}$ carried by $E$ is explicit.

\paragraph{Bifiltrations and their graphical representation.} Now we want to show that the information about the space of sections $E$ of a fixed-point sheaf and about the $N$-tuples of the twisting parameters $\bf{m}$, $\bf{n}$ can be encoded together in the form of a bifiltration of subspaces of $\mathbb{C}^N$. A graphical representation of these bifiltrations will provide us a recipe of how to combine the Young diagrams in the families.
\par Let us remind that a non-increasing bifiltration $B$ of subspaces of $\mathbb{C}^N$ is a set of spaces $B_{i,j} \subseteq \mathbb{C}^N$ enumerated with two indices and ordered with respect to both of them, so that $B_{i,j}\subseteq B_{i-1,j}$ and  $B_{i,j}\subseteq B_{i,j-1}$, satisfying the conditions for the maximal space  $B_{i \ll 0, j \ll 0}=\mathbb{C}^N$ and the minimal space $B_{i \gg 0, j \gg 0}=0$.
\par 
In our case the bifiltration arises from a decomposition of the space $E$ into isotypical representation of the twisted $\mathbb{C}^{*2}$,
$$
E=\bigoplus_{(i,j)\in\mathbb{Z}^2}B_{i,j},
$$
where $B_{ij}$ is a subspace of $E$ transforming under the action of the twisted $\mathbb{C}^{*2}$ with the weight $i\epsilon_1+j\epsilon_2$.  From (\ref{chiAlpha}) and (\ref{ideal}) we read
\begin{equation}\label{Bexplicit}
B_{i,j}=\{(p_{m_u-i+1,n_u-j+1})|u=1,\ldots,N,\, (m_u-i+1,n_u-j+1)\notin Y_u\}.
\end{equation}
 Note that by construction the dimensions $\dim B_{i,j}$ are the multiplicities of the irreducible representations appearing in $E$. In other words, the array $\{\dim B_{i,j}\}_{i,j\in\mathbb{Z}}$ characterizes $E$ as a vector space carrying a representation of $\mathbb{C}^{*2}$ completely. 
 \par Now we introduce the linear operators
 \begin{equation}\label{x-y}
 x,y: E\longrightarrow E,
 \end{equation}
 $$
  x(p)_{u}=(x\cdot p)_{u}, \qquad y(p)_u=(y\cdot p)_{u},
 $$
 where $\cdot$ is the usual product of polynomials. We see by definitions (\ref{Bexplicit}) and (\ref{ideal}) that 
 \begin{equation}\label{subset}
 xB_{i,j}\subset  B_{i-1,j}
\quad \mathrm{and} \quad
 y B_{i,j}\subset  B_{i,j-1}.
 \end{equation}
By construction, the maps $x$ and $y$ are injective, therefore they define isomorphisms of $B_{i,j}$ with its images in $B_{i-1,j}$ and $B_{i,j-1}$:
 \begin{equation}\label{iso}
 B_{i,j}\cong x B_{i,j}
\quad \mathrm{and} \quad
 B_{i,j}\cong y B_{i,j}.
 \end{equation}
Finally, all these isomorphisms are compatible in the sense that they commute with each other. So, we can identify the isomorphic vector spaces
 \begin{equation}\label{Bidentify}
 B_{i,j}=xB_{i,j}\subset B_{i-1,j}\quad  B_{i,j}=yB_{i,j}\subset B_{i,j-1}.    
 \end{equation}
 Then $B_{i,j}$ becomes a non-increasing bifiltration\footnote{The operators $x$ and $y$ play an important role, because they make $E$ into a $\mathbb{C}[x,y]$-module, without which the original sheaf can not be reconstructed. After the identification (\ref{Bidentify}), this information is encoded in relative alignment of the spaces $B_{i,j}$.}.
 \par 
 The only structure yet not described in terms of bifiltrations is  the twisted $\mathbb{C}^{*N}$ action. As the action of $\mathbb{C}^{*N}$ on $E$ commutes with $x$ and $y$, it is compatible with the identification (\ref{Bidentify}).  Then it is enough to specify how $\mathbb{C}^{*N}$ acts on the maximal space of bifiltration $B_{i \ll 0,j \ll 0}$. From (\ref{Bexplicit}) we see that
 $$B_{i\ll 0,j\ll 0}=\bigoplus_{u=1}^{N}E^{[u]}=\mathbb{C}^N,$$
 where $E^{[u]}$ with $u=1,\ldots,N$ is a one-dimensional space transforming with the weight $e^{i\alpha_u}$ under the action of $\mathbb{C}^{*N}$. Then, with the identification (\ref{Bidentify}) the explicit expression (\ref{Bidentify}) takes the form
    \begin{equation}
B_{i,j}=\bigoplus_{\substack{u:i\leq m_u,j\leq n_u,\\
 (m_u-i+1,n_u-j+1)\notin Y_u}}E^{[u]}.       \label{BexplicitId}
    \end{equation} 
 Therefore $B_{i,j}$ is a bifiltration consisting of not just any subspaces of $\mathbb{C}^N$, but exclusively of direct sums of $E^{[u]}$. The bifiltration (\ref{BexplicitId}) contains all information about the space of sections $E$ of a fixed-point sheaf and about the $N$-tuples of the twisting parameters $\bf{m}$, $\bf{n}$.

\par It is useful to introduce edge filtrations of a bifiltration. We define them as follows
\begin{equation} \label{edgefiltrations}
B^{(1)}_i = B_{i,j  \ll 0}, \qquad  B^{(2)}_j = B_{i \ll 0,j}.
\end{equation}
Due to (\ref{BexplicitId}) we see that
 \begin{equation} \label{edgefiltrationscons}
B^{(1)}_i =\bigoplus_{u:i\leq m_u}E^{[u]}, \qquad  B^{(2)}_j=\bigoplus_{u:j\leq n_u}E^{[u]}.
\end{equation}
Note that if the twisting parameters $\textbf{m}$, $\textbf{n}$ are ordered alike ($m_{i_1}>m_{i_2}>\ldots>m_{i_N}$ and $n_{i_1}>n_{i_2}>\ldots>n_{i_N}$), than the subspaces of the edge filtrations $B^{(1)}_i$, $B^{(2)}_j$ coincide.

\par Let us now look at the graphical representation of bifiltrations.
\par To begin with, we consider a simple case of a reflexive fixed-point sheaf, which is a sheaf with a space of section containing all the polynomials (\textit{i.e.} with all the Young diagrams $Y_u$ listing the missing monomials being empty). For such a sheaf and $N$-tuples of twisting parameters $\textbf{m}$, $\textbf{n}$ we construct a bifiltration according to (\ref{BexplicitId}).
\par Easy to see that the spaces of the bifiltration of a reflexive sheaf are simply the intersections of its edge filtrations
\begin{equation}
B_{i,j}^{({\rm ref})}=B^{(1)}_i\cap B^{(2)}_j.
\label{reflexiveH}    
\end{equation}
\par Let us look at some examples, always in the $N=2$ case (generalisation will be straightforward).
\par
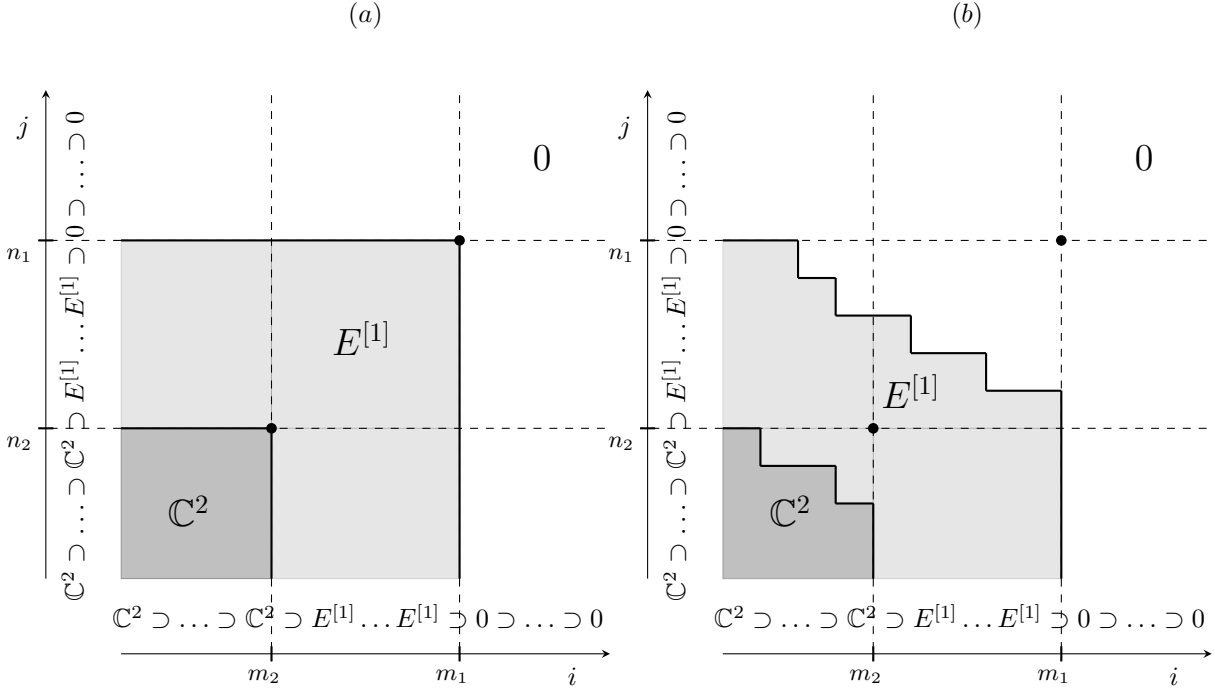
\begin{figure}
\centering
\begin{tikzpicture}

\draw[black,  dashed] (2.5,0.5)--(2.5,7);
\draw[black,  dashed] (5,0.5)--(5,7);
\draw[black,  dashed] (0.5,2.5)--(7,2.5);
\draw[black,  dashed] (0.5,5)--(7,5);
\draw [-stealth](0.5,-0.5) -- (7,-0.5);
\node[] at (6.5,-0.8) {{$i$}};
\draw[black,  dashed] (2.5,0.5)--(2.5,-0.6);
\draw[black,  thick]
(2.5,-0.6)--(2.5,-0.4);
\node[] at (2.4,-0.8) {\small{$m_2$}};
\draw[black,  dashed] (5,0.5)--(5,-0.6);
\draw[black,  thick] (5,-0.6)--(5,-0.4);
\node[] at (4.9,-0.8) {\small{$m_1$}};
\fill (5,5) circle[radius=2pt];
\fill (2.5,2.5) circle[radius=2pt];

\draw [-stealth](-0.5,0.5) -- (-0.5,7);
 \node[] at (-0.8,6.5) {{$j$}};
\draw[black,  dashed] (0.5,2.5)--(-0.6,2.5);
\draw[black,  thick]
(-0.6,2.5)--(-0.4,2.5);
\node[] at (-0.85,2.3) {\small{$n_2$}};
\draw[black,  dashed] (0.5,5)--(-0.6,5);
\draw[black,  thick] (-0.6,5)--(-0.4,5);
\node[] at (-0.85,4.8) {\small{$n_1$}};

\draw[black,  thick] (2.5,0.5)--(2.5,2.5);
\draw[black,  thick] (2.5,2.5)--(0.5,2.5);
\draw[black,  thick] (5,0.5)--(5,5);
\draw[black,  thick] (5,5)--(0.5,5);

\node[] at (3.75,8) {{$(a)$}};

\draw [draw=black, fill=black, opacity=0.25]
       (0.5,0.5) -- (2.5,0.5) -- (2.5,2.5) --(0.5,2.5) -- cycle;
\draw [draw=black, fill=black, opacity=0.1]
       (2.5,0.5) -- (2.5,2.5) --(5,2.5) --(5,0.5) -- cycle;
\draw [draw=black, fill=black, opacity=0.1]
        (2.5,2.5) --(5,2.5) --(5,5) --(2.5,5)-- cycle;
\draw [draw=black, fill=black, opacity=0.1]
        (2.5,2.5) --(2.5,5) --(0.5,5) --(0.5,2.5)-- cycle;

\node[] at (3.65,0) {{$\mathbb{C}^2\supset\ldots\supset\mathbb{C}^2\supset E^{[1]}\ldots E^{[1]}\supset 0 \supset\ldots\supset 0$}};

\node[] at (0,0.2) {{$\begin{rotate}{90}$\mathbb{C}^2\supset\ldots\supset\mathbb{C}^2\supset E^{[1]}\ldots E^{[1]} \supset 0 \supset\ldots\supset 0$\end{rotate}$}};

\node[] at (1.4,1.4) {\Large{$\mathbb{C}^2$}};
\node[] at (3.7,3.7) {\Large{$E^{[1]}$}};
\node[] at (6.1,6.1) {\Large{$0$}};

\draw[black,  dashed] (10.5,0.5)--(10.5,7);
\draw[black,  dashed] (13,0.5)--(13,7);
\draw[black,  dashed] (8.5,2.5)--(15,2.5);
\draw[black,  dashed] (8.5,5)--(15,5);
\draw[black,  thick] (10.5,0.5)--(10.5,1.5);
\draw[black,  thick] (8.5,2.5)--(9,2.5);
\draw[black,  thick] (8.5,5)--(9.5,5);
\draw[black,  thick] (13,0.5)--(13,3);
\fill (13,5) circle[radius=2pt];
\fill (10.5,2.5) circle[radius=2pt];

\draw [-stealth](8.5,-0.5) -- (15,-0.5);
\node[] at (14.5,-0.8) {{$i$}};
\draw[black,  dashed] (10.5,0.5)--(10.5,-0.6);
\draw[black,  thick]
(10.5,-0.6)--(10.5,-0.4);
\node[] at (10.4,-0.8) {\small{$m_2$}};
\draw[black,  dashed] (13,0.5)--(13,-0.6);
\draw[black,  thick] (13,-0.6)--(13,-0.4);
\node[] at (12.9,-0.8) {\small{$m_1$}};
 \draw [-stealth](7.5,0.5) -- (7.5,7);
 \node[] at (7.2,6.5) {{$j$}};
\draw[black,  dashed] (8.5,2.5)--(7.4,2.5);
\draw[black,  thick]
(7.4,2.5)--(7.6,2.5);
\node[] at (7.15,2.3) {\small{$n_2$}};
\draw[black,  dashed] (8.5,5)--(7.4,5);
\draw[black,  thick] (7.4,5)--(7.6,5);
\node[] at (7.15,4.8) {\small{$n_1$}};

\node[] at (11.75,8) {{$(b)$}};
\draw [draw=black, fill=black, opacity=0.25]
       (8.5,0.5) -- (10.5,0.5) --(10.5,1.5)--(10,1.5)--(10,2)--(9,2) --(9,2.5)--(8.5,2.5) -- cycle;
\draw [draw=black, fill=black, opacity=0.1]
       (10.5,1.5)--(10,1.5)--(10,2)--(9,2) --(9,2.5)--(8.5,2.5) -- (10.5,2.5)--  cycle;       
\draw [draw=black, fill=black, opacity=0.1]
       (10.5,0.5) -- (10.5,2.5) --(13,2.5) --(13,0.5) -- cycle;
\draw [draw=black, fill=black, opacity=0.1]
        (10.5,2.5) --(13,2.5)--(13,3)--(12,3)--(12,3.5) --(11,3.5) --(11,4)--(10.5,4) -- cycle;
\draw [draw=black, fill=black, opacity=0.1]
        (10.5,2.5) --(10.5,4) --(10,4)--(10,4.5)--(9.5,4.5)--(9.5,5)--(8.5,5) --(8.5,2.5)-- cycle;

\node[] at (11.65,0) {{$\mathbb{C}^2\supset\ldots\supset\mathbb{C}^2\supset E^{[1]}\ldots E^{[1]} \supset 0 \supset\ldots\supset 0$}};

\node[] at (8,0.2) {{$\begin{rotate}{90}$\mathbb{C}^2\supset\ldots\supset\mathbb{C}^2\supset E^{[1]} \ldots E^{[1]}\supset 0 \supset\ldots\supset 0$\end{rotate}$}};

\node[] at (9.4,1.4) {\Large{$\mathbb{C}^2$}};
\node[] at (11,3) {\Large{$E^{[1]}$}};
\node[] at (14.1,6.1) {\Large{$0$}};

\draw[black,  thick] (9.5,5)--(9.5,4.5);
\draw[black,  thick] (9.5,4.5)--(10,4.5);
\draw[black,  thick] (10,4.5)--(10,4);
\draw[black,  thick] (10,4)--(11,4);
\draw[black,  thick] (11,4)--(11,3.5);
\draw[black,  thick] (11,3.5)--(12,3.5);
\draw[black,  thick] (12,3.5)--(12,3);
\draw[black,  thick] (12,3)--(13,3);

\draw[black,  thick] (10.5,1.5)--(10,1.5);
\draw[black,  thick] (10,1.5)--(10,2);
\draw[black,  thick] (10,2)--(9,2);
\draw[black,  thick] (9,2)--(9,2.5);

\end{tikzpicture}
	\caption{Bifiltration corresponding to the twisted parameters ordered alike and (a) empty Young diagrams; \\ (b) non empty Young diagrams.} \label{bifrefandgen}
\end{figure}

 We now assume that the twisting parameters are ordered as $m_1>m_2$ and $n_1>n_2$. In this case according to (\ref{edgefiltrationscons}) the subspaces appearing in the both edge filtrations coincide and the bifiltration can be represented graphically as in Fig. \ref{bifrefandgen} (a). This and further pictures should be read as follows. Each space $B_{i,j}$ is represented by a cell with right-top coordinates $(i,j)$ on the plane. All cells belonging to a region bonded by solid lines correspond to the same space (in Fig. \ref{bifrefandgen} (a) these are the zero space, the one-dimensional space $B^{[1]}$ and the whole $\mathbb{C}^2$). The colours of regions show the dimensions of the corresponding space. Namely, the two-dimensional subspaces are coloured with dark grey, the one-dimensional subspaces are coloured with light grey, and the empty spaces are shown by white cells. The edge filtrations $B^{(\ell)}_i$ are written along the axes for convenience.
\par 
In general a space of sections of a fixed-point sheaf does not contain all polynomials, so $B_{i,j} \subset B_{i,j}^{({\rm ref})}$. In other words, a bifiltration $B$ can be obtained by cutting out some subspaces from $B_{i,j}^{({\rm ref})}$. From (\ref{BexplicitId}) we see that the set of cut out subspaces has the shape of the Young diagrams $Y_u$ and the origins of the cut out Young diagrams are located at the points $(m_u, n_u)$. 
\par 
An example of a general bifiltration is shown on Fig. \ref{bifrefandgen} (b). We again take $N=2$ and order the twisting parameters as $m_1>m_2$, $n_1>n_2$. By heavy points we mark the origins of the cut out Young diagrams located at $(m_1, n_1)$, $(m_2, n_2)$.
\par
\begin{figure}
\centering
\begin{tikzpicture}
\begin{scope}[shift={(-8,0)}]

\draw[black,  dashed] (10.5,0.5)--(10.5,7);
\draw[black,  dashed] (13,0.5)--(13,7);
\draw[black,  dashed] (8.5,2.5)--(15,2.5);
\draw[black,  dashed] (8.5,5)--(15,5);
\draw[black,  thick] (10.5,0.5)--(10.5,5);
\draw[black,  thick] (10.5,5)--(8.5,5);
\draw[black,  thick] (8.5,2.5)--(13,2.5);
\draw[black,  thick] (13,0.5)--(13,2.5);
\fill (10.5,5) circle[radius=2pt];
\fill (13,2.5) circle[radius=2pt];

\draw [-stealth](8.5,-0.5) -- (15,-0.5);
\node[] at (14.5,-0.8) {{$i$}};
\draw[black,  dashed] (10.5,0.5)--(10.5,-0.6);
\draw[black,  thick]
(10.5,-0.6)--(10.5,-0.4);
\node[] at (10.4,-0.8) {\small{$m_2$}};
\draw[black,  dashed] (13,0.5)--(13,-0.6);
\draw[black,  thick] (13,-0.6)--(13,-0.4);
\node[] at (12.9,-0.8) {\small{$m_1$}};
 \draw [-stealth](7.5,0.5) -- (7.5,7);
 \node[] at (7.2,6.5) {{$j$}};
\draw[black,  dashed] (8.5,2.5)--(7.4,2.5);
\draw[black,  thick]
(7.4,2.5)--(7.6,2.5);
\node[] at (7.15,2.3) {\small{$n_1$}};
\draw[black,  dashed] (8.5,5)--(7.4,5);
\draw[black,  thick] (7.4,5)--(7.6,5);
\node[] at (7.15,4.8) {\small{$n_2$}};

\node[] at (11.75,8) {{$(a)$}};
\draw [draw=black, fill=black, opacity=0.25]
       (8.5,0.5) -- (10.5,0.5) --(10.5,2.5)--(8.5,2.5) -- cycle;
\draw [draw=black, fill=black, opacity=0.1]
       (10.5,0.5) -- (10.5,2.5) --(13,2.5) --(13,0.5) -- cycle;

\draw [draw=black, fill=black, opacity=0.1]
        (10.5,2.5) --(10.5,5) --(8.5,5)--(8.5,2.5) -- cycle;

\node[] at (11.65,0) {{$\mathbb{C}^2\supset\ldots\supset\mathbb{C}^2\supset E^{[1]}\ldots E^{[1]}\supset 0 \supset\ldots\supset 0$}};

\node[] at (8,0.2) {{$\begin{rotate}{90}$\mathbb{C}^2\supset\ldots\supset\mathbb{C}^2\supset\, E^{[2]}\ldots \, E^{[2]}\supset 0 \supset\ldots\supset 0$\end{rotate}$}};

\node[] at (9.4,1.4) {\Large{$\mathbb{C}^2$}};
\node[] at (9.4,3.9) {\Large{$E^{[2]}$}};
\node[] at (11.9,1.4) {\Large{$E^{[1]}$}};
\node[] at (14.1,6.1) {\Large{$0$}};

\end{scope}


\draw[black,  dashed] (10.5,0.5)--(10.5,7);
\draw[black,  dashed] (13,0.5)--(13,7);
\draw[black,  dashed] (8.5,2.5)--(15,2.5);
\draw[black,  dashed] (8.5,5)--(15,5);
\draw[black,  thick] (10.5,0.5)--(10.5,3.5);
\draw[black,  thick] (9.5,5)--(8.5,5);
\draw[black,  thick] (8.5,2.5)--(11.5,2.5);
\draw[black,  thick] (13,0.5)--(13,1.5);
\fill (13,2.5) circle[radius=2pt];
\fill (10.5,5) circle[radius=2pt];

\draw[black,  thick]
(9.5,5)--(9.5,4.5)--(10,4.5)--(10,3.5)--(10.5,3.5);
\draw[black,  thick]
(11.5,2.5)--(11.5,2)--(12.5,2)--(12.5,1.5)--(13,1.5);

\draw [-stealth](8.5,-0.5) -- (15,-0.5);
\node[] at (14.5,-0.8) {{$i$}};
\draw[black,  dashed] (10.5,0.5)--(10.5,-0.6);
\draw[black,  thick]
(10.5,-0.6)--(10.5,-0.4);
\node[] at (10.4,-0.8) {\small{$m_2$}};
\draw[black,  dashed] (13,0.5)--(13,-0.6);
\draw[black,  thick] (13,-0.6)--(13,-0.4);
\node[] at (12.9,-0.8) {\small{$m_1$}};
 \draw [-stealth](7.5,0.5) -- (7.5,7);
 \node[] at (7.2,6.5) {{$j$}};
\draw[black,  dashed] (8.5,2.5)--(7.4,2.5);
\draw[black,  thick]
(7.4,2.5)--(7.6,2.5);
\node[] at (7.15,2.3) {\small{$n_1$}};
\draw[black,  dashed] (8.5,5)--(7.4,5);
\draw[black,  thick] (7.4,5)--(7.6,5);
\node[] at (7.15,4.8) {\small{$n_2$}};

\node[] at (11.75,8) {{$(b)$}};
\draw [draw=black, fill=black, opacity=0.25]
       (8.5,0.5) -- (10.5,0.5) --(10.5,2.5)--(8.5,2.5) -- cycle;
\draw [draw=black, fill=black, opacity=0.1]
       (10.5,0.5) -- (10.5,2.5) -- (11.5,2.5)--(11.5,2)--(12.5,2)--(12.5,1.5)--(13,1.5)--(13,0.5)-- cycle;
\draw [draw=black, fill=black, opacity=0.1]
        (10.5,2.5) --(10.5,3.5)--(10,3.5)--(10,4.5)--(9.5,4.5)--(9.5,5)--(8.5,5)--(8.5,2.5) -- cycle;

\node[] at (11.65,0) {{$\mathbb{C}^2\supset\ldots\supset\mathbb{C}^2\supset E^{[1]}\ldots E^{[1]}\supset 0 \supset\ldots\supset 0$}};

\node[] at (8,0.2) {{$\begin{rotate}{90}$\mathbb{C}^2\supset\ldots\supset\mathbb{C}^2\supset\, 
E^{[2]}\ldots \, E^{[2]}\supset 0 \supset\ldots\supset 0$\end{rotate}$}};

\node[] at (9.4,1.4) {\Large{$\mathbb{C}^2$}};
\node[] at (9.4,3.9) {\Large{$E^{[2]}$}};
\node[] at (11.9,1.4) {\Large{$E^{[1]}$}};
\node[] at (14.1,6.1) {\Large{$0$}};

\end{tikzpicture}
	\caption{Bifiltration corresponding to the twisted parameters ordered in the opposite way and (a) empty Young diagrams; (b) non empty Young diagrams.} \label{bifrefandgen2}
\end{figure}
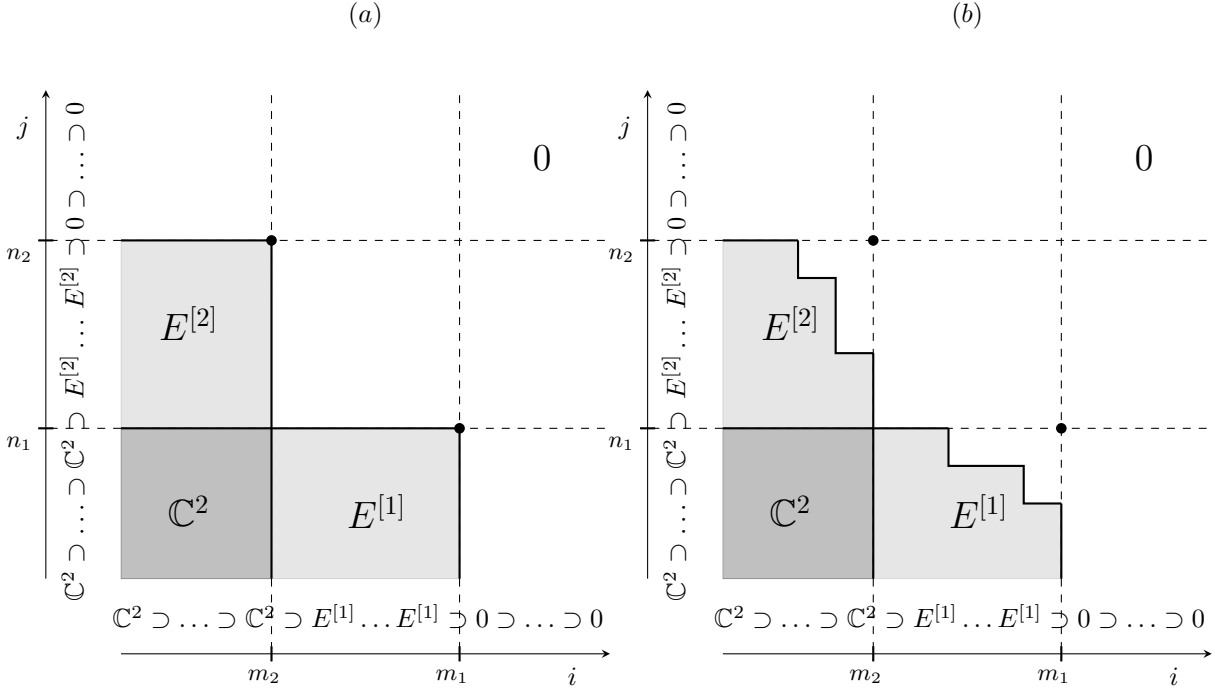

 If we order the twisting parameters differently, for example as $m_1>m_2$, $n_1<n_2$, then the one-dimensional subspaces of the edge filtrations (\ref{edgefiltrationscons}) do not coincide. A bifiltration corresponding to a reflexive sheaf with the twisting parameters ordered like this can be seen in Fig. \ref{bifrefandgen2} (a), and an example of a general bifiltration with this ordering of the twisting parameters can be seen in Fig.\ref{bifrefandgen2} (b).

\paragraph{Duality.} 

Finally we have all we need to propose the refinement of the relation (\ref{PartialWeyl}).
\par We want to establish a correspondence between the value of the partition function at the points which differ by the partial Weyl permutations. As agreed, we interpret the argument of the partition function as the twisting of the groups, and the twisting parameters in our graphical representation are the origins of cut out Young diagrams. Therefore in terms of the bifiltrations we want to see some kind of correspondence between the bifiltrations with the coordinates of the origins of two of the Young diagrams being partially permuted.
\par Let us for a time being concentrate on the case of $N=2$. In this case  $T^{(12)}=\mathbb{C}^{*2}\times \mathbb{C}^{*}$. The gauge $\mathbb{C}^{*}$ factor acts diagonally on the whole space $E$ and thus can be ignored. Therefore we expect that contributions of the fixed point sheaves producing identical representations of the geometric group $\mathbb{C}^{*2}$ are related in the limit $\alpha_{12}\rightarrow 0$. 
\par
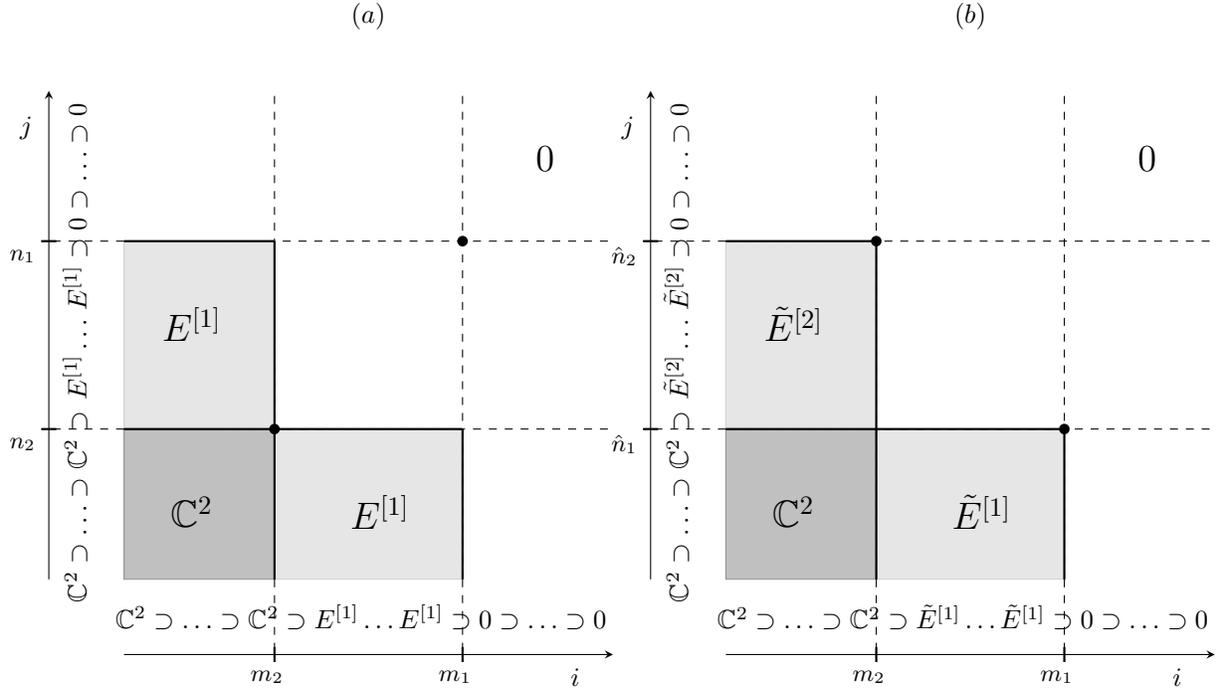
\begin{figure}
\centering
\begin{tikzpicture}
\begin{scope}[shift={(-8,0)}]

\draw[black,  dashed] (10.5,0.5)--(10.5,7);
\draw[black,  dashed] (13,0.5)--(13,7);
\draw[black,  dashed] (8.5,2.5)--(15,2.5);
\draw[black,  dashed] (8.5,5)--(15,5);
\draw[black,  thick] (10.5,0.5)--(10.5,5);
\draw[black,  thick] (10.5,5)--(8.5,5);
\draw[black,  thick] (8.5,2.5)--(13,2.5);
\draw[black,  thick] (13,0.5)--(13,2.5);
\fill (10.5,2.5) circle[radius=2pt];
\fill (13,5) circle[radius=2pt];

\draw [-stealth](8.5,-0.5) -- (15,-0.5);
\node[] at (14.5,-0.8) {{$i$}};
\draw[black,  dashed] (10.5,0.5)--(10.5,-0.6);
\draw[black,  thick]
(10.5,-0.6)--(10.5,-0.4);
\node[] at (10.4,-0.8) {\small{$m_2$}};
\draw[black,  dashed] (13,0.5)--(13,-0.6);
\draw[black,  thick] (13,-0.6)--(13,-0.4);
\node[] at (12.9,-0.8) {\small{$m_1$}};
 \draw [-stealth](7.5,0.5) -- (7.5,7);
 \node[] at (7.2,6.5) {{$j$}};
\draw[black,  dashed] (8.5,2.5)--(7.4,2.5);
\draw[black,  thick]
(7.4,2.5)--(7.6,2.5);
\node[] at (7.15,2.3) {\small{$n_2$}};
\draw[black,  dashed] (8.5,5)--(7.4,5);
\draw[black,  thick] (7.4,5)--(7.6,5);
\node[] at (7.15,4.8) {\small{$n_1$}};

\node[] at (11.75,8) {{$(a)$}};
\draw [draw=black, fill=black, opacity=0.25]
       (8.5,0.5) -- (10.5,0.5) --(10.5,2.5)--(8.5,2.5) -- cycle;
\draw [draw=black, fill=black, opacity=0.1]
       (10.5,0.5) -- (10.5,2.5) --(13,2.5) --(13,0.5) -- cycle;

\draw [draw=black, fill=black, opacity=0.1]
        (10.5,2.5) --(10.5,5) --(8.5,5)--(8.5,2.5) -- cycle;

\node[] at (11.65,0) {{$\mathbb{C}^2\supset\ldots\supset\mathbb{C}^2\supset {E}^{[1]}\ldots {E}^{[1]}\supset 0 \supset\ldots\supset 0$}};

\node[] at (8,0.2) {{$\begin{rotate}{90}$\mathbb{C}^2\supset\ldots\supset\mathbb{C}^2\supset\, {E}^{[1]}\ldots  \, {E}^{[1]}\supset 0 \supset\ldots\supset 0$\end{rotate}$}};

\node[] at (9.4,1.4) {\Large{$\mathbb{C}^2$}};
\node[] at (9.4,3.9) {\Large{${E}^{[1]}$}};
\node[] at (11.9,1.4) {\Large{${E}^{[1]}$}};
\node[] at (14.1,6.1) {\Large{$0$}};

\end{scope}


\draw[black,  dashed] (10.5,0.5)--(10.5,7);
\draw[black,  dashed] (13,0.5)--(13,7);
\draw[black,  dashed] (8.5,2.5)--(15,2.5);
\draw[black,  dashed] (8.5,5)--(15,5);
\draw[black,  thick] (10.5,0.5)--(10.5,5);
\draw[black,  thick] (10.5,5)--(8.5,5);
\draw[black,  thick] (8.5,2.5)--(13,2.5);
\draw[black,  thick] (13,0.5)--(13,2.5);
\fill (13,2.5) circle[radius=2pt];
\fill (10.5,5) circle[radius=2pt];

\draw [-stealth](8.5,-0.5) -- (15,-0.5);
\node[] at (14.5,-0.8) {{$i$}};
\draw[black,  dashed] (10.5,0.5)--(10.5,-0.6);
\draw[black,  thick]
(10.5,-0.6)--(10.5,-0.4);
\node[] at (10.4,-0.8) {\small{$m_2$}};
\draw[black,  dashed] (13,0.5)--(13,-0.6);
\draw[black,  thick] (13,-0.6)--(13,-0.4);
\node[] at (12.9,-0.8) {\small{$m_1$}};
 \draw [-stealth](7.5,0.5) -- (7.5,7);
 \node[] at (7.2,6.5) {{$j$}};
\draw[black,  dashed] (8.5,2.5)--(7.4,2.5);
\draw[black,  thick]
(7.4,2.5)--(7.6,2.5);
\node[] at (7.15,2.3) {\small{$\hat{n}_1$}};
\draw[black,  dashed] (8.5,5)--(7.4,5);
\draw[black,  thick] (7.4,5)--(7.6,5);
\node[] at (7.15,4.8) {\small{$\hat{n}_2$}};

\node[] at (11.75,8) {{$(b)$}};
\draw [draw=black, fill=black, opacity=0.25]
       (8.5,0.5) -- (10.5,0.5) --(10.5,2.5)--(8.5,2.5) -- cycle;
\draw [draw=black, fill=black, opacity=0.1]
       (10.5,0.5) -- (10.5,2.5) --(13,2.5) --(13,0.5) -- cycle;

\draw [draw=black, fill=black, opacity=0.1]
        (10.5,2.5) --(10.5,5) --(8.5,5)--(8.5,2.5) -- cycle;

\node[] at (11.65,0) {{$\mathbb{C}^2\supset\ldots\supset\mathbb{C}^2\supset \tilde{E}^{[1]}\ldots \tilde{E}^{[1]}\supset 0 \supset\ldots\supset 0$}};

\node[] at (8,0.2) {{$\begin{rotate}{90}$\mathbb{C}^2\supset\ldots\supset\mathbb{C}^2\supset\, \tilde{E}^{[2]}\ldots \, \tilde{E}^{[2]}\supset 0 \supset\ldots\supset 0$\end{rotate}$}};

\node[] at (9.4,1.4) {\Large{$\mathbb{C}^2$}};
\node[] at (9.4,3.9) {\Large{$\tilde{E}^{[2]}$}};
\node[] at (11.9,1.4) {\Large{$\tilde{E}^{[1]}$}};
\node[] at (14.1,6.1) {\Large{$0$}};

\end{tikzpicture}
	\caption{An example of two dual bifiltrations. (a) Bifiltration with $m_1>m_2$, $n_1>n_2$ and one rectangular Young diagram; (b) bifiltration with $m_1>m_2$, $\hat{n}_1<\hat{n}_2$ and empty Young diagrams.} \label{bifEFref}
\end{figure}

 
Let us compare bifiltrations with the twisting parameters ordered alike and oppositely. We will denote the latter bifiltration by ${B}$ and the former by $\tilde{B}$. As we just saw, the edge filtrations of $B$ have identical one-dimensional subspaces and the edge filtrations of $\tilde{B}$ are different, therefore $B$ and $\tilde{B}$ are for sure two different bifiltrations. However, for certain spaces of sections the dimensions of the spaces $B_{ij}$ and $\tilde{B}_{ij}$ of the corresponding bifiltrations can coincide. The simplest example is shown on Fig. \ref{bifEFref}, where $\tilde{B}$ has both Young diagram empty, while $B$ have one rectangular Young diagram $(m_2-m_1)\times(n_2-n_1)$ and one empty (we assume that $m_1>m_2$, $n_1>n_2$, $n_1=\hat{n}_2>\hat{n}_1=n_2$.).

\par 
\par Now if we recall that the dimensions $\dim B_{i,j}$ determine completely the representation of the geometric group $\mathbb{C}^{*2}$ carried by a bifiltration, we see that $B$ and $\tilde{B}$ are isomorphic as representations of  $\mathbb{C}^{*2}$.
In fact we will see that
\begin{equation} \label{dual1}
 \lim_{a \rightarrow 0}\frac{\mathcal{Z}_{(\square,\o)}(\alpha +\epsilon_{m,n})}{\mathcal{Z}_{(\o,\o)}(\alpha +\epsilon_{m,-n})}=-{\rm Sign}(\epsilon_1),
\end{equation}
where $\o$ stands for the empty Young diagram.
\par
\begin{figure}
\centering
\begin{tikzpicture}
\begin{scope}[shift={(-8,0)}]
\draw[black,  dashed] (10.5,0.5)--(10.5,7);
\draw[black,  dashed] (13,0.5)--(13,7);
\draw[black,  dashed] (8.5,2.5)--(15,2.5);
\draw[black,  dashed] (8.5,5)--(15,5);
\draw[black,  thick] (10.5,0.5)--(10.5,2.5);
\draw[black,  thick] (8.5,2.5)--(11.5,2.5);
\draw[black,  thick] (8.5,5)--(9,5);
\draw[black,  thick] (13,0.5)--(13,1.5);
\fill (13,5) circle[radius=2pt];
\fill (10.5,2.5) circle[radius=2pt];

\draw [-stealth](8.5,-0.5) -- (15,-0.5);
\node[] at (14.5,-0.8) {{$i$}};
\draw[black,  dashed] (10.5,0.5)--(10.5,-0.6);
\draw[black,  thick]
(10.5,-0.6)--(10.5,-0.4);
\node[] at (10.4,-0.8) {\small{$m_2$}};
\draw[black,  dashed] (13,0.5)--(13,-0.6);
\draw[black,  thick] (13,-0.6)--(13,-0.4);
\node[] at (12.9,-0.8) {\small{$m_1$}};
 \draw [-stealth](7.5,0.5) -- (7.5,7);
 \node[] at (7.2,6.5) {{$j$}};
\draw[black,  dashed] (8.5,2.5)--(7.4,2.5);
\draw[black,  thick]
(7.4,2.5)--(7.6,2.5);
\node[] at (7.15,2.3) {\small{$n_2$}};
\draw[black,  dashed] (8.5,5)--(7.4,5);
\draw[black,  thick] (7.4,5)--(7.6,5);
\node[] at (7.15,4.8) {\small{$n_1$}};

\node[] at (11.75,8) {{$(a)$}};
\draw [draw=black, fill=black, opacity=0.25]
       (8.5,0.5) -- (10.5,0.5) --(10.5,2)--(9.5,2)--(9.5,2.5)--(8.5,2.5) -- cycle;
\draw [draw=black, fill=black, opacity=0.1]
       (10.5,0.5)--(10.5,2)--(9.5,2)--(9.5,2.5) --  (11.5,2.5)--(11.5,2)--(12.5,2)--(12.5,1.5)--(13,1.5)--(13,0.5)-- cycle;
\draw [draw=black, fill=black, opacity=0.1]
        (9.5,2.5)--(9.5,4.5)--(9,4.5)--(9,5)--(8.5,5)--(8.5,2.5) -- cycle;

\node[] at (11.65,0) {{$\mathbb{C}^2\supset\ldots\supset\mathbb{C}^2\supset {E}^{[1]}\ldots {E}^{[1]} \supset 0 \supset\ldots\supset 0$}};

\node[] at (8,0.2) {{$\begin{rotate}{90}$\mathbb{C}^2\supset\ldots\supset\mathbb{C}^2\supset E^{[1]} \ldots {E}^{[1]}\supset 0 \supset\ldots\supset 0$\end{rotate}$}};

\node[] at (9.4,1.4) {\Large{$\mathbb{C}^2$}};
\node[] at (9,3.9) {\Large{${E}^{[1]}$}};
\node[] at (11.9,1.4) {\Large{${E}^{[1]}$}};
\node[] at (10.1,2.3) {${E}^{[2]}$};
\node[] at (14.1,6.1) {\Large{$0$}};

\draw[black,  thick]
(9,5)--(9,4.5)--(9.5,4.5)--(9.5,2)--(10.5,2);
\draw[black,  thick]
(11.5,2.5)--(11.5,2)--(12.5,2)--(12.5,1.5)--(13,1.5);
\end{scope}


\draw[black,  dashed] (10.5,0.5)--(10.5,7);
\draw[black,  dashed] (13,0.5)--(13,7);
\draw[black,  dashed] (8.5,2.5)--(15,2.5);
\draw[black,  dashed] (8.5,5)--(15,5);
\draw[black,  thick] (10.5,0.5)--(10.5,2);
\draw[black,  thick] (9,5)--(8.5,5);
\draw[black,  thick] (8.5,2.5)--(11.5,2.5);
\draw[black,  thick] (13,0.5)--(13,1.5);
\fill (13,2.5) circle[radius=2pt];
\fill (10.5,5) circle[radius=2pt];

\draw[black,  thick]
(9,5)--(9,4.5)--(9.5,4.5)--(9.5,2)--(10.5,2);
\draw[black,  thick]
(11.5,2.5)--(11.5,2)--(12.5,2)--(12.5,1.5)--(13,1.5);

\draw [-stealth](8.5,-0.5) -- (15,-0.5);
\node[] at (14.5,-0.8) {{$i$}};
\draw[black,  dashed] (10.5,0.5)--(10.5,-0.6);
\draw[black,  thick]
(10.5,-0.6)--(10.5,-0.4);
\node[] at (10.4,-0.8) {\small{$m_2$}};
\draw[black,  dashed] (13,0.5)--(13,-0.6);
\draw[black,  thick] (13,-0.6)--(13,-0.4);
\node[] at (12.9,-0.8) {\small{$m_1$}};
 \draw [-stealth](7.5,0.5) -- (7.5,7);
 \node[] at (7.2,6.5) {{$j$}};
\draw[black,  dashed] (8.5,2.5)--(7.4,2.5);
\draw[black,  thick]
(7.4,2.5)--(7.6,2.5);
\node[] at (7.15,2.3) {\small{$\hat{n}_1$}};
\draw[black,  dashed] (8.5,5)--(7.4,5);
\draw[black,  thick] (7.4,5)--(7.6,5);
\node[] at (7.15,4.8) {\small{$\hat{n}_2$}};

\node[] at (11.75,8) {{$(b)$}};
\draw [draw=black, fill=black, opacity=0.25]
       (8.5,0.5) -- (10.5,0.5) --(10.5,2)--(9.5,2)--(9.5,2.5)--(8.5,2.5) -- cycle;
\draw [draw=black, fill=black, opacity=0.1]
       (10.5,0.5)--(10.5,2)--(9.5,2)--(9.5,2.5) --  (11.5,2.5)--(11.5,2)--(12.5,2)--(12.5,1.5)--(13,1.5)--(13,0.5)-- cycle;
\draw [draw=black, fill=black, opacity=0.1]
        (9.5,2.5)--(9.5,4.5)--(9,4.5)--(9,5)--(8.5,5)--(8.5,2.5) -- cycle;

\node[] at (11.65,0) {{$\mathbb{C}^2\supset\ldots\supset\mathbb{C}^2\supset {E}^{[1]}\ldots {E}^{[1]}\supset 0 \supset\ldots\supset 0$}};

\node[] at (8,0.2) {{$\begin{rotate}{90}$\mathbb{C}^2\supset\ldots\supset\mathbb{C}^2\supset\, {E}^{[2]}\ldots \, {E} ^{[2]}\supset 0 \supset\ldots\supset 0$\end{rotate}$}};

\node[] at (9.4,1.4) {\Large{$\mathbb{C}^2$}};
\node[] at (9,3.9) {\Large{$E^{[2]}$}};
\node[] at (11.9,1.4) {\Large{${E}^{[1]}$}};
\node[] at (14.1,6.1) {\Large{$0$}};

\end{tikzpicture}
	\caption{Example of dual bifiltrations with all the cut out Young diagrams being non empty. (a) $m_1>m_2$, $n_1>n_2$ ; (b) $m_1>m_2$, $\hat{n}_1<\hat{n}_2$.} \label{bifdual2}
\end{figure}
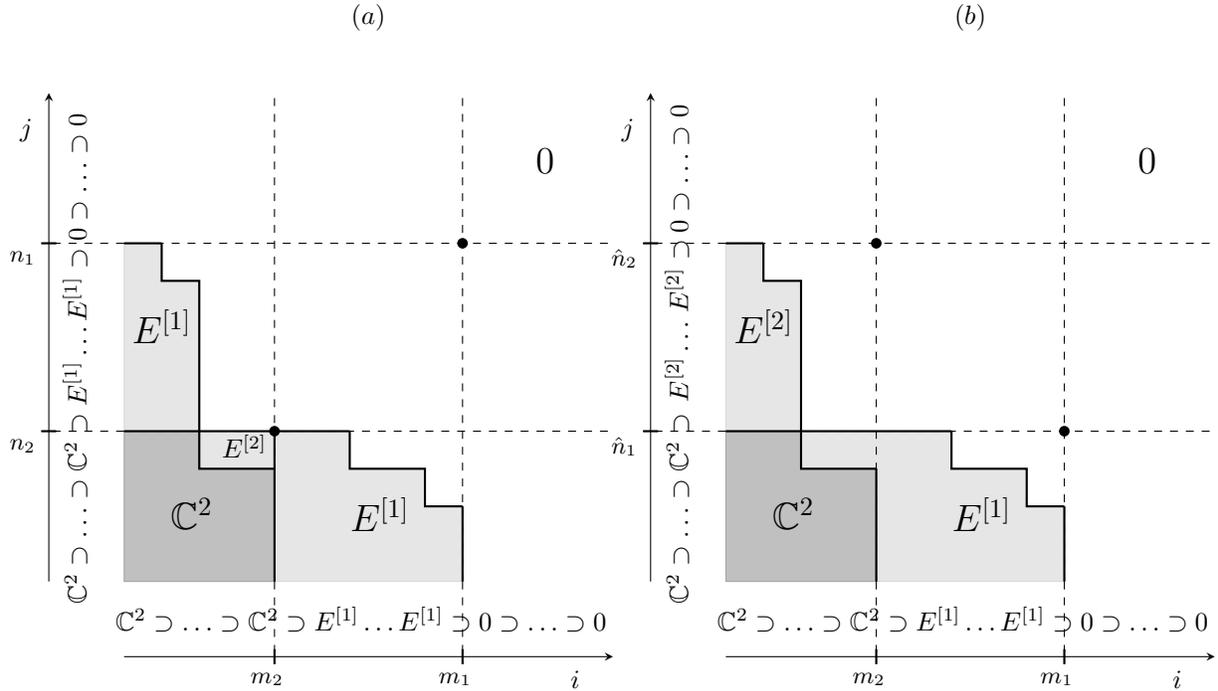

 On Fig. \ref{bifdual2} is shown a non-trivial example of dual bifiltrations with all the Young diagrams being non empty. We will see that for such couples of the cut out Young diagrams again holds the relation of the type (\ref{dual1}).

\par One could expect that the duality holds for all the bifiltrations $B$, $\tilde{B}$ such that ${\dim} \, B_{i,j}={\dim} \, \tilde{B}_{i,j}$ for all $i,j\in\mathbb{N}$.  However, in general there could be many bifiltrations $B$ satisfying
\begin{equation} \label{dimBij}
{\rm dim} \, B_{i,j}=d_{i,j}
\end{equation}
for some fixed numbers $d_{i,j}$ and the refinement (\ref{veryrefined}) should be formulated in terms of the dual families, and not in terms of single contributions.
 \par The rule is the following: a family is formed by all the bifiltrations with coinciding subspaces of the edge filtrations and with dimensions of the subspaces of the bifiltrations satisfying (\ref{dimBij}). Two families are dual if they have different one-dimensional subspaces of the edge filtrations, but the dimensions of the subspaces of the bifiltrations of the both families satisfy (\ref{dimBij}) with the same $d_{i,j}$. All bifiltrations in a family and its dual family are isomorphic as representations of twisted $\mathbb{C}^{*2}$. An example of a family is provided on Fig. \ref{fam} and its dual family is shown on Fig. \ref{dualfam}. The contours of the Young diagrams are shown by the lines of triangles and crosses.
 \par For families defined in this way we will indeed see that

\begin{equation} \label{dual2}
\underset{a \rightarrow 0}{\rm lim}\frac{\sum_{(Y_1,Y_2)\in\mathcal{F}} \mathcal{Z}_{(Y_1,Y_2)}(\alpha +\epsilon_{m,n})}{\sum_{(\tilde{Y}_1,\tilde{Y}_2)\in\mathcal{\tilde{F}}} \mathcal{Z}_{(\tilde{Y}_1,\tilde{Y}_2)}(\alpha +\epsilon_{m,-n})}=-{\rm Sign}(\epsilon_1).
\end{equation}

\begin{figure}
\centering
\begin{tikzpicture}

\draw[black,  dashed] (2.5,0.5)--(2.5,7);
\draw[black,  dashed] (5,0.5)--(5,7);
\draw[black,  dashed] (0.5,2.5)--(7,2.5);
\draw[black,  dashed] (0.5,5)--(7,5);
\draw [-stealth](0.5,-0.5) -- (7,-0.5);
\fill (2.5,2.5) circle[radius=2pt];
\fill (5,5) circle[radius=2pt];
\node[] at (6.5,-0.8) {{$i$}};
\draw[black,  dashed] (2.5,0.5)--(2.5,-0.6);
\draw[black,  thick]
(2.5,-0.6)--(2.5,-0.4);
\node[] at (2.4,-0.8) {\small{$m_2$}};
\draw[black,  dashed] (5,0.5)--(5,-0.6);
\draw[black,  thick] (5,-0.6)--(5,-0.4);
\node[] at (4.9,-0.8) {\small{$m_1$}};

\draw [-stealth](-0.5,0.5) -- (-0.5,7);
 \node[] at (-0.8,6.5) {{$j$}};
\draw[black,  dashed] (0.5,2.5)--(-0.6,2.5);
\draw[black,  thick]
(-0.6,2.5)--(-0.4,2.5);
\node[] at (-0.85,2.3) {\small{$n_2$}};
\draw[black,  dashed] (0.5,5)--(-0.6,5);
\draw[black,  thick] (-0.6,5)--(-0.4,5);
\node[] at (-0.85,4.8) {\small{$n_1$}};

\node[] at (3.75,8) {{$(a)$}};

\draw[black,  thick] (0.75,5)--(0.75,4.5);
\draw[black,  thick] (0.75,4.5)--(0.875,4.5);
\draw[black,  thick] (0.875,4.5)--(0.875,3.5);
\draw[black,  thick] (0.875,3.5)--(1.125,3.5);
\draw[black,  thick] (1.125,3.5)--(1.125,2);
\draw[black,  thick] (1.125,2)--(1.375,2);
\draw[black,  thick] (1.375,2)--(1.375,1.5);
\draw[black,  thick] (1.375,1.5)--(2,1.5);
\draw[black,  thick] (2,1.5)--(2,1.25);
\draw[black,  thick] (2,1.25)--(3.5,1.25);
\draw[black,  thick] (3.5,1.25)--(3.5,1);
\draw[black,  thick] (3.5,1)--(5,1);

\draw[black,  thick] (1.75,2.5)--(1.75,2);
\draw[black,  thick] (1.75,2)--(2.25,2);
\draw[black,  thick] (2.25,2)--(2.25,1.75);
\draw[black,  thick] (2.25,1.75)--(2.5,1.75);

\draw[black,  thick] (2.5,0.5)--(2.5,1.75);
\draw[black,  thick] (0.5,2.5)--(1.75,2.5);
\draw[black,  thick] (5,0.5)--(5,1);
\draw[black,  thick] (0.5,5)--(0.75,5);

\draw[decorate, decoration={triangles, segment length=3mm}]
 (0.75,5)--(0.75,4.5)--(0.875,4.5)--(0.875,3.5)--(1.125,3.5)--(1.125,2)--(1.375,2)--(1.375,1.5)--(2,1.5)--(2,1.25)--(3.5,1.25)--(3.5,1)--(5,1)--(5,5)--cycle;
 
\draw[decorate, decoration={crosses, segment length=2mm}]
(2.5,2.5)--(1.75,2.5)--(1.75,2)--(2.25,2)--(2.25,1.75)--(2.5,1.75)--cycle;

\draw [draw=black, fill=black, opacity=0.25]
       (0.5,0.5)--(0.5,2.5)--(1.125,2.5)--(1.125,2)--(1.375,2)--(1.375,1.5)--(2,1.5)--(2,1.25)--(2.5,1.25)-- (2.5,0.5) -- cycle;
\draw [draw=black, fill=black, opacity=0.1]
       (2.5,0.5) -- (2.5,1.25) --(3.5,1.25)--(3.5,1)--(5,1)--(5,0.5) -- cycle;

\draw [draw=black, fill=black, opacity=0.1]
        (0.5,2.5)--(0.5,5) --(0.75,5)--(0.75,4.5)--(0.875,4.5)--(0.875,3.5)--(1.125,3.5)--(1.125,2.5)-- cycle;

\draw [draw=black, fill=black, opacity=0.1]
(1.125,2.5)-- (1.125,2)--(1.375,2)--(1.375,1.5)--(2,1.5)--(2,1.25)--(2.5,1.25)--(2.5,1.75)--(2.25,1.75)--(2.25,2) --(1.75,2)-- (1.75,2.5)--cycle;

\node[] at (3.65,0) {{$\mathbb{C}^2\supset\ldots\supset\mathbb{C}^2\supset E^{[1]}\ldots E^{[1]}\supset 0 \supset\ldots\supset 0$}};

\node[] at (0,0.2) {{$\begin{rotate}{90}$\mathbb{C}^2\supset\ldots\supset\mathbb{C}^2\supset E^{[1]}\ldots E^{[1]}\supset 0 \supset\ldots\supset 0$\end{rotate}$}};

\node[] at (0.9,0.9) {\Large{$\mathbb{C}^2$}};
\node[] at (1.75,1.75) {{$E^{[2]}$}};
\node[] at (0.77,3) {{$E^{[1]}$}};
\node[] at (3,1) {{$E^{[1]}$}};
\node[] at (3.7,3.7) {\Large{$0$}};

\begin{scope}[shift={(8,0)}]
    
\draw[black,  dashed] (2.5,0.5)--(2.5,7);
\draw[black,  dashed] (5,0.5)--(5,7);
\draw[black,  dashed] (0.5,2.5)--(7,2.5);
\draw[black,  dashed] (0.5,5)--(7,5);
\draw [-stealth](0.5,-0.5) -- (7,-0.5);
\fill (2.5,2.5) circle[radius=2pt];
\fill (5,5) circle[radius=2pt];
\node[] at (6.5,-0.8) {{$i$}};
\draw[black,  dashed] (2.5,0.5)--(2.5,-0.6);
\draw[black,  thick]
(2.5,-0.6)--(2.5,-0.4);
\node[] at (2.4,-0.8) {\small{$m_2$}};
\draw[black,  dashed] (5,0.5)--(5,-0.6);
\draw[black,  thick] (5,-0.6)--(5,-0.4);
\node[] at (4.9,-0.8) {\small{$m_1$}};

\draw [-stealth](-0.5,0.5) -- (-0.5,7);
 \node[] at (-0.8,6.5) {{$j$}};
\draw[black,  dashed] (0.5,2.5)--(-0.6,2.5);
\draw[black,  thick]
(-0.6,2.5)--(-0.4,2.5);
\node[] at (-0.85,2.3) {\small{$n_2$}};
\draw[black,  dashed] (0.5,5)--(-0.6,5);
\draw[black,  thick] (-0.6,5)--(-0.4,5);
\node[] at (-0.85,4.8) {\small{$n_1$}};

\node[] at (3.75,8) {{$(b)$}};

\draw[black,  thick] (0.75,5)--(0.75,4.5);
\draw[black,  thick] (0.75,4.5)--(0.875,4.5);
\draw[black,  thick] (0.875,4.5)--(0.875,3.5);
\draw[black,  thick] (0.875,3.5)--(1.125,3.5);
\draw[black,  thick] (1.125,3.5)--(1.125,2);
\draw[black,  thick] (1.125,2)--(1.375,2);
\draw[black,  thick] (1.375,2)--(1.375,1.5);
\draw[black,  thick] (1.375,1.5)--(2,1.5);
\draw[black,  thick] (2,1.5)--(2,1.25);
\draw[black,  thick] (2,1.25)--(3.5,1.25);
\draw[black,  thick] (3.5,1.25)--(3.5,1);
\draw[black,  thick] (3.5,1)--(5,1);

\draw[black,  thick] (1.75,2.5)--(1.75,2);
\draw[black,  thick] (1.75,2)--(2.25,2);
\draw[black,  thick] (2.25,2)--(2.25,1.75);
\draw[black,  thick] (2.25,1.75)--(2.5,1.75);

\draw[black,  thick] (2.5,0.5)--(2.5,1.75);
\draw[black,  thick] (0.5,2.5)--(1.75,2.5);
\draw[black,  thick] (5,0.5)--(5,1);
\draw[black,  thick] (0.5,5)--(0.75,5);

\draw[decorate, decoration={triangles, segment length=3mm}]
 (0.75,5)--(0.75,4.5)--(0.875,4.5)--(0.875,3.5)--(1.125,3.5)--(1.125,2.5)--(1.75,2.5)--(1.75,2)--(2.25,2)--(2.25,1.75)--(2.5,1.75)--(2.5,1.25)--(3.5,1.25)--(3.5,1)--(5,1)--(5,5)--cycle;

\draw[decorate, decoration={crosses, segment length=2mm}]
(2.5,2.5)--(1.125,2.5)--(1.125,2)--(1.375,2)--(1.375,1.5)--(2,1.5)--(2,1.25)--(2.5,1.25)--cycle;

\draw [draw=black, fill=black, opacity=0.25]
       (0.5,0.5)--(0.5,2.5)--(1.125,2.5)--(1.125,2)--(1.375,2)--(1.375,1.5)--(2,1.5)--(2,1.25)--(2.5,1.25)-- (2.5,0.5) -- cycle;
\draw [draw=black, fill=black, opacity=0.1]
       (2.5,0.5) -- (2.5,1.25) --(3.5,1.25)--(3.5,1)--(5,1)--(5,0.5) -- cycle;

\draw [draw=black, fill=black, opacity=0.1]
        (0.5,2.5)--(0.5,5) --(0.75,5)--(0.75,4.5)--(0.875,4.5)--(0.875,3.5)--(1.125,3.5)--(1.125,2.5)-- cycle;

\draw [draw=black, fill=black, opacity=0.1]
(1.125,2.5)-- (1.125,2)--(1.375,2)--(1.375,1.5)--(2,1.5)--(2,1.25)--(2.5,1.25)--(2.5,1.75)--(2.25,1.75)--(2.25,2) --(1.75,2)-- (1.75,2.5)--cycle;

\node[] at (3.65,0) {{$\mathbb{C}^2\supset\ldots\supset\mathbb{C}^2\supset E^{[1]}\ldots E^{[1]}\supset 0 \supset\ldots\supset 0$}};

\node[] at (0,0.2) {{$\begin{rotate}{90}$\mathbb{C}^2\supset\ldots\supset\mathbb{C}^2\supset E^{[1]}\ldots E^{[1]}\supset 0 \supset\ldots\supset 0$\end{rotate}$}};

\node[] at (0.9,0.9) {\Large{$\mathbb{C}^2$}};
\node[] at (1.75,1.75) {{$E^{[1]}$}};
\node[] at (0.77,3) {{$E^{[1]}$}};
\node[] at (3,1) {{$E^{[1]}$}};
\node[] at (3.7,3.7) {\Large{$0$}};

\end{scope}

\end{tikzpicture}
	\caption{Example of a family of  bifiltrations.} \label{fam}
\end{figure}
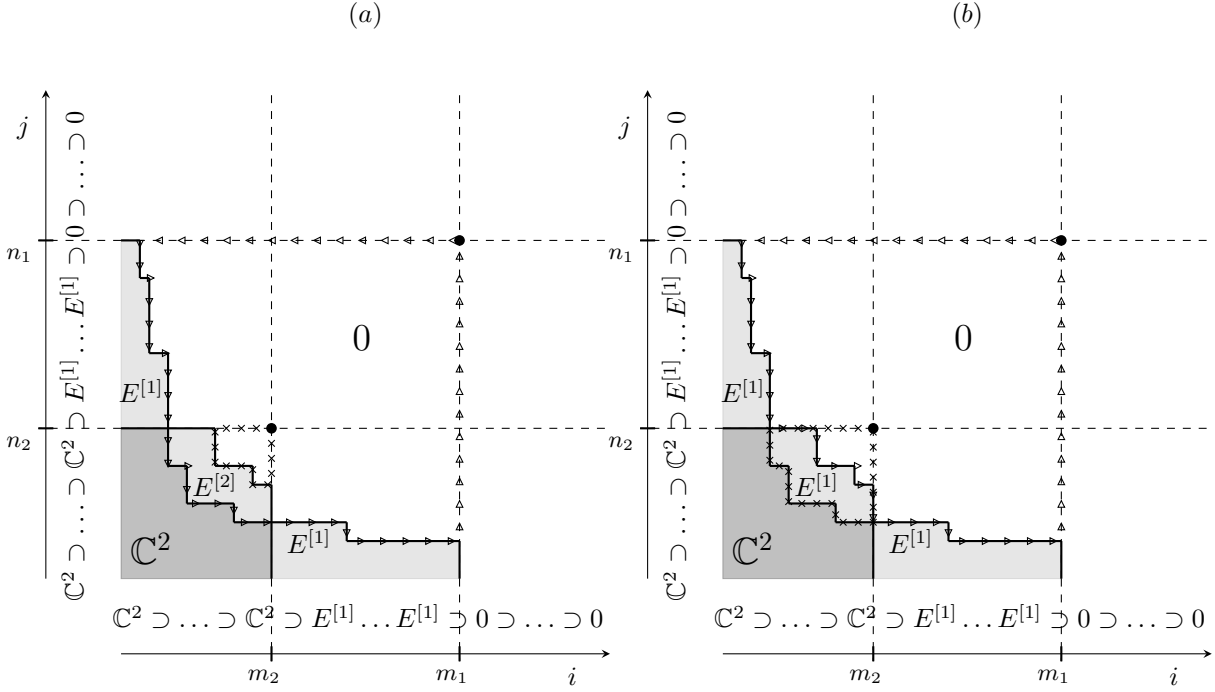
\begin{figure}
\centering
\begin{tikzpicture}

\draw[black,  dashed] (2.5,0.5)--(2.5,7);
\draw[black,  dashed] (5,0.5)--(5,7);
\draw[black,  dashed] (0.5,2.5)--(7,2.5);
\draw[black,  dashed] (0.5,5)--(7,5);

\draw [-stealth](0.5,-0.5) -- (7,-0.5);
\fill (2.5,5) circle[radius=2pt];
\fill (5,2.5) circle[radius=2pt];
\node[] at (6.5,-0.8) {{$i$}};
\draw[black,  dashed] (2.5,0.5)--(2.5,-0.6);
\draw[black,  thick]
(2.5,-0.6)--(2.5,-0.4);
\node[] at (2.4,-0.8) {\small{$m_2$}};
\draw[black,  dashed] (5,0.5)--(5,-0.6);
\draw[black,  thick] (5,-0.6)--(5,-0.4);
\node[] at (4.9,-0.8) {\small{$m_1$}};

\draw [-stealth](-0.5,0.5) -- (-0.5,7);
 \node[] at (-0.8,6.5) {{$j$}};
\draw[black,  dashed] (0.5,2.5)--(-0.6,2.5);
\draw[black,  thick]
(-0.6,2.5)--(-0.4,2.5);
\node[] at (-0.85,2.3) {\small{$\hat{n}_1$}};
\draw[black,  dashed] (0.5,5)--(-0.6,5);
\draw[black,  thick] (-0.6,5)--(-0.4,5);
\node[] at (-0.85,4.8) {\small{$\hat{n}_2$}};

\node[] at (3.75,8) {{$(a)$}};

\draw[black,  thick] (0.75,5)--(0.75,4.5);
\draw[black,  thick] (0.75,4.5)--(0.875,4.5);
\draw[black,  thick] (0.875,4.5)--(0.875,3.5);
\draw[black,  thick] (0.875,3.5)--(1.125,3.5);
\draw[black,  thick] (1.125,3.5)--(1.125,2);
\draw[black,  thick] (1.125,2)--(1.375,2);
\draw[black,  thick] (1.375,2)--(1.375,1.5);
\draw[black,  thick] (1.375,1.5)--(2,1.5);
\draw[black,  thick] (2,1.5)--(2,1.25);
\draw[black,  thick] (2,1.25)--(3.5,1.25);
\draw[black,  thick] (3.5,1.25)--(3.5,1);
\draw[black,  thick] (3.5,1)--(5,1);

\draw[black,  thick] (1.75,2.5)--(1.75,2);
\draw[black,  thick] (1.75,2)--(2.25,2);
\draw[black,  thick] (2.25,2)--(2.25,1.75);
\draw[black,  thick] (2.25,1.75)--(2.5,1.75);

\draw[black,  thick] (2.5,0.5)--(2.5,1.75);
\draw[black,  thick] (0.5,2.5)--(1.75,2.5);
\draw[black,  thick] (5,0.5)--(5,1);
\draw[black,  thick] (0.5,5)--(0.75,5);

\draw[decorate, decoration={triangles, segment length=3mm}]
 (0.75,5)--(0.75,4.5)--(0.875,4.5)--(0.875,3.5)--(1.125,3.5)--(1.125,2)--(1.375,2)--(1.375,1.5)--(2,1.5)--(2,1.25)--(2.5,1.25)--(2.5,5)--cycle;
 
\draw[decorate, decoration={crosses, segment length=2mm}]
(5,2.5)--(1.75,2.5)--(1.75,2)--(2.25,2)--(2.25,1.75)--(2.5,1.75)--(2.5,1.25)--(3.5,1.25)--(3.5,1)--(5,1)--cycle;

\draw [draw=black, fill=black, opacity=0.25]
       (0.5,0.5)--(0.5,2.5)--(1.125,2.5)--(1.125,2)--(1.375,2)--(1.375,1.5)--(2,1.5)--(2,1.25)--(2.5,1.25)-- (2.5,0.5) -- cycle;
\draw [draw=black, fill=black, opacity=0.1]
       (2.5,0.5) -- (2.5,1.25) --(3.5,1.25)--(3.5,1)--(5,1)--(5,0.5) -- cycle;

\draw [draw=black, fill=black, opacity=0.1]
        (0.5,2.5)--(0.5,5) --(0.75,5)--(0.75,4.5)--(0.875,4.5)--(0.875,3.5)--(1.125,3.5)--(1.125,2.5)-- cycle;

\draw [draw=black, fill=black, opacity=0.1]
(1.125,2.5)-- (1.125,2)--(1.375,2)--(1.375,1.5)--(2,1.5)--(2,1.25)--(2.5,1.25)--(2.5,1.75)--(2.25,1.75)--(2.25,2) --(1.75,2)-- (1.75,2.5)--cycle;

\node[] at (3.65,0) {{$\mathbb{C}^2\supset\ldots\supset\mathbb{C}^2\supset \tilde{E}^{[1]}\ldots \tilde{E}^{[1]}\supset 0 \supset\ldots\supset 0$}};

\node[] at (0,0.2) {{$\begin{rotate}{90}$\mathbb{C}^2\supset\ldots\supset\mathbb{C}^2\supset \tilde{E}^{[2]}\ldots \tilde{E}^{[2]}\supset 0 \supset\ldots\supset 0$\end{rotate}$}};

\node[] at (0.9,0.9) {\Large{$\mathbb{C}^2$}};
\node[] at (1.75,1.75) {{$\tilde{E}^{[1]}$}};
\node[] at (0.77,3) {{$\tilde{E}^{[2]}$}};
\node[] at (3,1) {{$\tilde{E}^{[1]}$}};
\node[] at (3.7,3.7) {\Large{$0$}};

\begin{scope}[shift={(8,0)}]

\draw[black,  dashed] (2.5,0.5)--(2.5,7);
\draw[black,  dashed] (5,0.5)--(5,7);
\draw[black,  dashed] (0.5,2.5)--(7,2.5);
\draw[black,  dashed] (0.5,5)--(7,5);

\draw [-stealth](0.5,-0.5) -- (7,-0.5);
\fill (2.5,5) circle[radius=2pt];
\fill (5,2.5) circle[radius=2pt];
\node[] at (6.5,-0.8) {{$i$}};
\draw[black,  dashed] (2.5,0.5)--(2.5,-0.6);
\draw[black,  thick]
(2.5,-0.6)--(2.5,-0.4);
\node[] at (2.4,-0.8) {\small{$m_2$}};
\draw[black,  dashed] (5,0.5)--(5,-0.6);
\draw[black,  thick] (5,-0.6)--(5,-0.4);
\node[] at (4.9,-0.8) {\small{$m_1$}};

\draw [-stealth](-0.5,0.5) -- (-0.5,7);
 \node[] at (-0.8,6.5) {{$j$}};
\draw[black,  dashed] (0.5,2.5)--(-0.6,2.5);
\draw[black,  thick]
(-0.6,2.5)--(-0.4,2.5);
\node[] at (-0.85,2.3) {\small{$\hat{n}_1$}};
\draw[black,  dashed] (0.5,5)--(-0.6,5);
\draw[black,  thick] (-0.6,5)--(-0.4,5);
\node[] at (-0.85,4.8) {\small{$\hat{n}_2$}};

\node[] at (3.75,8) {{$(b)$}};

\draw[black,  thick] (0.75,5)--(0.75,4.5);
\draw[black,  thick] (0.75,4.5)--(0.875,4.5);
\draw[black,  thick] (0.875,4.5)--(0.875,3.5);
\draw[black,  thick] (0.875,3.5)--(1.125,3.5);
\draw[black,  thick] (1.125,3.5)--(1.125,2);
\draw[black,  thick] (1.125,2)--(1.375,2);
\draw[black,  thick] (1.375,2)--(1.375,1.5);
\draw[black,  thick] (1.375,1.5)--(2,1.5);
\draw[black,  thick] (2,1.5)--(2,1.25);
\draw[black,  thick] (2,1.25)--(3.5,1.25);
\draw[black,  thick] (3.5,1.25)--(3.5,1);
\draw[black,  thick] (3.5,1)--(5,1);

\draw[black,  thick] (1.75,2.5)--(1.75,2);
\draw[black,  thick] (1.75,2)--(2.25,2);
\draw[black,  thick] (2.25,2)--(2.25,1.75);
\draw[black,  thick] (2.25,1.75)--(2.5,1.75);

\draw[black,  thick] (2.5,0.5)--(2.5,1.75);
\draw[black,  thick] (0.5,2.5)--(1.75,2.5);
\draw[black,  thick] (5,0.5)--(5,1);
\draw[black,  thick] (0.5,5)--(0.75,5);

\draw[decorate, decoration={triangles, segment length=3mm}]
 (2.5,5)--(0.75,5)--(0.75,4.5)--(0.875,4.5)--(0.875,3.5)--(1.125,3.5)--(1.125,2.5)--(1.75,2.5)--(1.75,2)--(2.25,2)--(2.25,1.75)--(2.5,1.75)--cycle;
  
\draw[decorate, decoration={crosses, segment length=2mm}]
(5,2.5)--(1.125,2.5)--(1.125,2)--(1.375,2)--(1.375,1.5)--(2,1.5)--(2,1.25)--(3.5,1.25)--(3.5,1)--(5,1)--cycle;

\draw [draw=black, fill=black, opacity=0.25]
       (0.5,0.5)--(0.5,2.5)--(1.125,2.5)--(1.125,2)--(1.375,2)--(1.375,1.5)--(2,1.5)--(2,1.25)--(2.5,1.25)-- (2.5,0.5) -- cycle;
\draw [draw=black, fill=black, opacity=0.1]
       (2.5,0.5) -- (2.5,1.25) --(3.5,1.25)--(3.5,1)--(5,1)--(5,0.5) -- cycle;

\draw [draw=black, fill=black, opacity=0.1]
        (0.5,2.5)--(0.5,5) --(0.75,5)--(0.75,4.5)--(0.875,4.5)--(0.875,3.5)--(1.125,3.5)--(1.125,2.5)-- cycle;

\draw [draw=black, fill=black, opacity=0.1]
(1.125,2.5)-- (1.125,2)--(1.375,2)--(1.375,1.5)--(2,1.5)--(2,1.25)--(2.5,1.25)--(2.5,1.75)--(2.25,1.75)--(2.25,2) --(1.75,2)-- (1.75,2.5)--cycle;

\node[] at (3.65,0) {{$\mathbb{C}^2\supset\ldots\supset\mathbb{C}^2\supset \tilde{E}^{[1]}\ldots \tilde{E}^{[1]}\supset 0 \supset\ldots\supset 0$}};

\node[] at (0,0.2) {{$\begin{rotate}{90}$\mathbb{C}^2\supset\ldots\supset\mathbb{C}^2\supset \tilde{E}^{[2]}\ldots \tilde{E}^{[2]}\supset 0 \supset\ldots\supset 0$\end{rotate}$}};

\node[] at (0.9,0.9) {\Large{$\mathbb{C}^2$}};
\node[] at (1.75,1.75) {{$\tilde{E}^{[2]}$}};
\node[] at (0.77,3) {{$\tilde{E}^{[2]}$}};
\node[] at (3,1) {{$\tilde{E}^{[1]}$}};
\node[] at (3.7,3.7) {\Large{$0$}};

\end{scope}

\end{tikzpicture}
	\caption{Example of a family of  bifiltrations dual to the family on the Fig. \ref{fam}.} \label{dualfam}
\end{figure}
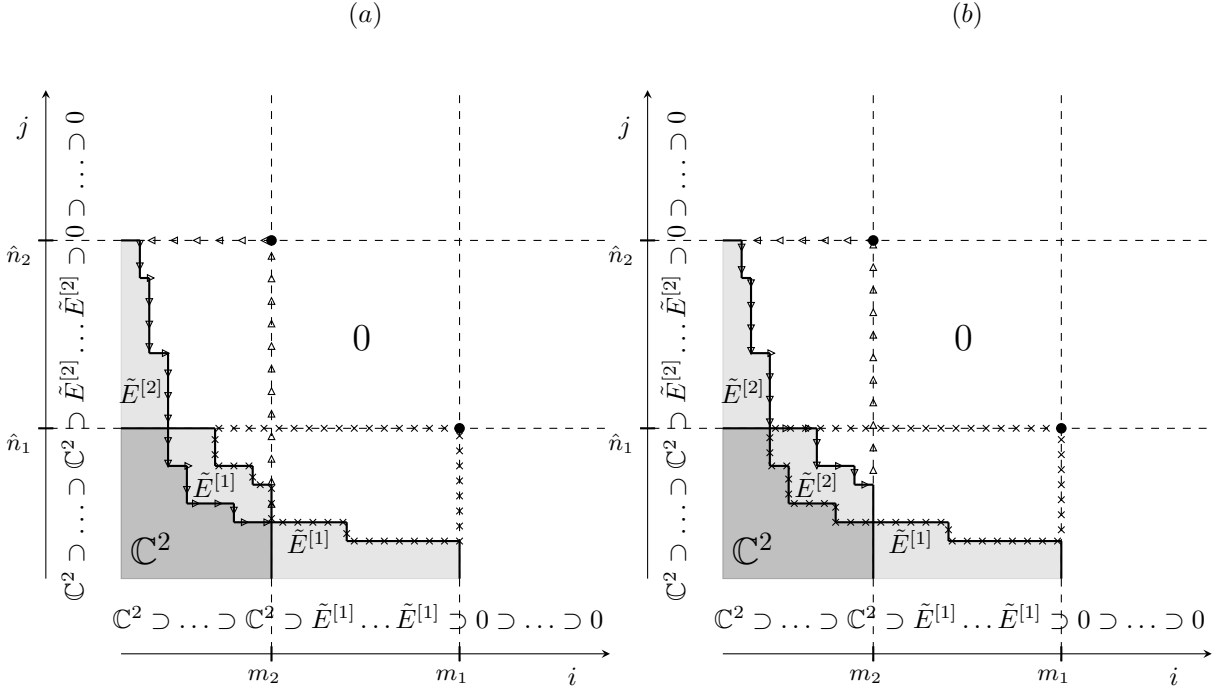
 
 \par Generalisation to the higher rank is straightforward. For a fixed couple $u,v\in\{1,\ldots,N\}$ we expect that the families consist of the sheaves producing identical representations of the group $T^{(uv)}$. We
 can decompose the space $E$ into two parts, 
 $$
 E=E'+E'', \, E'=E^{[u]}\oplus E^{[v]}, \, E''=\bigoplus_{w\neq u,v} E^{[w]},
 $$
 and note that representation of $T$ carried by $E''$ can be uniquely reconstructed from the representation of $T^{(uv)}$ carried by $E''$, since the broken factor by construction does not affect $E''$. For $E'$ the action of the gauge factor $C^{*(N-1)}$ is diagonal and fixed and hence only the geometric group $\mathbb{C}^{*2}$ is relevant. We set 
 $$
 B_{i,j}=B'_{i,j}+B''_{i,j},\, B'_{i,j}=B_{i,j}\cap E'\, B''_{i,j}=B_{i,j}\cap E''
 $$
 and conclude that the members of the same family (and its dual) should have identical bifiltrations $B''_{i,j}$ and dimensions $\dim B'_{i,j}$. In other words, $Y_u$ and $Y_v$ should be related in the same way as in $N=2$ case and the rest $N-2$ of the Young diagrams should coincide in a family and its dual.

\paragraph{Remark.}

 The bifiltration $B_{i,j}$ is very similar to the ones appearing in Klyachko classification of equivariant sheaves on compact toric varieties \cite{Klyachko, KnutsonSharp}. It is not a coincidence.  In fact, in this classification the sheaves on $\mathbb{CP}^2$ are described by triples of bifiltrations, one for each fixed point of $\mathbb{CP}^2$ with respect to $\mathbb{C}^{*2}$. But since we consider framed sheaves, which are trivial on one of the divisors, only one of the bifiltration survives. The twisting (\ref{twist}) then prescribes a non-trivial equivariantization of this trivial sheaf. However the decomposition (\ref{BexplicitId}) determining the framing does not appear in the Klyachko's construction.
 \par 
 The similarity of our approach with Klyachko classification has yet another interpretation. 
 The equivariant torsion-free sheaves on compact toric varieties are the fixed point in the geometric approach to the computation of $\mathcal{N}=2$ gauge theory partition function. At the same time there are a lot of indications that this partition function can expressed via products of the shifted $\mathbb{C}^2$ partition functions \cite{GNY, Bonelli, Bershtein}. Apparently there is a correspondence between the bifiltrations, describing the equivariant torsion free sheaves on the compact toric variety, and bifiltrations, assigned to the terms of shifted partition function as above.     
 \par 
 Finally, it is worth noting that the relation between the representations of the symmetry group acting on the tangent space of the moduli space of sheaves at a fixed point and on the space of sections of the fixed point sheaves, very similar to the ones lying behind our reasoning, was found in \cite{Klyachko}.
 
\subsection{Proof via the dual families of Young diagrams} \label{rigorousproof}

In this subsection we rigorously formulate the refinement (\ref{veryrefined}) and prove it and hence (\ref{zinstrelation}). We start with the pure theory and then add the matter hypermultiplets to the consideration.

\paragraph{Taking the integrals.} First let us look at the integral form of $Z_k^{({\rm 0})}$ (\ref{Zk}) and define the variables of integration around the poles
\begin{equation}
    \xi_I=\phi_I-\Phi_I,
\end{equation}
so the integrals with respect to $\xi_I$ go around the zeros. The contours are chosen to be  circles with centres at the origin and radii $r_I=\alpha_I \delta_1 + \beta_I \delta_2$, where $(\alpha_I,\beta_I)$ are the coordinates of the cell $I$ as defined below (\ref{statpoints}), and $\delta_1/\delta_2 > k$ or the other way round. In that way the contours of integration over $\xi_I$, $\xi_J$ with $I,J$ belonging to the same diagram do not intersect, while the integration over $\xi_I$, $\xi_J$ with $I,J$ belonging to the different diagrams is independent since there are no poles with respect to $\xi_{IJ}$.
\par We start the proof with the $N=2$ case. We set $a_{12}=a$, $\hat{a}^{(12)}_{12}=\hat{a}$ and denote by $Z_k(Y_1,Y_2)$ a contribution to $Z^{({\rm 0})}_k$ coming from the pole marked by the Young diagrams $Y_1,Y_2$. 
\begin{eqnarray}
    Z_k (Y_1,Y_2)=\frac{\epsilon^k}{(2\pi {\rm i} \epsilon_1 \epsilon_2)^k} \oint \prod_{J \in Y_1}\prod_{I \in Y_2} {\rm d}\xi_J{\rm d}\xi_I f_J(0,\xi_J)f_J(a,\xi_J) f_I(-a, \xi_I) f_I(0, \xi_I) \cdot \\
   \mathcal{W}_{JI} (a,\xi_{JI}) \left[ \prod_{\underset{T \neq J}{T \in Y_1}} \mathcal{W}_{JT} (0,\xi_{JT}) \right] \left[ \prod_{\underset{T \neq I}{T\in Y_2}} \mathcal{W}_{TI} (0,\xi_{TI}) \right], \nonumber
\end{eqnarray}
where
\begin{flalign}
 &f_I(a, \xi_I)=[(a-\epsilon_1(\alpha_I-1)-\epsilon_2(\beta_I-1)+\xi_I)(-a+ \epsilon_1\alpha_I+\epsilon_2\beta_I-\xi_I)]^{-1} \nonumber
\end{flalign}
and
\begin{flalign}
   &\mathcal{W}_{JI}(a_{JI},  \xi_{JI})  =\frac{\phi_{JI}^2(\phi_{JI}^2-\epsilon^2)}{(\phi_{JI}^2-\epsilon_1)^2(\phi_{JI}^2-\epsilon_2)^2},  \\
    &\phi_{JI} =a_{JI}-\epsilon_1(\alpha_J-\alpha_I)-\epsilon_2(\beta_J-\beta_I)+\xi_{JI}. \nonumber
\end{flalign}
Note for future that we can shift the indices and the arguments simultaneously
\begin{equation} \label{fsym}
     f_{(\alpha_I,\beta_I)}(a,\xi_I)= f_{(\alpha_I+m,\beta_I+n)}(a+m\epsilon_1+n\epsilon_2,\xi_I),
\end{equation}
\begin{eqnarray} \label{wsym}
   \mathcal{W}_{(\alpha_J,\beta_J)(\alpha_I,\beta_I)}(a,  \xi_{JI}) &=&\mathcal{W}_{(\alpha_J+m,\beta_J+n)(\alpha_I,\beta_I)}(a+m\epsilon_1+n\epsilon_2,  \xi_{JI})\\
    &=& \mathcal{W}_{(\alpha_J,\beta_J)(\alpha_I+m,\beta_I+n)}(a-m\epsilon_1-n\epsilon_2,  \xi_{JI})\nonumber
\end{eqnarray}
We will further refer to factors $\mathcal{W}_{IJ}$ as the interaction factors. There is interaction between every pair $I \neq J$ of $k$ cells.
\par The poles with respect to all $\xi_I$ are simple. To see that let us look at the integrals with respect to $\xi_I$ with $I$ running through the cells of one of the Young diagrams. We evaluate the integrals one by one, starting with the contour closest to zero. There is a simple pole with respect to $\xi_{(1,1)}$ coming from $f_{(1,1)}(0,\xi_{(1,1)})$, while the interaction factors do not contain any poles since all the rest of the variables $\xi_J$ are separated from the origin.
\par As soon as we compute the first integral we set $\xi_{(1,1)}=0$ in the interaction factors, and the poles with respect to two more variables come from the interaction, namely $\xi_{(2,1)}$ and $\xi_{(1,2)}$. The poles appear to be simple again and we can easily take the integrals.
\par When it comes to the integration over $\xi_{(2,2)}$, we see that there is a double zero in the denominator coming from the interaction with the boxes $(1,2)$ and $(2,1)$ and a zero in the numerator coming from the interaction with the box $(1,1)$. Therefore the pole is simple again.
\par The pattern repeats on the next steps and we always see that there are single poles with respect to the variables $\xi_{(1,K)}$ and $\xi_{(K,1)}$ coming from the interaction with the boxes $(1,K-1)$ and $(K-1,1)$ correspondingly, while with respect to the variables $\xi_{(L>1,K>1)}$ we have simple poles combined from the interactions with $(L,K-1)$, $(L-1,K)$ in denominator and $(L-1,K-1)$ in numerator. Thus taking all the integrals in (\ref{zky1y2}) with respect to $\xi_I$ marking the boxes in both Young diagrams we just get
\begin{eqnarray} \label{zky1y2}
    Z_k (Y_1,Y_2)=\frac{\epsilon^k}{(\epsilon_1 \epsilon_2)^k} \prod_{J \in Y_1}\prod_{I \in Y_2} f_J(a, 0)\bar{f}_J(0, 0) \bar{f}_I(0, 0) f_I(-a, 0) \\ \nonumber
   \mathcal{W}_{JI} (a,0) \left[ \overline{\prod}_{\underset{T \neq J}{T \in Y_1}} \mathcal{W}_{JT} (0,0) \right] \left[ \overline{\prod}_{\underset{T \neq I}{T\in Y_2}} \mathcal{W}_{TI} (0,0) \right],
\end{eqnarray}
where $\bar{f}_I(0,0)$ stands for $f_I(0,0)$ with the omitted multiplier $(-\epsilon_1(\alpha_I-1)-\epsilon_2(\beta_I-1))^{-1}$ with $I=(1,1)$ and $\overline{\prod}$ stands for the product with all the zeros in numerator and denominator omitted. (Note that as soon as we omit some factors in $\mathcal{W}_{JT}$, it is no longer symmetrical with respect to the indices permutation and one has to keep the order of indices in agreement with the integration over $\xi_I$.)
\par Since the dependence on $\xi_I$ vanished after the integration we will further omit the second argument of $f_I$ and $\mathcal{W}_{IJ}$.

\paragraph{The main idea of the proof.} As we will see soon $Z_k(Y_1,Y_2)$ has a pole at $a=\bar{a}=\epsilon_{m,n}$ only if $Y_1$ contain the box $(m,n)$. So let us look at the simplest nontrivial case, namely $Y_2,\, \tilde{Y}_1,\, \tilde{Y}_2= \o$ and $Y_1$ be a rectangle of size $m \times n$, which we denote as $Y_1 = \square$ for convenience. We will denote the dual point as $\hat{\bar{a}}=\epsilon_{m,-n}$. We would like to show that
\begin{equation} \label{rel1}
    {\rm Res}_{a=\bar{a}} Z_k (Y_1,Y_2)= \frac{1}{\mathcal{P}_{2}(m,n)}Z_{k-mn} (\tilde{Y}_1,\tilde{Y}_2)|_{a=\hat{\bar{a}}},
\end{equation}
We omitted the argument $a$ of $\mathcal{P}_{2}(m,n)$ since in the simplest case $N=2$ it is actually a numerical coefficient and not a polynomial.
\par As one can see $Z_{k-mn} (\tilde{Y}_1,\tilde{Y}_2)=Z_{0} (\o,\o)=1$ and from (\ref{zky1y2})
\begin{eqnarray} \label{zksquare}
    Z_k (Y_1,Y_2)=Z_{mn} (\square,\o)=\frac{\epsilon^{mn}}{(\epsilon_1 \epsilon_2)^{mn}}\prod_{I \in \square} \bar{f}_I(0) f_I(a)
   \left[ \overline{\prod}_{\underset{T \neq I}{T\in \square}} \mathcal{W}_{TI} (0) \right],
\end{eqnarray}
The pole with respect to $a$ is simple and taking the residue one gets    
\begin{equation}
    {\rm Res}_{a=\bar{a}} Z_k (Y_1,Y_2)=\prod_{i=-m+1}^{m}\sideset{}{'} \prod_{j=-n+1}^{n}  \left(-\epsilon_{i,j} \right)^{-1}= \prod_{i=-m}^{m-1}\sideset{}{'} \prod_{j=-n}^{n-1} \left(\epsilon_{i,j} \right)^{-1},
\end{equation}
which is exactly $\mathcal{P}_{2}(m,n)^{-1}$, so (\ref{rel1}) is verified.

\par
We can always separate in (\ref{zky1y2}) the factors combining in $Z_{mn}(\square,\o)$. Although $Z_k(Y_1,Y_2)$ in general can have a higher order pole at $a=\bar{a}$, we will show soon that we can group all $(Y_1,Y_2)$ in families in such a way that the sum of $Z_k(Y_1,Y_2)$ over the family $\mathcal{F}$ has only a simple pole at this point and it appears in the factor $Z_{mn}(\square,\o)$. Therefore \begin{equation}
    {\rm Res}_{a=\bar{a}} \sum_{Y_1,Y_2 \in \mathcal{F}} Z_k (Y_1,Y_2)= \frac{1}{\mathcal{P}_{2}(m,n)} \frac{\epsilon^{k-mn}}{(\epsilon_1 \epsilon_2)^{k-mn}} \cdot \Sigma_1,
\end{equation}
\begin{flalign} \nonumber
    &\Sigma_1=\bigg( \sum_{Y_1,Y_2 \in \mathcal{F}}\prod_{I \in Y_1 \setminus \square} \prod_{J \in Y_2}  f_I(0) f_I(\bar{a}+\alpha)f_J(-\bar{a}-\alpha)\bar{f}_J(0)\cdot \\
   \mathcal{W}_{JI} (\bar{a}+\alpha) &  \left[ \overline{\prod}_{\underset{T \neq J}{T \in Y_2}} \mathcal{W}_{JT} (0) \right] \left[ \overline{\prod}_{\underset{T \neq I}{T\in Y_1 \setminus \square}} \mathcal{W}_{TI} (0) \right]\left[ \overline{\prod}_{T\in \square} \mathcal{W}_{IT} (0) \mathcal{W}_{JT} (\bar{a}+\alpha) \right] \bigg)\bigg|_{\alpha=0}  \nonumber 
\end{flalign}
Note that $f_I(0)$, $I \in Y_1 \setminus \square$ does not have any factors omitted since the box $(1,1)$ is already included in $Z_{mn}(\square,\o)$.
\par On the other hand we will group all $Z_{k-mn}(\tilde{Y}_1,\tilde{Y}_2)$ in corresponding sums over dual families $\bar{\mathcal{F}}$ and show that a sum over a dual family $\bar{\mathcal{F}}$ is regular at $\hat{\bar{a}}$ (although an individual term $Z_{k-mn}(\tilde{Y}_1,\tilde{Y}_2)$ can be singular at this point)
\begin{equation}
    \sum_{\tilde{Y}_1,\tilde{Y}_2 \in \bar{\mathcal{F}}}Z_{k-mn}(\tilde{Y}_1,\tilde{Y}_2)|_{a=\hat{\bar{a}}}= \frac{\epsilon^{k-mn}}{(\epsilon_1 \epsilon_2)^{k-mn}} \Sigma_2
\end{equation}
\begin{flalign} \nonumber
    \Sigma_2= \bigg(\sum_{\tilde{Y}_1,\tilde{Y}_2 \in \bar{\mathcal{F}}} \prod_{I\in \tilde{Y}_1} \prod_{J\in \tilde{Y}_2}    \, \bar{f}_I(0) f_I(\hat{\bar{a}}+\alpha) f_J(-\hat{\bar{a}}-\alpha)\bar{f}_J(0) \cdot \\ \mathcal{W}_{JI}(\hat{\bar{a}}+\alpha) \left[ \overline{\prod}_{\underset{I \neq T}{T \in \tilde{Y}_1}} \mathcal{W}_{TJ}(0) \right]\left[ \overline{\prod}_{\underset{J \neq T}{T \in \tilde{Y}_2}} \mathcal{W}_{TI}(0) \right] \bigg) \bigg|_{\alpha=0} . \nonumber 
\end{flalign}
We will show then that for a dual pair $\mathcal{F}$, $\bar{\mathcal{F}}$ the sums coincide, $\Sigma_1=\Sigma_2$
and thus will prove (\ref{zinstrelation}) for $N=2$ case.

\paragraph{Families and dual families of Young diagrams.}
 The way of grouping the pairs of Young diagrams in families is dictated by the correspondence formulated in the end of Subsection \ref{Refined}. We gather in a family all such pairs that their corresponding bifiltrations have the same dimension at all positions. 
 \par
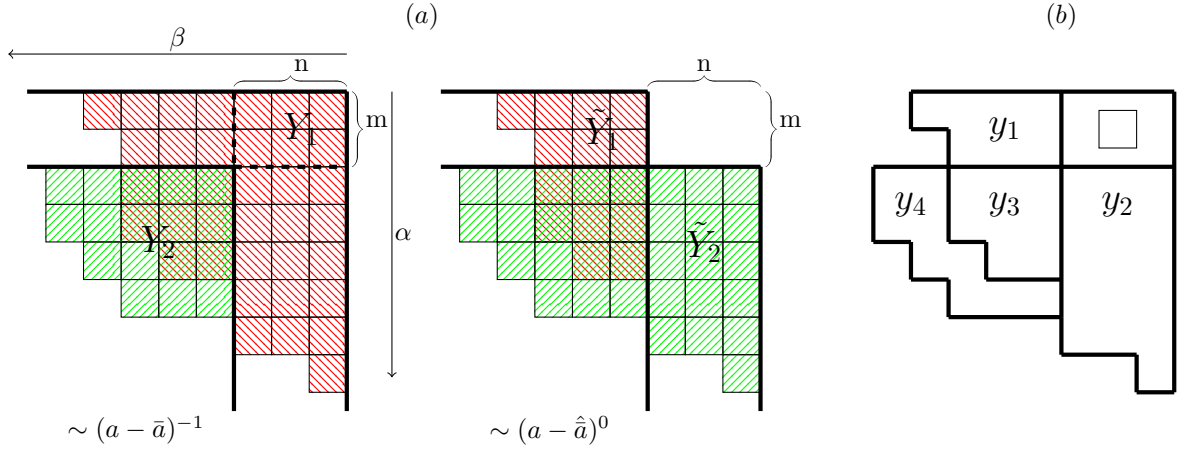
\begin{figure}
\centering
\begin{tikzpicture}

\node[] at (1,1) {{$(a)$}};
\drawYT{color=black,opacity=1, pattern=north west lines, pattern color=red, thin}{0}{0}{-0.5}{-0.5}{7,6,6,6,5,3,3,1}
\drawYT{color=black,opacity=1, pattern=north east lines, pattern color=green, thin}{-1.5}{-1}{-0.5}{-0.5}{5,5,4,3} 
 \draw[black, ultra thick] (0,0)--(0,-4.25);
  \draw[black, ultra thick] (0,0)--(-4.25,0);
   \draw[black, ultra thick] (-1.5,-1)--(-1.5,-4.25);
  \draw[black, ultra thick] (-1.5,-1)--(-4.25,-1);
        \draw[black, ultra thick,dashed](-1.5,0)--(-1.5,-1);
      \draw[black, ultra thick,dashed](-1.5,-1)--(0,-1);
     \draw[->](0,0.5)--(-4.5,0.5);
     \node[] at (-2.25,0.7) {$\beta$};
      \draw[->](0.6,0)--(0.6,-3.8);
     \node[] at (0.75,-1.9) {$\alpha$};
      
      \draw [decorate, 
decoration = {calligraphic brace, raise = 2pt, amplitude = 4pt}] (-1.5,0) --  (0,0);
\node[] at (-0.6,0.35) {n};
\draw [decorate, decoration = {calligraphic brace, raise = 2pt, amplitude = 4pt}] (0,0)--(0,-1);
\node[] at (0.4,-0.4) {m};
  \node[] at (-0.6,-0.5) {\Large{$Y_1$}};
\node[] at (-2.5,-2) {\Large{$Y_2$}};

\node[] at (-2.8,-4.5) {{$\sim(a-\bar{a})^{-1}$}};
\node[] at (2.7,-4.5) {{$\sim(a-\hat{\bar{a}})^{0}$}};

\begin{scope}[shift={(-2.5,0)}]
\drawYT{color=black,opacity=1, pattern=north west lines, pattern color=red, thin}{6.5}{0}{-0.5}{-0.5}{4,3,3,3,2}
\drawYT{color=black,opacity=1, pattern=north east lines, pattern color=green, thin}{8}{-1}{-0.5}{-0.5}{8,8,7,6,3,1} 
 \draw[black, ultra thick] (6.5,0)--(6.5,-4.25);
  \draw[black, ultra thick] (6.5,0)--(3.75,0);
   \draw[black, ultra thick] (8,-1)--(8,-4.25);
  \draw[black, ultra thick] (8,-1)--(3.75,-1);
        
      \draw [decorate, 
decoration = {calligraphic brace, raise = 2pt, amplitude = 4pt}] (6.5,0) --  (8,0);
\node[] at (7.25,0.35) {n};
\draw [decorate, decoration = {calligraphic brace, raise = 2pt, amplitude = 4pt}] (8,0)--(8,-1);
\node[] at (8.4,-0.4) {m};
  \node[] at (5.9,-0.5) {\Large{$\tilde{Y}_1$}};
\node[] at (7.25,-2) {\Large{$\tilde{Y}_2$}};
\end{scope}

\node[] at (9.5,1) {{$(b)$}};
\begin{scope}[shift={(11,0)}]

\draw[black, ultra thick] (-1.5,0)--(-1.5,-3);
 \draw[black, ultra thick] (-3,-1)--(0,-1);
 \draw[black, ultra thick] (-3.5,0)--(-3.5,-0.5);
 \draw[black, ultra thick] (-3.5,-0.5)--(-3,-0.5);
\draw[black, ultra thick] (-3,-0.5)--(-3,-2);
\draw[black, ultra thick] (-3,-2)--(-2.5,-2);
\draw[black, ultra thick] (-2.5,-2)--(-2.5,-2.5);
\draw[black, ultra thick] (-2.5,-2.5)--(-1.5,-2.5);
\draw[black, ultra thick] (-1.5,-3)--(-1.5,-3.5);
\draw[black, ultra thick] (-1.5,-3)--(-3,-3);
\draw[black, ultra thick] (-3,-3)--(-3,-2.5);
\draw[black, ultra thick] (-3,-2.5)--(-3.5,-2.5);
\draw[black, ultra thick] (-3.5,-2.5)--(-3.5,-2);
\draw[black, ultra thick] (-3.5,-2)--(-4,-2);
\draw[black, ultra thick] (-4,-2)--(-4,-1);
\draw[black, ultra thick] (-4,-1)--(-3,-1);
\draw[black, ultra thick] (-1.5,-3.5)--(-0.5,-3.5);
\draw[black, ultra thick] (-0.5,-3.5)--(-0.5,-4);
\draw[black, ultra thick] (-0.5,-4)--(0,-4);
\draw[black, ultra thick] (0,-4)--(0,0);
\draw[black, ultra thick] (0,0)--(-3.5,0);
 \draw[black] (-1,-0.25) rectangle (-0.5,-0.75);
\node[] at (-2.25,-0.5) {\Large{$y_1$}};
\node[] at (-0.75,-1.5) {\Large{$y_2$}};
\node[] at (-2.25,-1.5) {\Large{$y_3$}};
\node[] at (-3.5,-1.5) {\Large{$y_4$}};

\end{scope}

\end{tikzpicture}
	\caption{(a) Example of a family of one member with the correspondent exponent of $(a-\bar{a})$ and its dual family with the correspondent exponent of $(a-\hat{\bar{a}})$. There is an overlap, but no blinking group. (b) Overline of subregions of the diagrams.} \label{overlap}
\end{figure}

 In terms of diagrams the general recipe of combining pairs $(Y_1,Y_2)$ in families is the following.
 We shift the origin of diagram $Y_2$ on $m$ cells in positive vertical direction and $n$ cells in positive horizontal direction with respect to the origin of $Y_1$ (see Fig.\ref{overlap} (a)). We introduce an occupation number for a cell, which is $0$ if a cell does not belong to any diagram, $1$ if it belongs to one diagram, and $2$ if it belongs to both. Then a family is formed by all the pairs $(Y_1,Y_2)$ which have the same occupation numbers for all the cells.

\par
\begin{figure}

\begin{tikzpicture}

\drawYT{color=black,opacity=1, pattern=north west lines, pattern color=red, thin}{0}{0}{-0.5}{-0.5}{7,6,6,6,5,3,1}
\drawYT{color=black,opacity=1, pattern=north east lines, pattern color=green, thin}{-1.5}{-1}{-0.5}{-0.5}{4,2,1,1,1} 
 \draw[black, ultra thick] (0,0)--(0,-3.75);
  \draw[black, ultra thick] (0,0)--(-3.75,0);
   \draw[black, ultra thick] (-1.5,-1)--(-1.5,-3.75);
  \draw[black, ultra thick] (-1.5,-1)--(-3.75,-1);
\node[] at (-0.5,-0.8) {\Large{$Y_1$}};
\node[] at (-2,-1.5) {\Large{$Y_2$}};

\drawYT{color=black,opacity=1, pattern=north west lines, pattern color=red, thin}{4}{0}{-0.5}{-0.5}{7,6,6,5,5,3,1}
\drawYT{color=black,opacity=1, pattern=north east lines, pattern color=green, thin}{2.5}{-1}{-0.5}{-0.5}{4,3,1,1,1} 
 \draw[black, ultra thick] (4,0)--(4,-3.75);
  \draw[black, ultra thick] (4,0)--(0.25,0);
   \draw[black, ultra thick] (2.5,-1)--(2.5,-3.75);
  \draw[black, ultra thick] (2.5,-1)--(0.25,-1);
  \node[] at (3.5,-0.8) {\Large{$Y_1$}};
\node[] at (2,-1.5) {\Large{$Y_2$}};

  \drawYT{color=black,opacity=1, pattern=north west lines, pattern color=red, thin}{8}{0}{-0.5}{-0.5}{7,6,6,6,4,3,1}
    \drawYT{color=black,opacity=1, pattern=north east lines, pattern color=green, thin}{6.5}{-1}{-0.5}{-0.5}{4,2,2,1,1}
 \draw[black, ultra thick] (8,0)--(8,-3.75);
  \draw[black, ultra thick] (8,0)--(4.25,0);
   \draw[black, ultra thick] (6.5,-1)--(6.5,-3.75);
  \draw[black, ultra thick] (6.5,-1)--(4.25,-1);
  \node[] at (7.5,-0.8) {\Large{$Y_1$}};
\node[] at (6,-1.5) {\Large{$Y_2$}};
  
   \drawYT{color=black,opacity=1, pattern=north west lines, pattern color=red, thin}{12}{0}{-0.5}{-0.5}{7,6,6,5,4,3,1}
    \drawYT{color=black,opacity=1, pattern=north east lines, pattern color=green, thin}{10.5}{-1}{-0.5}{-0.5}{4,3,2,1,1}
 \draw[black, ultra thick] (12,0)--(12,-3.75);
  \draw[black, ultra thick] (12,0)--(8.25,0);
   \draw[black, ultra thick] (10.5,-1)--(10.5,-3.75);
  \draw[black, ultra thick] (10.5,-1)--(8.25,-1);
 \node[] at (11.5,-0.8) {\Large{$Y_1$}};
\node[] at (10,-1.5) {\Large{$Y_2$}}; 
  
  \drawYT{color=black,opacity=1, pattern=north west lines, pattern color=red, thin}{-1.5}{-5}{-0.5}{-0.5}{4,3,3,3,2,1,1}
\drawYT{color=black,opacity=1, pattern=north east lines, pattern color=green, thin}{0}{-6}{-0.5}{-0.5}{7,5,4,3,1} 
 \draw[black, ultra thick] (-1.5,-5)--(-3.75,-5);
  \draw[black, ultra thick] (-1.5,-5)--(-1.5,-8.75);
   \draw[black, ultra thick] (0,-6)--(-3.75,-6);
  \draw[black, ultra thick] (0,-6)--(0,-8.75);
\node[] at (-2,-5.5) {\Large{$\tilde{Y_1}$}};
\node[] at (-0.6,-6.5) {\Large{$\tilde{Y_2}$}};

\node[] at (-2.8,-4) {{$\sim(a-\bar{a})^{-5}$}};
\node[] at (1.2,-4) {{$\sim(a-\bar{a})^{-5}$}};
\node[] at (5.2,-4) {{$\sim(a-\bar{a})^{-5}$}};
\node[] at (9.2,-4) {{$\sim(a-\bar{a})^{-5}$}};

  \drawYT{color=black,opacity=1, pattern=north west lines, pattern color=red, thin}{2.5}{-5}{-0.5}{-0.5}{4,3,3,2,2,1,1}
\drawYT{color=black,opacity=1, pattern=north east lines, pattern color=green, thin}{4}{-6}{-0.5}{-0.5}{7,6,4,3,1} 
 \draw[black, ultra thick] (2.5,-5)--(0.25,-5);
  \draw[black, ultra thick] (2.5,-5)--(2.5,-8.75);
   \draw[black, ultra thick] (4,-6)--(0.25,-6);
  \draw[black, ultra thick] (4,-6)--(4,-8.75);  
  \node[] at (2,-5.5) {\Large{$\tilde{Y_1}$}};
\node[] at (3.4,-6.5) {\Large{$\tilde{Y_2}$}};
  
    \drawYT{color=black,opacity=1, pattern=north west lines, pattern color=red, thin}{6.5}{-5}{-0.5}{-0.5}{4,3,3,3,1,1,1}
\drawYT{color=black,opacity=1, pattern=north east lines, pattern color=green, thin}{8}{-6}{-0.5}{-0.5}{7,5,5,3,1} 
 \draw[black, ultra thick] (6.5,-5)--(4.25,-5);
  \draw[black, ultra thick] (6.5,-5)--(6.5,-8.75);
   \draw[black, ultra thick] (8,-6)--(4.25,-6);
  \draw[black, ultra thick] (8,-6)--(8,-8.75); 
  \node[] at (6,-5.5) {\Large{$\tilde{Y_1}$}};
\node[] at (7.4,-6.5) {\Large{$\tilde{Y_2}$}};
  
      \drawYT{color=black,opacity=1, pattern=north west lines, pattern color=red, thin}{10.5}{-5}{-0.5}{-0.5}{4,3,3,2,1,1,1}
\drawYT{color=black,opacity=1, pattern=north east lines, pattern color=green, thin}{12}{-6}{-0.5}{-0.5}{7,6,5,3,1} 
 \draw[black, ultra thick] (10.5,-5)--(8.25,-5);
  \draw[black, ultra thick] (10.5,-5)--(10.5,-8.75);
   \draw[black, ultra thick] (12,-6)--(8.25,-6);
  \draw[black, ultra thick] (12,-6)--(12,-8.75); 
\node[] at (10,-5.5) {\Large{$\tilde{Y_1}$}};
\node[] at (11.4,-6.5) {\Large{$\tilde{Y_2}$}};

\node[] at (-2.8,-9) {{$\sim(a-\hat{\bar{a}})^{-4}$}};
\node[] at (1.2,-9) {{$\sim(a-\hat{\bar{a}})^{-4}$}};
\node[] at (5.2,-9) {{$\sim(a-\hat{\bar{a}})^{-4}$}};
\node[] at (9.2,-9) {{$\sim(a-\hat{\bar{a}})^{-4}$}};

\end{tikzpicture}
	\caption{Example of a family and its dual family with the associated to each member exponent of $(a-\bar{a})$ and $(a-\hat{\bar{a}})$ correspondingly} \label{Generalfamilyanddual}
\end{figure}
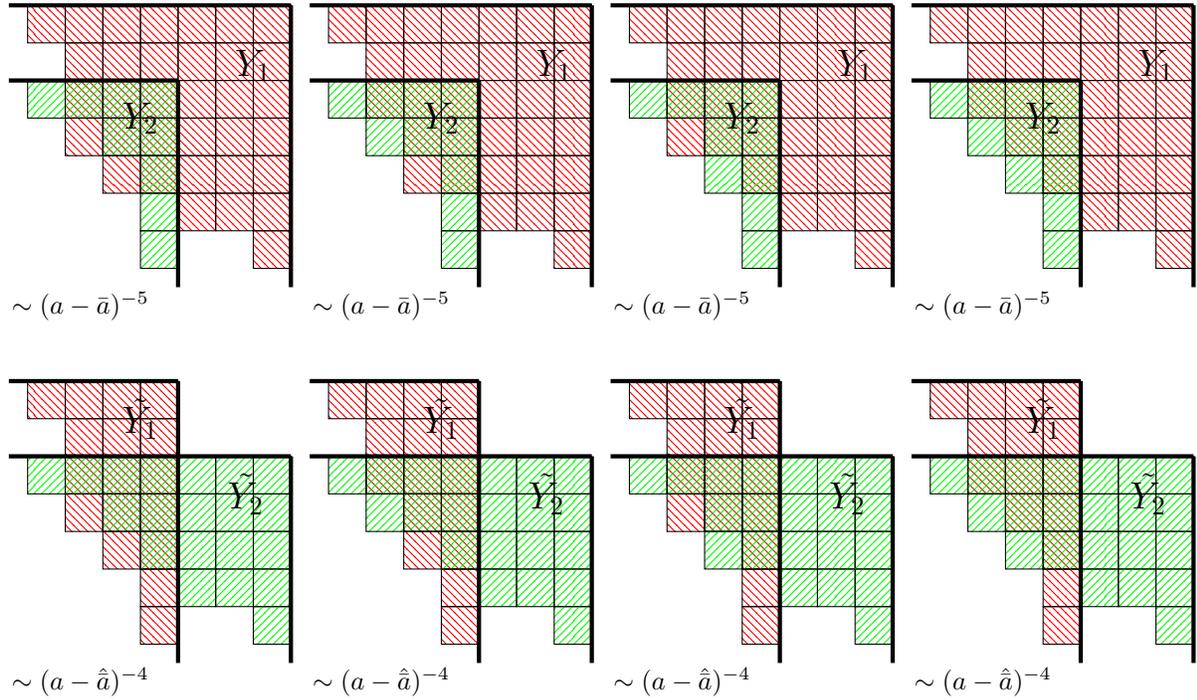

 We also define a dual family of pairs of Young diagrams $(\tilde{Y_1}$, $\tilde{Y_2})$. We shift the corner of diagram $\tilde{Y}_2$ on $m$ cells in positive vertical direction, while the origin of diagram $\tilde{Y}_1$ we shift on $n$ cells in positive horizontal direction with respect to the origin of $Y_1$. We introduce the dual occupation number based on belonging of a cell to the diagrams $\tilde{Y}_1$, $\tilde{Y}_2$. The families are dual if all the cells except the rectangle $m \times n$ at the origin of $Y_1$ have the same occupation number and dual occupation number. See Fig. \ref{Generalfamilyanddual} for example of a family with several members.

\par This construction is clearly in agreement with the conjecture of section \ref{Refined} since the occupation numbers of the cells in a family coincide with the dimension of spaces lacking in a bifiltration $B$ comparing to the bifiltration $B^{\rm (ref)}$ (see (\ref{reflexiveH})).  

\paragraph{Proving the relation between the families and their dual families.}
Let us first understand the order of the pole of $Z_k (Y_1,Y_2)$  at $a=\bar{a}$. The poles and zeroes at $\bar{a}$ are arising from several factors in $Z_k (Y_1,Y_2)$.

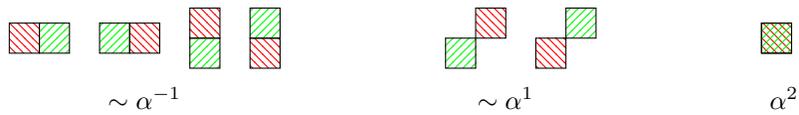
\begin{figure}
\centering
\begin{tikzpicture}

\drawYT{color=black,opacity=1, pattern=north west lines, pattern color=red, thin}{-6.2}{-0.2}{-0.4}{-0.4}{1}
\drawYT{color=black,opacity=1, pattern=north east lines, pattern color=green, thin}{-5.8}{-0.2}{-0.4}{-0.4}{1}
\drawYT{color=black,opacity=1, pattern=north west lines, pattern color=red, thin}{-4.6}{-0.2}{-0.4}{-0.4}{1}
\drawYT{color=black,opacity=1, pattern=north east lines, pattern color=green, thin}{-5}{-0.2}{-0.4}{-0.4}{1}
\drawYT{color=black,opacity=1, pattern=north west lines, pattern color=red, thin}{-3.8}{0}{-0.4}{-0.4}{1}
\drawYT{color=black,opacity=1, pattern=north east lines, pattern color=green, thin}{-3.8}{-0.4}{-0.4}{-0.4}{1}
\drawYT{color=black,opacity=1, pattern=north west lines, pattern color=red, thin}{-3}{-0.4}{-0.4}{-0.4}{1}
\drawYT{color=black,opacity=1, pattern=north east lines, pattern color=green, thin}{-3}{0}{-0.4}{-0.4}{1}
\node[align=left] at (-4.8,-1.2) {$\sim \alpha^{-1}$};

\drawYT{color=black,opacity=1, pattern=north west lines, pattern color=red, thin}{0}{0}{-0.4}{-0.4}{1}
\drawYT{color=black,opacity=1, pattern=north east lines, pattern color=green, thin}{-0.4}{-0.4}{-0.4}{-0.4}{1}

\drawYT{color=black,opacity=1, pattern=north west lines, pattern color=red, thin}{0.8}{-0.4}{-0.4}{-0.4}{1}
\drawYT{color=black,opacity=1, pattern=north east lines, pattern color=green, thin}{1.2}{0}{-0.4}{-0.4}{1}
\node[align=left] at (0,-1.2) {$\sim \alpha^1$};

\drawYT{color=black,opacity=1, pattern=north west lines, pattern color=red, thin}{3.8}{-0.2}{-0.4}{-0.4}{1}
\drawYT{color=black,opacity=1, pattern=north east lines, pattern color=green, thin}{3.8}{-0.2}{-0.4}{-0.4}{1}
\node[align=left] at (3.7,-1.2) {$\alpha^2$};

\end{tikzpicture}
	\caption{Zeroes and poles coming from interaction} \label{zerospolespfrominter}
\end{figure}

\par The first source of poles is $f_I(a)$, where $I$ marks the cells $(m,n)$ or $(m+1,n+1)$ in the diagram $Y_1$.
\par The second source is the interaction between the diagrams $Y_1$ and $Y_2$. Due to the shift of the origin of $Y_2$ with respect to the origin of $Y_1$ we have poles or zeros at $\bar{a}$ from the interaction of coinciding cells or from the nearest neighbours, but not from the separated cells. To be more precise, every pair of coinciding cells gives a double zero, every pair of cells sharing an edge gives a pole and every pair of cells having a common upper right or lower left angle brings a zero (see Fig. \ref{zerospolespfrominter}).
\par Keeping that in mind it is easy to see that if $Y_1$ and $Y_2$ overlap, the interaction of the whole overlapping region gives us a double zero at $\bar{a}$.
\par
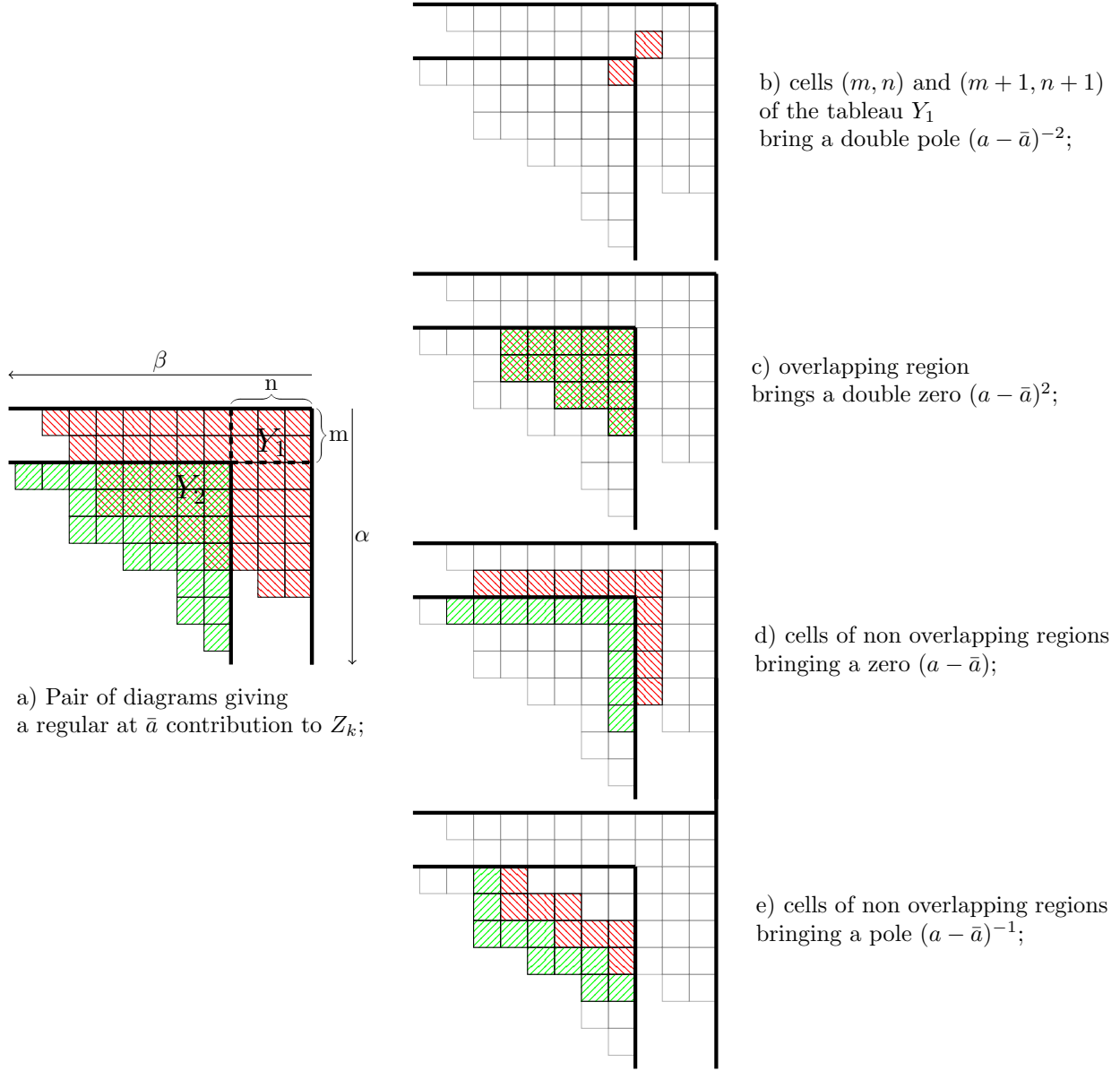
\begin{figure}
\centering
\begin{tikzpicture}

\drawYT{color=black,opacity=1, pattern=north west lines, pattern color=red, thin}{0}{0}{-0.4}{-0.4}{10,9,8,8,6,4,2}
\drawYT{color=black,opacity=1, pattern=north east lines, pattern color=green, thin}{-1.2}{-0.8}{-0.4}{-0.4}{8,6,6,4,2,2,1} 
 \draw[black, ultra thick] (0,0)--(0,-3.8);
  \draw[black, ultra thick] (0,0)--(-4.5,0);
   \draw[black, ultra thick] (-1.2,-0.8)--(-1.2,-3.8);
  \draw[black, ultra thick] (-1.2,-0.8)--(-4.5,-0.8);
      \draw[black, ultra thick,dashed](-1.2,0)--(-1.2,-0.8);
      \draw[black, ultra thick,dashed](-1.2,-0.8)--(0,-0.8);
     \draw[->](0,0.5)--(-4.5,0.5);
     \node[] at (-2.25,0.7) {$\beta$};
      \draw[->](0.6,0)--(0.6,-3.8);
     \node[] at (0.75,-1.9) {$\alpha$};
      
      \draw [decorate, 
decoration = {calligraphic brace, raise = 2pt, amplitude = 4pt}] (-1.2,0) --  (0,0);
\node[] at (-0.6,0.35) {n};
\draw [decorate, decoration = {calligraphic brace, raise = 2pt, amplitude = 4pt}] (0,0)--(0,-0.8);
\node[] at (0.4,-0.4) {m};
  \node[] at (-0.6,-0.5) {\Large{$Y_1$}};
\node[] at (-1.8,-1.2) {\Large{$Y_2$}};
  \node[align=left] at (-1.8,-4.5) {a) Pair of diagrams giving \\ a regular at $\bar{a}$ contribution to $Z_k$;};
 \draw[black, ultra thick] (0,0)--(0,-3.8);

 \drawYT{color=black,opacity=0.3,fill=white, thin}{6}{6}{-0.4}{-0.4}{10,9,8,8,6,4,2}
\drawYT{color=black,opacity=0.3, fill=white,  thin}{4.8}{5.2}{-0.4}{-0.4}{8,6,6,4,2,2,1} 
\drawYT{color=black,opacity=1, pattern=north west lines, pattern color=red, thin}{5.2}{5.6}{-0.4}{-0.4}{1}
\drawYT{color=black,opacity=1, pattern=north west lines, pattern color=red, thin}{4.8}{5.2}{-0.4}{-0.4}{1}
 \draw[black, ultra thick] (6,6)--(6,2.2);
  \draw[black, ultra thick] (6,6)--(1.5,6);
   \draw[black, ultra thick] (4.8,5.2)--(4.8,2.2);
  \draw[black, ultra thick] (4.8,5.2)--(1.5,5.2);
    \node[align=left] at (9.2,4.4) {b) cells $(m,n)$ and $(m+1,n+1)$ \\ of the tableau $Y_1$ \\ bring a double pole $(a-\bar{a})^{-2}$;};
  
  \drawYT{color=black,opacity=0.3,fill=white, thin}{6}{2}{-0.4}{-0.4}{10,9,8,8,6,4,2}
\drawYT{color=black,opacity=0.3, fill=white,  thin}{4.8}{1.2}{-0.4}{-0.4}{8,6,6,4,2,2,1} 
 \drawYT{color=black,opacity=1, pattern=north west lines, pattern color=red, thin}{4.8}{1.2}{-0.4}{-0.4}{5,5,3,1}
 \drawYT{color=black,opacity=1, pattern=north east lines, pattern color=green, thin}{4.8}{1.2}{-0.4}{-0.4}{5,5,3,1}
 \draw[black, ultra thick] (6,2)--(6,-1.8);
  \draw[black, ultra thick] (6,2)--(1.5,2);
   \draw[black, ultra thick] (4.8,1.2)--(4.8,-1.8);
  \draw[black, ultra thick] (4.8,1.2)--(1.5,1.2);
     \node[align=left] at (8.8,0.4) {c) overlapping region \\ brings a double zero  $(a-\bar{a})^2$;}; 
  
    \drawYT{color=black,opacity=0.3,fill=white, thin}{6}{-2}{-0.4}{-0.4}{10,9,8,8,6,4,2}
\drawYT{color=black,opacity=0.3, fill=white,  thin}{4.8}{-2.8}{-0.4}{-0.4}{8,6,6,4,2,2,1} 
 \drawYT{color=black,opacity=1, pattern=north west lines, pattern color=red, thin}{5.2}{-2.4}{-0.4}{-0.4}{7,1,1,1,1}
 \drawYT{color=black,opacity=1, pattern=north east lines, pattern color=green, thin}{4.8}{-2.8}{-0.4}{-0.4}{7,1,1,1,1}
 \draw[black, ultra thick] (6,-2)--(6,-5.8);
  \draw[black, ultra thick] (6,-2)--(1.5,-2);
   \draw[black, ultra thick] (4.8,-2.8)--(4.8,-5.8);
  \draw[black, ultra thick] (4.8,-2.8)--(1.5,-2.8);
 \node[align=left] at (9.2,-3.6) {d) cells of non overlapping regions \\ bringing a zero $(a-\bar{a})$;}; 
  
    \drawYT{color=black,opacity=0.3,fill=white, thin}{6}{-6}{-0.4}{-0.4}{10,9,8,8,6,4,2}
\drawYT{color=black,opacity=0.3, fill=white,  thin}{4.8}{-6.8}{-0.4}{-0.4}{8,6,6,4,2,2,1} 
  \drawYT{color=black,opacity=1, pattern=north west lines, pattern color=red, thin}{3.2}{-6.8}{-0.4}{-0.4}{1,1}
\drawYT{color=black,opacity=1, pattern=north east lines, pattern color=green, thin}{2.8}{-6.8}{-0.4}{-0.4}{1,1,1}
  \drawYT{color=black,opacity=1, pattern=north west lines, pattern color=red, thin}{4}{-7.2}{-0.4}{-0.4}{2,1}
    \drawYT{color=black,opacity=1, pattern=north east lines, pattern color=green, thin}{3.6}{-7.6}{-0.4}{-0.4}{2,1}
      \drawYT{color=black,opacity=1, pattern=north west lines, pattern color=red, thin}{4.8}{-7.6}{-0.4}{-0.4}{2,1}
    \drawYT{color=black,opacity=1, pattern=north east lines, pattern color=green, thin}{4.4}{-8}{-0.4}{-0.4}{2,1}
     \drawYT{color=black,opacity=1, pattern=north east lines, pattern color=green, thin}{4.8}{-8.4}{-0.4}{-0.4}{1}
 \draw[black, ultra thick] (6,-4)--(6,-9.8);
  \draw[black, ultra thick] (6,-6)--(1.5,-6);
   \draw[black, ultra thick] (4.8,-6.8)--(4.8,-9.8);
  \draw[black, ultra thick] (4.8,-6.8)--(1.5,-6.8);  
 \node[align=left] at (9.2,-7.6) {e) cells of non overlapping regions \\ bringing a pole $(a-\bar{a})^{-1}$;}; 
 
 \end{tikzpicture}
	\caption{Pair of diagrams giving a regular at $\bar{a}$ contribution to $Z_k$ and the sources of zeros and poles at $\bar{a}$} \label{nondual}
\end{figure}

 Counting the poles and zeroes coming from the interaction of overlapping and non overlapping regions, we see that they all cancel each other if the first line of $Y_2$ is longer than the $m$-th line of $Y_1$, the first column of $Y_2$ is longer than the $n$-th column of $Y_1$ and the edge of $Y_2$ does not touch the edge of $Y_1$ (see Fig \ref{nondual} for example). If any of these conditions is broken, the contribution of the pair of diagrams gains a pole.

From this immediately follows that  pairs of diagrams $Z_k(Y_1,Y_2)$ which cannot be drawn in the dual way are exactly the pairs giving regular contribution to $Z_k(Y_1,Y_2)$ at the point $a=\bar{a}$ and hence not contributing to the residue at this point. Therefore for our proof it is enough to consider only the families of diagrams $(Y_1, Y_2)$ which have dual families $(\tilde{Y}_1,\tilde{Y}_2)$. In particular, to contribute to the residue at $a=\bar{a}$ the diagram $Y_1$ must contain the box $(m,n)$.
\par Another thing which is easy to see from the counting of the poles is that $Z_k(Y_1, Y_2)$ is regular at  $\bar{a}=\epsilon_{m,n}$ if $m=n=0$.
\par Before considering the general case let us look at two simple examples.
\par The first one has an overlap of $Y_1$ and $Y_2$, but does not have blinking cells, \textit{i.e.} the cells which can belong either to $Y_1$ or to $Y_2$ (see Fig. \ref{overlap} (a) again). It means that there is a single member in the family and in the dual family, and we see that $Z_k(Y_1,Y_2)$ indeed has only a simple pole and the dual $Z_{k-mn}(\tilde{Y}_1,\tilde{Y}_2)$ is regular.

\par We intersect the diagrams into subregions $\square$, $y_1$, $y_2$, $y_3$ which can belong to different diagrams (the overline of the subregions is shown on Fig.\ref{overlap} (b)). The transformation of $\Sigma_1$ into $\Sigma_2$ goes as follows:

\begin{itemize}
    \item The interaction between the pairs of $y_i$ as parts of $\vec{Y}$ turns into the interaction between the same pairs of $y_i$ as parts of $\vec{\tilde{Y}}$ with the help of the shift of coordinates and arguments (\ref{wsym}).

    \item Factors $f_I$ with index $I$ associated with the cells of $y_i$ as a part of $\vec{Y}$ multiplied by the interaction between $y_i$ and $\square$ turn into factors $f_I$ with index $I$ associated with the cells of $y_i$ as a part of $\vec{\tilde{Y}}$ after a shift of coordinates and arguments (\ref{fsym}).
\end{itemize}
\par Checking the transformation one has to be careful with the  factors omitted due to integration over $\xi_I$ both in $\Sigma_1$ and $\Sigma_2$.
\par
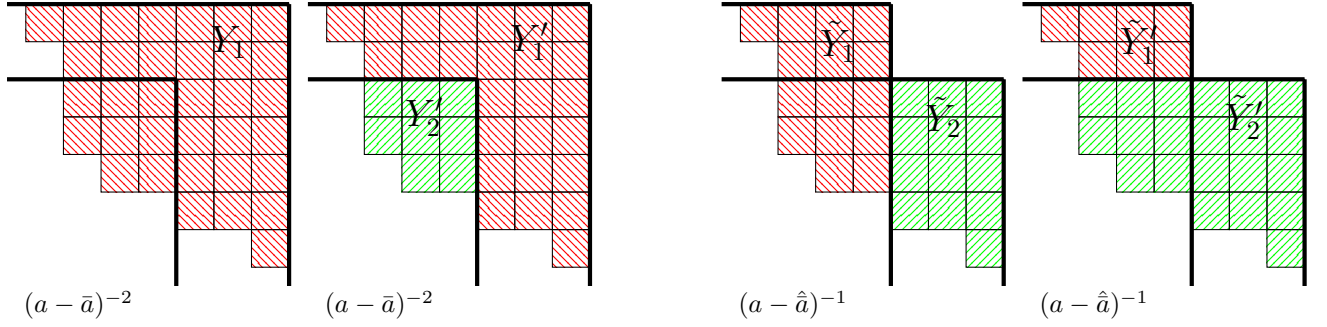
\begin{figure}
\centering
\begin{tikzpicture}

\drawYT{color=black,opacity=1, pattern=north west lines, pattern color=red, thin}{0}{0}{-0.5}{-0.5}{7,6,6,6,5,3,1}

 \draw[black, ultra thick] (0,0)--(0,-3.75);
  \draw[black, ultra thick] (0,0)--(-3.75,0);
   \draw[black, ultra thick] (-1.5,-1)--(-1.5,-3.75);
  \draw[black, ultra thick] (-1.5,-1)--(-3.75,-1);
\node[] at (-0.8,-0.5) {\Large{$Y_1$}};

\drawYT{color=black,opacity=1, pattern=north west lines, pattern color=red, thin}{4}{0}{-0.5}{-0.5}{7,6,3,3,3,3,1}
\drawYT{color=black,opacity=1, pattern=north east lines, pattern color=green, thin}{2.5}{-1}{-0.5}{-0.5}{3,3,2} 
 \draw[black, ultra thick] (4,0)--(4,-3.75);
  \draw[black, ultra thick] (4,0)--(0.25,0);
   \draw[black, ultra thick] (2.5,-1)--(2.5,-3.75);
  \draw[black, ultra thick] (2.5,-1)--(0.25,-1);
\node[] at (3.2,-0.5) {\Large{$Y_1'$}};
\node[] at (1.8,-1.5) {\Large{$Y_2'$}};

\node[] at (-2.8,-4) {{$(a-\bar{a})^{-2}$}};
\node[] at (1.2,-4) {{$(a-\bar{a})^{-2}$}};

\node[] at (6.7,-4) {{$(a-\hat{\bar{a}})^{-1}$}};
\node[] at (10.7,-4) {{$(a-\hat{\bar{a}})^{-1}$}};

\drawYT{color=black,opacity=1, pattern=north west lines, pattern color=red, thin}{8}{0}{-0.5}{-0.5}{4,3,3,3,2}
\drawYT{color=black,opacity=1, pattern=north east lines, pattern color=green, thin}{9.5}{-1}{-0.5}{-0.5}{3,3,3,3,1} 
 \draw[black, ultra thick] (8,0)--(8,-3.75);
  \draw[black, ultra thick] (8,0)--(5.75,0);
   \draw[black, ultra thick] (9.5,-1)--(9.5,-3.75);
  \draw[black, ultra thick] (9.5,-1)--(5.75,-1);
  \node[] at (7.3,-0.5) {\Large{$\tilde{Y}_1$}};
\node[] at (8.7,-1.5) {\Large{$\tilde{Y}_2$}};

\drawYT{color=black,opacity=1, pattern=north west lines, pattern color=red, thin}{12}{0}{-0.5}{-0.5}{4,3}
\drawYT{color=black,opacity=1, pattern=north east lines, pattern color=green, thin}{13.5}{-1}{-0.5}{-0.5}{6,6,5,3,1}
 \draw[black, ultra thick] (12,0)--(12,-3.75);
  \draw[black, ultra thick] (12,0)--(9.75,0);
   \draw[black, ultra thick] (13.5,-1)--(13.5,-3.75);
  \draw[black, ultra thick] (13.5,-1)--(9.75,-1);  
    \node[] at (11.3,-0.5) {\Large{$\tilde{Y}_1'$}};
\node[] at (12.7,-1.5) {\Large{$\tilde{Y}_2'$}};

\end{tikzpicture}
	\caption{Example of a family with a blinking group, but with no overlap, and its dual family with the associated to each member exponent of $(a-\bar{a})$ and $(a-\hat{\bar{a}})$ } \label{blinking}
\end{figure}

 The second example is the one with no overlap, but with a group of blinking cells (see Fig.\ref{blinking}). We denote the family members as $(Y_1, Y_2)$ and $(Y_1', Y_2')$. We again intersect the diagrams into subregions and their outline coincides with the outline of subregions $\square$, $y_1$, $y_2$, $y_3$ from Fig. \ref{overlap}(b).
\par Both $Z_k(Y_1, Y_2)$ and $Z_k(Y_1', Y_2')$ have double pole at $\bar{a}$, but the sum has only a simple pole.

\par The factors common for the members in the family of $\vec{Y}$ can be taken out of parenthesis in a sum over the family and they transform into the common factors of the members of the dual family $\vec{\tilde{Y}}$ exactly as in the previous example. Let us give a closer look to the different parts of the family members around the singularity $a=\bar{a}+\alpha$ and see how the cancellation of the extra singularity happens. The sum of the different parts is the following
\begin{equation}
    \Delta_1=\prod_{I\in y_3 \subset Y_1}f_{I}(0) f_{I}(\bar{a}+\alpha) \left[\overline{\prod}_{{T\in Y_1 \setminus y_3}} \mathcal{W}_{IT}(0)\right]+\prod_{I\in y_3 \subset Y_2'}f_{I}(-\bar{a}-\alpha)\bar{f}_{I}(0)\left[\prod_{{T\in Y_1'}} \mathcal{W}_{IT}(-\bar{a}-\alpha)\right]
\end{equation}
Both terms are singular at $\alpha=0$. Let us write explicitly the behaviour around the singularity
\begin{equation}
       f_{(m+1,n+1)}(\bar{a}+\alpha)=\frac{1}{\alpha} \bar{f}_{(m+1,n+1)}(\bar{a}+\alpha) \nonumber
\end{equation}
\begin{equation}
    \left[\underset{I=(1,1)\in Y_2'}{\prod_{{T\in Y_1'}}} \mathcal{W}_{IT}(-\bar{a}-\alpha)\right]=-\frac{1}{\alpha}\left[\underset{I=(1,1)\in Y_2'}{\overline{\prod}_{{T\in Y_1'}}} \mathcal{W}_{IT}(-\bar{a}-\alpha)\right].
\end{equation}
(Note that it is crucial here that as the interaction in the first term we have $\mathcal{W}_{IT}$ and not $\mathcal{W}_{TI}$ because of the order of integration with respect to $\xi_I$.)
\par Expanding all the factors of $\Delta_1$ around $\alpha=0$ we see that the poles cancel and
\begin{equation}
    \Delta_1= \frac{\partial}{\partial \alpha} \left( \prod_{I\in y_3 \subset Y_1}f_{I}(\alpha)\bar{f}_{I}(\bar{a}+\alpha)\left[\overline{\prod}_{{T\in Y_1 \setminus y_3}} \mathcal{W}_{IT}(\alpha)\right]\right)\bigg|_{\alpha=0}.
\end{equation}
\par On the other hand, the corresponding factors in $Z_{k-mn}(\tilde{Y}_1, \tilde{Y}_2)$ and $Z_{k-mn}(\tilde{Y}_1', \tilde{Y}_2')$ are the following
\begin{eqnarray} \label{Delta1}
    \Delta_2=\prod_{I\in y_3 \subset \tilde{Y}_1} f_I(0)f_I(\hat{\bar{a}}+\alpha)\left[\overline{\prod}_{T\in y_1 \subset\tilde{Y}_1} \mathcal{W}_{IT}(0)\right]\left[\prod_{T\in \tilde{Y}_2} \mathcal{W}_{IT}(\hat{\bar{a}}+\alpha)\right]+\\
    \prod_{I\in y_3 \subset \tilde{Y}_2} f_I(-\hat{\bar{a}}-\alpha)f_I(0)\left[\prod_{T\in \tilde{Y}_1} \mathcal{W}_{IT}(-\hat{\bar{a}}-\alpha)\right]\left[\overline{\prod}_{T\in y_2 \subset \tilde{Y}_2} \mathcal{W}_{IT}(0)\right].
\end{eqnarray}
Again, both terms are singular at $\hat{\bar{a}}$ with the singularity coming from the interaction of $y_3\in Y_i$ with $y_j\in Y_j$, $i\neq j$, but the sum is regular.
\par In the same way as before we get a regular expression
\begin{equation} \label{Delta2}
    \Delta_2=\frac{\partial}{\partial \alpha} \left( \prod_{I\in y_3 \subset \tilde{Y}_1} f_I(\alpha)f_I(\hat{\bar{a}}+\alpha)\left[\overline{\prod}_{T\in y_1 \subset\tilde{Y}_1} \mathcal{W}_{IT}(\alpha)\right]\left[\overline{\prod}_{T\in \tilde{Y}_2} \mathcal{W}_{IT}(\hat{\bar{a}}+\alpha)\right] \right) \bigg|_{\alpha=0}
\end{equation}
It is easy to check that (\ref{Delta1}) coincides with (\ref{Delta2}).
\par In general we have both the overlap and several blinking groups of cells (see again Fig. \ref{Generalfamilyanddual}). For $n$ blinking groups we will have $n$ extra poles which should be cancelled in a sum over $2^n$ family members. We already know how the common factors of a family transform into the common factors of its dual family. We now denote the region of overlapping by $y_3$, a region occupied by an $i$-th blinking group with two possible affiliations as $y_{3+i}$. If we now explicitly write the behaviour around the singularity we will get the sum of the different factors over $2^n$ members of the family
\begin{eqnarray}
    \Delta_1= \frac{1}{\alpha^n} \sum_{(Y_1, Y_2) \in \mathcal{F}} \left(  \prod_{i: \, y_i\in Y_1} \prod_{I \in y_i} f_I(0)f_I(\bar{a}+\alpha) \left[\overline{\prod}_{{T\in Y_1 \setminus y_i}} \mathcal{W}_{IT}(0)\right] \left[\overline{\prod}_{{T\in Y_2}} \mathcal{W}_{IT}(\bar{a}+\alpha)\right] \right)\cdot \\
    \left( \prod_{j: \, y_j\in Y_2} (-1) \prod_{J \in y_j}
    f_J(-\bar{a}-\alpha)f_J(0)\left[\overline{\prod}_{{T\in Y_1}} \mathcal{W}_{JT}(-\bar{a}-\alpha)\right] \left[\overline{\prod}_{{T\in Y_2 \setminus y_j}} \mathcal{W}_{JT}(0)\right]
    \right) \nonumber
\end{eqnarray}
Expanding all factors in terms of small parameter $\alpha$ we will see that all singular terms cancel and $\Delta_1$ can be written as
\begin{equation}
    \Delta_1= \frac{\partial^n}{\partial \alpha^n}  \left( \prod_{i=4}^{n+3} \prod_{I\in y_{i} \subset \overline{Y_1}}f_{I}(\alpha)f_{I}(\bar{a}+\alpha)\left[\overline{\prod}_{{T\in  \overline{Y_1} \setminus y_{i}}} \mathcal{W}_{IT}(\alpha)\right] \left[\overline{\prod}_{{T\in \underline{Y_2}}} \mathcal{W}_{IT}(\bar{a}+\alpha)\right]\right)\bigg|_{\alpha=0},
\end{equation}
where $\overline{Y_1}$ contains all the blinking parts $y_i$, $i=4,\ldots,n+3$ and $\underline{Y_2}$ does not have any.
\par On the other hand, a sum of different factors in $Z_{k-mn}$ treated in the same way gives us
\begin{equation} 
    \Delta_2=\frac{\partial^n}{\partial \alpha^n}  \left( \prod_{i=4}^{n+3} \prod_{I\in y_{i} \subset \overline{\tilde{Y}_1}} f_I(\alpha)f_I(\hat{\bar{a}}+\alpha)\left[\overline{\prod}_{T\in \overline{\tilde{Y}_1} \setminus y_i} \mathcal{W}_{IT}(\alpha)\right]\left[\overline{\prod}_{T\in  \underline{\tilde{Y}_2}} \mathcal{W}_{IT}(\hat{\bar{a}}+\alpha)\right] \right) \bigg|_{\alpha=0},
\end{equation}
where again $\overline{\tilde{Y}_1}$ contains all the blinking groups and $\underline{\tilde{Y}_2}$ does not contain any.
\par Both $\Delta_1$ and $\Delta_2$ are regular and one can make sure that they coincide.
\par By this we proved (\ref{zinstrelation}) for $N=2$ pure theory.
\par  Now we also can finally show that $Z(a)$ is regular at $\bar{a}=\epsilon_{m,n}$ if $m=0$ or $n=0$. In order to see it we should again form a family (with one of sides of the rectangular  being zero).  Again, although a single member of the family $Z_k(Y_1,Y_2)$ can be singular at this point, the sum over the family is regular.  To show that we repeat the steps we made above to prove the regularity of the sum over the dual family.

\paragraph{Higher rank case.}\par Let us generalise the proof for any $N$. To do that we just add $N-2$ diagrams to all families and the same $N-2$ diagrams to the dual families. Adding the diagrams we do not bring any new poles or zeroes with respect to $a_{12}$. Note that after the partial Weyl permutation (\ref{permutation}) of the coefficients in $a_1$, $a_2$, changes not only the difference $a_{12}$, but also all $a_{u1}$, $a_{u2}$.
\par Comparing to (\ref{zky1y2}) in the case of higher $N$  we have more interaction factors and we have more factors $f_I$ associated with every cell $I$ in  $Z_k$. Every cell brings us now
\begin{equation} \nonumber
    f_I(a_{I1})f_I(a_{I2})\ldots f_I(a_{IN})
\end{equation}
Let us compare $Z_k(Y_1, Y_2, Y_3,\ldots,Y_N)$ and $Z_{k-mn}(\tilde{Y}_1, \tilde{Y}_2, Y_3,\ldots,Y_N)$.
\begin{itemize}
    \item Self-interaction of diagrams $Y_3,...,Y_N$ and interaction among themselves appears identically in $Z_k$ and $Z_{k-mn}$.
    
    \item Interaction of diagrams $Y_3,...,Y_N$ with $Y_1, Y_2$ turns into their interaction with $\tilde{Y}_1, \tilde{Y}_2$ due to (\ref{wsym}).
    
    \item Factors  $f_I(a_{u1})f_I(a_{u2})$ at $a_{12}=\bar{a}$ turn into factors  $f_I(a_{u1})f_I(a_{u2})$ at $a_{12}=\hat{\bar{a}}$ due to the interaction with the rectangle.
    
    \item The factors $f_I$ coming from the cells $I\in \square$ contribute to $Z_k(\square,\o, \ldots, \o)$, \textit{i.e.} to the coefficient in (\ref{zinstrelation})
    
    \begin{eqnarray}
        \mathcal{P}^{(1 \, 2)}_{N}(m,n|\textbf{a})= \prod_{i=-m+1}^{m}\sideset{}{'} \prod_{j=-n+1}^{n} (-\epsilon_{i,j}) \cdot \prod_{v=3}^N \prod_{i=1}^m\prod_{j=1}^n \left[  (a_{1v}-\epsilon_{i-1,j-1})(-a_{1v}+\epsilon_{i,j}) \right] \\
        =  \prod_{i=-m}^{m-1}\sideset{}{'} \prod_{j=-n}^{n-1} \epsilon_{i,j}  \cdot \prod_{v=3}^N\prod_{i=1}^m\prod_{j=1}^n  \left[  (a_{2v}+\epsilon_{i,j})(-a_{1v}+\epsilon_{i,j}) \right]. \nonumber
    \end{eqnarray}
 
\end{itemize}
All the rest transforms exactly as in $N=2$ case. Therefore (\ref{zinstrelation}) for  the pure theory is proved.

\paragraph{Adding matter hypermultiplets.}
\par In a theory with a matter hypermultiplet the partition function gains an additional factor, but it does not affect our consideration of the order of the poles at $a_{uv}=\epsilon_{m,n}$. In the case of fundamental matter it is clear directly from (\ref{Zkfund}) and in the case of a theory with adjoint multiplet it is easier to see from (\ref{ZkadjNak}).
\par The additional factors associated with the cells $I\in \square$ contribute to the polynomials (\ref{polynomialPAdjmass}), (\ref{polynomialPFundmass}), and the factors associated with the rest of the cells transform into the factors associated with the corresponding cells in the dual family. 
\par In the case of an adjoint matter multiplet it is easy to see that the transformation of the additional factors goes exactly in the same way as in the pure theory, so we get
 \begin{eqnarray}
        \mathcal{P}^{(12)}_{N,{\rm adj}}(m,n|{\bf a})= \prod_{i=-m+1}^{m}\sideset{}{'} \prod_{j=-n+1}^{n} (-(\epsilon_{i,j}+M)) \cdot \prod_{v=3}^N \prod_{i=1}^m\prod_{j=1}^n \left[(a_{2v}+\epsilon_{i,j}+M)(-a_{1v}+\epsilon_{i,j}+M) \right]. 
\end{eqnarray}
\par If we are dealing with a theory with fundamental multiplets, the additional factor associated with a cell $I$ is
\begin{eqnarray}
    g_I(a_I)=\prod_{t=1}^{N_f}(a_I-\epsilon_1 (\alpha_I-1)-\epsilon_2 (\beta_I-1)-m_t) \\
    \cdot \prod_{t=1}^{N_a}(-a_I+\epsilon_1 \alpha_I+\epsilon_2 \beta_I+m_t). \nonumber
\end{eqnarray}
\par The additional factors $g_I(a_I)$ arising from the cells belonging to the diagrams $Y_3, \ldots, Y_N$ coincide in $Z^{({\rm fund})}_k$ and $Z^{({\rm fund})}_{k-mn}$.
\par To see the transformation of the factors associated with $Y_1$, $Y_2$ we have to recall that $\sum_{u=1}^N a_u =0$. Keeping this in mind we can write that
\begin{equation}
    a_1=\frac{1}{2}a_{12}-\frac{1}{2}\left(\sum_{w=3}^N a_w\right)
\end{equation}
and see that the cells marked by $I\in Y_1 \setminus \square$ transform into the cells of $\tilde{Y}_1$
\begin{equation}
    g_{(\alpha_I,\beta_I)}(a_1)=g_{(\alpha_I,\beta_I-n)}(\hat{a}_1)
\end{equation}
and the cells marked by $I\in Y_2$ transform into the cells of $\tilde{Y}_2$
\begin{equation}
    g_{(\alpha_I,\beta_I)}(a_2)=g_{(\alpha_I-m,\beta_I)}(\hat{a}_2).
\end{equation}

\par As for the factors $g_I$ coming from the cells marked by $I \in \square$, they contribute to the polynomial $\mathcal{P}^{(12)}_{N,{\rm fund}}(m,n|{\bf a})$.
\begin{eqnarray}
    \mathcal{P}^{(12)}_{N, {\rm fund}}(m,n|{\bf a})=\prod_{i=1}^m \prod_{j=1}^n g_{(i,j)}(a_1)=\prod_{i=1}^m \prod_{j=1}^n \Bigg[ \prod_{t=1}^{N_f}\left(\frac{1}{2}a_{12}-\epsilon_{i,j}+\epsilon-m_t -\frac{1}{2}\sum_{w=3}^N a_w \right) \\
    \cdot \prod_{t=1}^{N_{af}}\left(-\frac{1}{2}a_{12}+\epsilon_{i,j}+m_t +\frac{1}{2}\sum_{w=3}^N a_w \right) \Bigg] \nonumber
\end{eqnarray}
and hence (\ref{polynomialPFundmass}) immediately follows.
\par By this we completely proved (\ref{zinstrelation}).

\section{Zamolodchikov-like recurrence relations} \label{ZamRecRelSUN}
\subsection{Recurrence relations in terms of the variables \texorpdfstring{$a_{uv}$}{auv}}

Let us first write the recurrence relation for the  pure theory in terms of the variables $a_{uv}$. As it is clear from (\ref{ZkNak}) the partition function of the pure theory at infinity tends to $1$.
\par In the $SU(2)$ theory we get the Zamolodchikov recurrence relation
\begin{eqnarray} \label{Zamolodchikov}
    {\rm SU(2)}: \hspace{1 cm} {Z}^{({\rm 0})}(a)&=&1+\sum_{m,\, n=1}^{\infty} \frac{q^{mn} {Z}^{({\rm 
    0})}(\hat{a}^{(12)})\mathcal{P}^{(12)}_{2,{\rm 
    0}}(m,n)}{\mathcal{P}^{(12)}_{2}(m,n)}\left( \frac{1}{a-\epsilon_{m,n}}-\frac{1}{a+\epsilon_{m,n}}\right)  \nonumber \\ &=& 1+\sum_{m,\, n=1}^{\infty} \frac{q^{mn} {Z}^{({\rm 
    0})}(\epsilon_{m,-n})}{(a-\epsilon_{m,n})(a+\epsilon_{m,n})}\frac{2 \epsilon_{m,n}\mathcal{P}^{(12)}_{2,{\rm 0}}(m,n)}{\mathcal{P}^{(12)}_{2}(m,n)}.
\end{eqnarray}
\par In $SU(3)$ theory we should chose $N-1=2$ independent variables, for example, $a_{13}$ and $a_{23}$. Let us assume that $a_{23}$ is away from the poles. Then $Z^{({\rm 0})}({\bf a})$ has poles only with respect to $a_{13}$ at the points $a_{13}=\epsilon_{m,n}$ and $a_{13}=\epsilon_{m,n}+a_{23}$. Using (\ref{zinstrelation}) we can write immediately
\begin{eqnarray} \label{ZamolodchikovSU3}
 {\rm SU(3)}: \hspace{1 cm} {Z}^{({\rm 0})}({\bf a})=  1+\sum_{m,\, n=1}^{\infty} \frac{q^{mn} {Z}^{({\rm 0})}(\hat{\bf a}^{(13)})}{(a_{13}+\epsilon_{m,n})(a_{13}-\epsilon_{m,n})}\frac{2 \epsilon_{m,n}\mathcal{P}^{(13)}_{3,{\rm 0}}(m,n|{\bf a})}{\mathcal{P}^{(13)}_{3}(m,n|{\bf a})} \nonumber\\
 + \sum_{m,\, n=1}^{\infty} \frac{q^{mn} {Z}^{({\rm 
 0})}(\hat{\bf a}^{(12)})}{(a_{13}-a_{23}+\epsilon_{m,n})(a_{13}-a_{23}-\epsilon_{m,n})}\frac{2 \epsilon_{m,n}\mathcal{P}^{(12)}_{3,{\rm 0}}(m,n|{\bf a})}{\mathcal{P}^{(12)}_{3}(m,n|{\bf a})} .
\end{eqnarray}
By analytical continuation (\ref{ZamolodchikovSU3}) is valid everywhere on the domain of $Z({\bf a})$.
\par We can generalise the answer for $SU(N)$ theory. Let us chose $N-1$ independent variables to be $a_{uN}$, $u=1 \ldots N-1$ and assume that $a_{\hat{u}N}$, $\hat{u}=2 \ldots N-1$ are away from the poles as well as their differences $a_{\hat{u}N}-a_{\hat{v}N}$. Then $Z^{({\rm 0})}({\bf a})$ has poles only with respect to $a_{1N}$ at the points $a_{1N}=\epsilon_{m,n}$ and $a_{1N}=\epsilon_{m,n}+a_{\hat{u}N}$ and we can write
\begin{eqnarray} \label{ZamolodchikovSUN}
 {\rm SU(N)}: \hspace{1 cm} {Z}^{({\rm 0})}({\bf a})=  1+
 \sum_{w=2}^{N}\sum_{m,\, n=1}^{\infty} \frac{q^{mn} {Z}^{({\rm 0})}(\hat{\bf a}^{(1w)})}{(a_{1N}-a_{wN}+\epsilon_{m,n})(a_{1N}-a_{wN}-\epsilon_{m,n})} \nonumber\\
 \cdot\frac{2 \epsilon_{m,n}\mathcal{P}^{(1w)}_{N,{\rm 0}}(m,n|{\bf a})}{\mathcal{P}^{(1w)}_{N}(m,n|{\bf a})}.
\end{eqnarray}

In presence of matter hypermultiplets, however, defining asymptotic behaviour at infinity is a difficult problem. Indeed, according to our construction of the recurrence relation, we should seek for asymptotic behaviour with $a_{1N} \rightarrow \infty$ with all the rest of independent variables $a_{uv}$ being arbitrary. Such an asymmetric way to approach infinity results in non-trivial dependence on $a_{uv}$ at infinity. In the next subsection we will treat the problem in a symmetrical way and will be able to say more on the question.

\paragraph{Remark.} Function $Z^{({\rm 0})}({\bf{a}})$ depends on $N-1$ variables and can have a singularity of order up to $N(N-1)/2$. Function $Z^{({\rm 0})}(\hat{\bf{a}}^{(1w)})$ depends on $N-2$ variables and has one fixed parameter $\hat{a}_{1w}=\epsilon_{m,-n}$. Its order of singularity is up to $N(N-1)/2-1$. One can apply again the recurrence relation to $Z^{({\rm 0})}(\hat{\bf{a}}^{(1w)})$ and make up to $N-1$ steps in the reduction of the order of singularity. For $N>2$ one can never express $Z^{({\rm 0})}({\bf a})$ on the whole domain in terms of the value of $Z^{({\rm 0})}({\bf a})$ at its regular points. In particular in the case of $SU(3)$ theory one can write $Z^{({\rm 0})}({\bf a})$ on the whole domain through its values at the regular points and the points where the function is singular with respect to only one variable $a_{uv}$.

\subsection{Recurrence relations in terms of symmetric variables} \label{SymRecRel}
Although the generalisation of the Zamolodchikov relation provided above is very straightforward, it lacks obvious Weyl symmetry. For sure $Z^{({\rm 0})}(\bf{a})$ written as (\ref{ZamolodchikovSU3}) has no choice but to satisfy (\ref{zinstrelation}) when one takes the residue with respect to $a_{23}$, but showing it explicitly requires some additional computations. Another problem is the mentioned above difficulty with finding the asymptotic behaviour of the partition function.
\par To expose the Weyl symmetry of the recurrence relation and to write it for the theories with matter hypermultiplets we are going to rewrite it in terms of symmetric variables.
\paragraph{Symmetric variables.} One can find an elegant symmetrical form of $Z^{({\rm R})}$ for $SU(3)$ theory in \cite{Poghossian} given in terms of the parameters, which are nothing but a basis of symmetric functions of $a_u$ written upon a condition $\sum_i a_i =0$. Following this lead we introduce variables providing a basis for symmetric functions of $a_u$ in $SU(N)$ theory
\begin{eqnarray}
w_1&=& \sum_{i_1<i_2} a_{i_1} a_{i_2} \nonumber \\
   w_2&=& \sum_{i_1<i_2<i_3} a_{i_1} a_{i_2}a_{i_3}  \label{atosym} \\
   &\ldots&  \nonumber\\
   w_{N-1}&=&a_1 a_2 \ldots a_N. \nonumber
\end{eqnarray}
For further convenience we also introduce a vector composed of the first $N-2$ variables $w_i$
\begin{equation*}
    {\bm \omega} = (w_1, \ldots, w_{N-2}).
\end{equation*}
We will also use the notation ${\bf w}=( {\bm \omega},w_{N-1})$.
\par By (\ref{atosym}) we introduced a map $\bf{a}\mapsto \bf{w}$. In order to rewrite (\ref{zinstrelation}) in terms of the symmetric variables we will need also the inverse map $\bf{w}\mapsto \bf{a}$. This function is multivalued and has $N!$ branches corresponding to the Weyl permutations of $a_{u}$.

\paragraph{Poles of $Z^{({\rm R})}({\bf w})$.} Function $Z({\bf a})$ has poles at $a_{uv}=\epsilon_{mn}$ and is Weyl symmetric, hence all its singular terms can be grouped in such a way that the common denominator of these singularities is symmetrical and has the form
\begin{equation} \label{determinator}
    \Delta^{(m,n)}({\bf a})=\prod_{u\neq v}(a_{uv}^2-\epsilon^2_{m,n}).
\end{equation}
\par To find the poles in terms of the symmetric variables we need to write the denominator (\ref{determinator}) as a function of ${\bf w}$. To do that let us introduce a polynomial of $x$ with coefficients defined by ${\bf a}$ or equivalently by ${\bf w}$.
\begin{equation}
    Q(x|{\bf a})=(x-a_1)\ldots(x-a_N)=x^N+x^{N-2}w_1-  x^{N-3} w_2 +\ldots + (-1)^N w_{N-1} =  Q^{({\bf w})}(x)\, .
\end{equation}
Then the denominator (\ref{determinator}) can be written as
\begin{eqnarray}
    \Delta^{(m,n)}({\bf a})=(-1)^{\frac{N(N-1)}{2}}\frac{1}{\epsilon_{mn}^N}{\rm \bf res}(Q(x|{\bf a}),Q(x+\epsilon_{m,n}|{\bf a}))= \nonumber \\
   (-1)^{\frac{N(N-1)}{2}}\frac{1}{\epsilon_{mn}^N}{\rm \bf res}(Q(x)^{({\bf w})},Q(x+\epsilon_{m,n})^{({\bf w})}) = \Delta^{(m,n)}({\bf w}),
\end{eqnarray}
where ${\rm \bf res}(A(x),B(y))$ is the resultant, and for normalised polynomials $A(x)$, $B(y)$ it is defined as
\begin{equation}
    {\rm \bf res}(A(x),B(y))=\prod_{(\bar{x},\bar{y}):A(\bar{x})=0,B(\bar{y})=0}(\bar{x}-\bar{y}) \, .
\end{equation}
The resultant ${\rm \bf res}(A(x),B(y))$ can be written as a determinant of the Sylvester matrix with components defined by the coefficients of the polynomials $A(x)$, $B(y)$ \cite{Gelfand}, but in order to compute it in any particular case one can simply use the Euclidean algorithm described below or a builtin function of a computer  algebra system.
\par Therefore the poles of $Z^{({\rm R})}({\bf w})$ are located at ${\bf \bar{w}}^{(k|m,n)}=({\bm \omega},\bar{w}^{(k|m,n)}_{N-1})$, $m \cdot n >0$, where ${\bm \omega}$ is arbitrary parameters and $\bar{w}^{(k|m,n)}_{N-1}$ are roots of the equation
\begin{equation} \label{wroots}
    \Delta^{(m,n)}({\bf {w}})=0.
\end{equation}
 The equation (\ref{wroots}) on $\bar{w}_{N-1}^{(k|m,n)}$ is of order $N-1$ and $k$ marks the roots.
 \par Note that since by construction the map $\bf{w} \mapsto \bf{a}$ has $N!$ branches, every one of $N-1$ roots $\bar{w}^{(k|m,n)}_{N-1}({\bm \omega})$ describes the poles with respect to all $a_{uv}$ at the points $\bar{a}_{uv}=\pm \epsilon_{m,n}$. In order to rewrite (\ref{zinstrelation}) in terms of the symmetric variables we will choose one branch of the inverse map, but as long as the final relations are written in terms of single-valued functions of $\bf{w}$, this intermediate choice will not ruin the Weyl symmetry.

\paragraph{Dual point.} A residue of $Z^{({\rm R})}$ is proportional to its value at the dual point. While in \cite{Poghossian} the dual point was taken from the AGT approach, we are appealing to the statement proven in a previous section that in terms of the variables ${\bf a}$ a residue of $Z^{({\rm R})}$ with respect to $a_{uv}$ at a point $\bar{a}_{uv}$ is proportional to $Z^{({\rm R})}$ at the point ${\bf \hat{a}^{(uv)}}$ with partial Weyl permutation performed in $\bar{a}_u$, $\bar{a}_v$ and the rest of the variables left unchanged.
\par Let us choose the branch such that $\bar{a}_{12}=\epsilon_{m,n}$ and its dual point ${\bf \hat{a}}=(\hat{a}_1, \hat{a}_2, a_3\ldots, a_N)$ with $\hat{a}_1$, $\hat{a}_2$ related with $\bar{a}_1$, $\bar{a}_2$ by the partial Weyl permutation.
\par We introduce two polynomials
\begin{eqnarray}
\bar{Q}(x) = Q(x|{\bf\bar{a}})=Q^{({\bf \bar{w}}^{(k|m,n)})}(x), \nonumber \\
\hat{Q}(x)= Q(x|{\bf\hat{a}})=Q^{({\bf \hat{w}}^{(k|m,n)})}(x).
\end{eqnarray}
The point ${\bf \hat{w}}^{(k|m,n)}$ is the wanted dual to ${\bf \bar{w}}^{(k|m,n)}$.
\par The normalised difference of these polynomials is a polynomial of degree $N-2$ with the roots $a_i$, $i=3\ldots N$.
\begin{eqnarray} \label{DeltaQ}
    \Delta Q(x) = \frac{1}{\Delta w_1^{(m,n)}} \left(\bar{Q}(x)-\hat{Q}(x)\right)  \nonumber \\ = a^{N-2} -  a^{N-3} \frac{\Delta w_2^{(k|m,n)}}{\Delta w_1^{(m,n)}} +\ldots + (-1)^N \frac{\Delta w_{N-1}^{(k|m,n)}}{\Delta w_1^{(m,n)}}=(x-a_3)\ldots(x-a_N),
\end{eqnarray}
where
\begin{eqnarray*}
    \Delta w_1^{(m,n)} &=& w_1-\hat{w}_1^{(m,n)} \, ,\\
    \Delta w_i^{(k|m,n)} &=& w_i-\hat{w}_i^{(k|m,n)} \, , \quad\quad i=2,\ldots,N-2 \\
    \Delta w_{N-1}^{(k|m,n)} &=&\bar{w}^{(k|m,n)}_{N-1}-\hat{w}^{(k|m,n)}_{N-1}.
\end{eqnarray*}
From (\ref{DeltaQ}) we see immediately that
\begin{eqnarray}
    \Delta w_1^{(m,n)} = (x-\bar{a}_1)(x-\bar{a}_2)-(x-\hat{a}_1)(x-\hat{a}_2) \nonumber \\ =\bar{a}_1\bar{a}_2-\hat{a}_1\hat{a}_2=-m \, n \, \epsilon_1 \epsilon_2.
\end{eqnarray}
Polynomial $\bar{Q}(x)$ is dividable by $\Delta Q(x)$, and the quotient is a polynomial of degree 2 with the roots $\bar{a}_1$, $\bar{a}_2$. 
\begin{equation}
   \frac{\bar{Q}(x)}{\Delta Q(x)} =x^2 +x \frac{\Delta w_2^{(k|m,n)}}{\Delta w_1^{(m,n)}}+\bar{w}_1-\frac{\Delta w_3^{(k|m,n)}}{\Delta w_1^{(m,n)}} +\frac{(\Delta w_2^{(k|m,n)})^2}{(\Delta w_1^{(m,n)})^2}=(x-\bar{a}_1)(x-\bar{a}_2).
\end{equation}
The discriminant of this polynomial is $\bar{a}_{12}^2=\epsilon_{m,n}^2$, and together with the requirement that the remainder of division of the polynomials $\bar{Q}(x)$, $\Delta Q(x)$ vanish in all orders of $x$ we get a system of $N-1$ recurrent equations
\begin{equation} \label{recsys}
    w_{i-1}-\frac{\Delta w_{i+1}^{(k|m,n)}}{\Delta w_1^{(m,n)}}+\frac{\Delta w_i^{(k|m,n)} \Delta w_2^{(k|m,n)} }{(\Delta w_1^{(m,n)} )^2}+\frac{1}{4}\frac{\Delta w_{i-1}^{(k|m,n)} }{\Delta w_1^{(m,n)}}\left(\epsilon_{m,n}^2-\frac{(\Delta w_2^{(k|m,n)} )^2}{(\Delta w_1^{(m,n)})^2} \right)=0,
\end{equation}
The boundary conditions are $\Delta w_N^{(k|m,n)}  = \Delta w_{N+1}^{(k|m,n)}  =0$. Solving the system (\ref{recsys}) and using that $\bar{Q}(\bar{a}_1)=\bar{Q}(\bar{a}_2)=0$ we can find all $\Delta w_i^{(k|m,n)} $ in terms of $\Delta w_2^{(k|m,n)} $ as follows
\begin{equation}
    \frac{\Delta w_i^{(k|m,n)} }{\Delta w_1^{(m,n)} }=\frac{1}{\epsilon_{m,n}}\left( \sum_{t=i}^{N-2} w_t \left( \left(-\bar{a}_1 \right)^{-t+i-1}-\left(\bar{a}_2\right)^{-t+i-1} \right)+\bar{w}_{N-1}^{(k)}\left( \left(-\bar{a}_1 \right)^{-N+i}-\left(\bar{a}_2\right)^{-N+i} \right) \right), 
\end{equation}
where
\begin{eqnarray}
    \bar{a}_1(\Delta w_2^{(k|m,n)} )=\frac{1}{2} \left( \epsilon_{m,n}- \frac{\Delta w_2^{(k|m,n)} }{\Delta w_1^{(m,n)} }\right) \\
    \bar{a}_2(\Delta w_2^{(k|m,n)} )=-\frac{1}{2} \left(\epsilon_{m,n}+\frac{\Delta w_2^{(k|m,n)} }{\Delta w_1^{(m,n)} } \right) \nonumber.
\end{eqnarray}
For the shift $\Delta w_{N-1}^{(k|m,n)} $ we always get
\begin{equation} \label{branchconnection}
    \Delta w_{N-1}^{(k|m,n)} =\frac{4 \bar{w}_{N-1}^{(k|m,n)}}{\frac{(\Delta w_2^{(k|m,n)} )^2}{(\Delta w_1^{(m,n)} )^2}-\epsilon_{m,n}^2}.
\end{equation}
One can note that $\hat{w}_{N-1}^{(k|m,n)}$ satisfies an equation
\begin{equation} \label{wshiftedroots}
    \Delta^{(m,-n)}(\hat{\bm \omega}^{(k|m,n)},\hat{w}_{N-1}^{(k|m,n)})=0.
\end{equation}
\par Finally to find $\Delta w_2^{(k|m,n)} $ we use the fact that polynomial $\bar{Q}(x)$ has the roots $\bar{a}_1$, $\bar{a}_2$ with the difference $\epsilon_{m,n}$. In other words we want to impose a condition that the polynomials $\bar{Q}(x)$ and $\bar{Q}(x+\bar{a}_{12})$ have the greatest common divisor ${\rm\bf gcd}(\bar{Q}(x),\bar{Q}(x+\epsilon_{mn}))=(x-a_2)$. To do it we apply the Euclidean algorithm.
\par The algorithm is based on the fact that if we divide a polynomial $A(x)$ by a polynomial $B(x)$
\begin{equation*}
    A(x)=q(x)B(x)+r(x),
\end{equation*}
where $q(x)$ is the quotient and $r(x)$ is the remainder, then
\begin{equation*}
    {\rm \bf gcd}(A(x),B(x))= {\rm \bf gcd}(B(x),r(x)).
\end{equation*}
So on the first two steps we write
\begin{eqnarray}
     \bar Q(x+\epsilon_{m,n})=q_1(x)\bar Q(x)+r_1(x), \nonumber \\
     \bar Q(x)=q_2(x) r_1(x) +r_2(x)
\end{eqnarray}
and then we proceed with division
\begin{equation}
    r_{i-2}(x)=q_{i-1}(x)r_{i-1}(x)+r_{i}(x).
\end{equation}
\par In the general case without degeneration after $N$ steps we get a constant remainder $r_N$ proportional to $\Delta^{(mn)}({\bf \bar{w}}^{(k|m,n)})$ and hence $r_N=0$. It expresses the fact that the polynomial $\bar{Q}(x)$ indeed has two roots with the difference $\epsilon_{m,n}$. On the previous step on the other hand we get a linear polynomial $r_{N-1}(x)$ which is the wanted greatest common divisor $(x-a_2)$, so  
\begin{equation*}
r_{N-1}(a_2(\Delta w_2^{(k|m,n)} ))=0,
\end{equation*}
which gives us a linear equation for $\Delta w_2^{(k|m,n)} $ with coefficients depending on $({\bm \omega}, \bar{w}_{N-1}^{(k|m,n)})$.
\par Although the algorithm is very straightforward, it is difficult to write an explicit form of the resulting equation for $\Delta w_2^{(k|m,n)} $ in the general case of $SU(N)$.
\par In such a way we find all the $\Delta w_i^{(k|m,n)} $ and hence the point ${\bf \hat{w}}^{(k|m,n)}$ dual to the pole ${\bf \bar{w}}^{(k|m,n)}$.

\paragraph{Polynomials $\mathcal{P}$ via symmetric variables.} The polynomials connecting residue of $Z^{({\rm R})}$ with its value at the dual point can be easily expressed through the symmetric variables. Indeed,
   \begin{eqnarray} \label{Psym}
       && \mathcal{P}^{(1 \, 2)}_{N}(m,n|\textbf{a})
        =  \prod_{i=-m}^{m-1}\sideset{}{'} \prod_{j=-n}^{n-1} \epsilon_{i,j}  \cdot \prod_{k=3}^N\prod_{i=1}^m\prod_{j=1}^n  \left[  (a_{2k}+\epsilon_{i,j})(-a_{1k}+\epsilon_{i,j}) \right] = \\
        &&\prod_{i=-m}^{m-1}\sideset{}{'} \prod_{j=-n}^{n-1} \epsilon_{i,j}  \cdot \prod_{i=1}^m\prod_{j=1}^n  \left[  (-1)^N \Delta Q(a_2(\Delta w_2^{(k|m,n)} )+\epsilon_{i,j})\Delta Q(a_1(\Delta w_2^{(k|m,n)} )-\epsilon_{i,j}) \right] \triangleq \mathcal{P}_{N}^{(m,n)}({\bm \omega}, \bar{w}_{N-1}^{(k|m,n)}). \nonumber
\end{eqnarray}
In the same manner we see
\begin{eqnarray}  \label{Psymadj}
  &&\mathcal{P}^{(uv)}_{N,{\rm adj}}(m,n|{\bf a})=  \prod_{i=-m}^{m-1}\prod_{j=-n}^{n-1} (\epsilon_{i,j}-M)  \\ && \cdot   \prod_{i=1}^{m} \prod_{j=1}^{n}\left[(-1)^N\Delta Q(a_2(\Delta w_2^{(k|m,n)} )+\epsilon_{i,j}+M) \Delta Q(a_1(\Delta w_2^{(k|m,n)} )-\epsilon_{i,j}-M)\right] \triangleq \mathcal{P}_{N,{\rm adj}}^{(m,n)}({\bm \omega}, \bar{w}_{N-1}^{(k|m,n)}). \nonumber
\end{eqnarray}
In the case of presence of fundamental matter we get
\begin{eqnarray} \label{Psymfund}
    \mathcal{P}^{(12)}_{N, {\rm fund}}(m,n|{\bf a})=
    \prod_{i=1}^{m}\prod_{j=1}^{n} \Bigg[ \prod_{t=1}^{N_f}   \left(-\frac{1}{2}\epsilon_{m,n}+\epsilon_{i,j}-m_t-\frac{1}{2}\sum_{w=3}^N a_w \right) \nonumber \\
 \cdot  \prod_{t=1}^{N_a}  \left(-\frac{1}{2}\epsilon_{m,n}+\epsilon_{i,j}+m_t+\frac{1}{2}\sum_{w=3}^N a_w \right) \Bigg] \nonumber \\
=\prod_{i=1}^{m}\prod_{j=1}^{n} \Bigg[ \prod_{t=1}^{N_f}   \left(-\frac{1}{2}\epsilon_{m,n}+\epsilon_{i,j}-m_t-\frac{1}{2}\frac{\Delta w_2^{(k|m,n)} }{\Delta w_1^{(m,n)} }  \right) \\
\cdot  \prod_{t=1}^{N_a}  \left(-\frac{1}{2}\epsilon_{m,n}+\epsilon_{i,j}+m_t+\frac{1}{2}\frac{\Delta w_2^{(k|m,n)} }{\Delta w_1^{(m,n)} }  \right) \Bigg] \triangleq \mathcal{P}_{N, {\rm fund}}^{(m,n)}({\bm \omega}, \bar{w}_{N-1}^{(k|m,n)}). \nonumber
\end{eqnarray}

\paragraph{Jacobian.} The last missing piece we need to rewrite the residue of $Z^{({\rm R})}$ with respect to $a_{12}$ in terms of a residue with respect to $w_{N-1}$ is the Jacobian in  terms of the symmetric variables. We will denote it as
\begin{equation}
   J^{(m,n)}({\bm \omega},\bar{w}_{N-1}^{(k|m,n)}){\rm Res}_{a_{12}=\bar{a}_{12}}Z^{({\rm R})}({\bf a})={\rm  Res}_{w_{N-1}=\bar{w}_{N-1}^{(k|m,n)}}Z^{({\rm R})}({\bf w}) 
\end{equation}
where
\begin{equation}
 J^{(m,n)}({\bm \omega},\bar{w}_{N-1}^{(k|m,n)})=\left( \frac{\partial \Delta^{(m,n)}({\bf w})}{\partial w_{N-1}}\right)^{-1} \bigg|_{w_{N-1}= \bar{w}_{N-1}^{(k|m,n)}} \cdot \left( \frac{\partial \Delta^{(m,n)} ({\bf a})}{\partial a_{12}}\right) \bigg|_{a_{12}=\bar{a}_{12}=\epsilon_{mn}}.
\end{equation}
The last factor can be expressed via the same polynomials $Q(x)$, $\Delta Q(x)$ again.
\begin{eqnarray}
    \left( \frac{\partial \Delta^{(m,n)} ({\bf a})}{\partial a_{12}}\right) \bigg|_{a_{12}=\bar{a}_{12}=\epsilon_{m,n}}= 2 \epsilon_{m,n} \left(\prod_{k=3}^{N} (a_{1k}^2-\epsilon_{m,n}^2) \right)\left(\prod_{k=3}^{N} (a_{2k}^2-\epsilon_{m,n}^2)\right)\left(\prod_{\underset{k \neq l}{k,l=3}}^{N} (a_{kl}^2-\epsilon_{m,n}^2) \right) \nonumber\\ 
    \label{dDletaDa12}
= 2 \epsilon_{m,n}\Delta Q(a_1+\epsilon_{m,n})\Delta Q(a_1-\epsilon_{m,n}) \Delta Q(a_2+\epsilon_{m,n})\Delta Q(a_2-\epsilon_{m,n}) \\  \cdot \frac{(-1)^{\frac{(N-2)(N-3)}{2}}  }{\epsilon_{m,n}^{N-2}} {\rm {\bf res}} (\Delta Q(x),\Delta Q(x+\epsilon_{m,n})). \nonumber 
\end{eqnarray}
Therefore
\begin{eqnarray} \label{jacobian}
    J^{(m,n)}({\bm \omega},\bar{w}_{N-1}^{(k|m,n)})&=&(-1)2\epsilon_{m,n}^3 {\rm \bf res}(\Delta Q(x),\Delta Q(x+\epsilon_{m,n})) \left(\frac{\partial{\rm \bf res} \,\, ( Q^{({\bf w})}(x), Q^{({\bf w})}(x+\epsilon_{m,n})) }{\partial w_{N-1}} \right)^{-1}\bigg|_{ w_{N-1}= \bar{w}_{N-1}^{(k|m,n)}} \nonumber \\ &\cdot& \Delta Q\left(-\frac{1}{2}\frac{\Delta w_2^{(k|m,n)} }{\Delta w_1^{(m,n)} }+\frac{3}{2}\epsilon_{m,n}\right)\Delta Q\left(-\frac{1}{2}\frac{\Delta w_2^{(k|m,n)} }{\Delta w_1^{(m,n)} }-\frac{3}{2}\epsilon_{m,n}\right) \nonumber\\ &\cdot&\Delta Q\left(-\frac{1}{2}\frac{\Delta w_2^{(k|m,n)} }{\Delta w_1^{(m,n)} }-\frac{1}{2}\epsilon_{m,n}\right)\Delta Q\left(-\frac{1}{2}\frac{\Delta w_2^{(k|m,n)} }{\Delta w_1^{(m,n)} }+\frac{1}{2}\epsilon_{m,n}\right).
\end{eqnarray}
The residue formula in terms of the symmetric variables has the form
\begin{equation} \label{residuesym}
    {\rm Res}_{w_{N-1}=\bar{w}_{N-1}^{(k|m,n)}} Z^{({\rm R})}( {\bm \omega},w_{N-1})=q^{m n}J^{(m,n)}({\bm \omega},\bar{w}_{N-1}^{(k|m,n)}) \frac{\mathcal{P}^{(mn)}_{N, {\rm R}}({\bm \omega},\bar{w}_{N-1}^{(k|m,n)})}{\mathcal{P}^{(mn)}_{N}({\bm \omega},\bar{w}_{N-1}^{(k|m,n)})}Z^{\rm (R)}({\bf \hat{w}}^{(k|m,n)}).
\end{equation}
As expected, this is a relation between single-valued functions of $(\bm{\omega},\bar{w}_{N-1}^{(k|m,n)})$, any trace of the intermediate choice $\bar{a}_{12}={\epsilon}_{m,n}$ disappeared, and hence (\ref{residuesym}) is Weyl symmetric.

\paragraph{Asymptotic behaviour at infinity.} To construct the recurrence relation in terms of the symmetric variables we have to send $w_{N-1}$ to infinity while keeping all the rest of $w_i$ finite. In this case
\begin{equation}
    Q(x|{\bf a})\rightarrow x^N + (-1)^N w_{N-1}
\end{equation}
and since the roots of the polynomial $Q(x|{\bf a})$ are $a_u$ we see, that in terms of the variables $a_u$ the correct way to approach infinity is to place them at the vertices of a regular $N$-sided polygon and send its diameter to infinity, so
\begin{equation}
    a_u = a_N \, \varepsilon^u, \quad \varepsilon =e^{\frac{2\pi {\rm i}u}{N}}, \quad |a_N| \rightarrow \infty.
\end{equation}
With this symmetric approach we are able to analyse the asymptotic behaviour both in the pure theory and in a theory with matter hypermultiplet.

\begin{itemize}
    \item Pure theory. For the pure theory we see from (\ref{ZkNak}) that
\begin{equation}
    Z_k^{({\rm 0})} \underset{w_{N-1} \rightarrow \infty}{\longrightarrow } \frac{1}{a_N^{2k(N-1)}},
\end{equation}
so the only non-vanishing at infinity contribution is $Z_0$ and thus
\begin{equation}
    Z^{({\rm 0})} \underset{w_{N-1} \rightarrow \infty}{\longrightarrow } 1.
\end{equation}

\item Adjoint matter. In the case of a theory with adjoint matter hypermultiplet it is easy to see both from (\ref{Zkadj}) and from (\ref{ZkadjNak}) that interaction between the Young diagrams simply turns into a factor $1$, and thus the asymptotic behaviour of $Z^{({\rm adj})}$ is factorised
\begin{eqnarray} \label{Zadjass}
    Z^{({\rm adj})}\underset{w_{N-1} \rightarrow \infty}{\longrightarrow } \sum_k q^k \sum_{\underset{|\vec{Y}|=k}{\vec{Y}}}\prod_{u=1}^N \prod_{(i,j)\in Y_u}\frac{f_{(i,j)}(M)}{f_{(i,j)}(0)}= \left( \sum_Y \prod_{(i,j)\in Y}\frac{f_{(i,j)}(M)}{f_{(i,j)}(0)} \right)^N,
\end{eqnarray}
where the last sum runs over single Young diagrams $Y$ and
\begin{equation}
    f_{(i,j)}(M)=\left(\frac{\epsilon_1}{\epsilon_2}(i-\tilde{l}_{Y_u,j})-(j-1-l_{Y_u,i})+\frac{M}{\epsilon_2}\right)\left(-(i-1-\tilde{l}_{Y_u,j})+\frac{\epsilon_2}{\epsilon_1}(j-l_{Y_u,i})+\frac{M}{\epsilon_1}\right).
\end{equation}
Therefore the asymptotic behaviour of $Z^{({\rm adj})}$ is a universal constant to the power of $N$.
\par In the simplest case of $M=0$ the result is easy to get
\begin{eqnarray}
    Z^{({\rm adj})} \underset{w_{N-1} \rightarrow \infty}{\longrightarrow } \left( \sum_Y \prod_{(i,j)\in Y} 1\right)^N =  \left( \prod_{k=1}^{\infty} (1-q^k)^{-1}\right)^N =  \left( q^{-\frac{1}{24}}\eta(q)\right)^{-N}. 
\end{eqnarray}
If $\epsilon_1=-\epsilon_2=\tilde{\epsilon}$ we have
\begin{equation}
 f_{(i,j)}(M)=\left(h_{(i,j)}-\frac{M}{\tilde{\epsilon}}\right)\left(h_{(i,j)}+\frac{M}{\tilde{\epsilon}})\right),   
\end{equation}
where $h_{(i,j)}$ is the hook length of the cell $(i,j)\in Y$
\begin{equation}
    h_{(i,j)}=l_{Y,i}+\tilde{l}_{Y,j}-i-j+1.
\end{equation}
For this case the product in (\ref{Zadjass}) was computed in \cite{Han2} with combinatorical calculations, whereas in \cite{NekrasovOkunkov} the asymptotic behaviour of $Z^{({\rm adj})}$ was analysed in $U(1)$ theory with the guage theory approach. The result obtained in these papers is
\begin{eqnarray}
    Z^{({\rm adj})} \underset{w_{N-1} \rightarrow \infty}{\longrightarrow } \left( \sum_Y \prod_{(i,j)\in Y} \frac{\left(h_{(i,j)}-\frac{M}{\tilde{\epsilon}}\right)\left(h_{(i,j)}+\frac{M}{\tilde{\epsilon}})\right)}{h_{(i,j)}^2}\right)^N =  \left( q^{-\frac{1}{24}}\eta(q)\right)^{-N\left(1-\frac{M^2}{\tilde{\epsilon}^2} \right)}. 
\end{eqnarray}
For $\epsilon_1 \neq -\epsilon_2$ the product in (\ref{Zadjass}) was not rigorously computed yet, but in \cite{Poghossian2}, \cite{Poghossian} a suggestion has been made for $SU( 2)$ and $SU(3)$ theories which appears to be correct. Embracing this conjecture we get
\begin{eqnarray}
Z^{({\rm adj})}\underset{w_{N-1} \rightarrow \infty}{\longrightarrow }\left( \sum_Y \prod_{(i,j)\in Y}\frac{f_{(i,j)}(M)}{f_{(i,j)}(0)} \right)^N= \left( q^{-\frac{1}{24}}\eta(q)\right)^{-N\left(1+\frac{M}{\epsilon_1} \right)\left(1+\frac{M}{\epsilon_2} \right)}.
\end{eqnarray}

\item Fundamental and anti-fundamental matter. In the case of fundamental and anti-fundamental hypermultiplets we see from (\ref{ZkfundNak}) that the leading term of $Z_k^{({\rm fund})}$ is 
\begin{equation}
    Z^{({\rm fund})}_k \underset{w_{N-1} \rightarrow \infty}{\longrightarrow } \frac{a_N^{k(N_f+N_a)}}{a_N^{2k(N-1)}}c_k,
\end{equation}
where $c_k$ is some constant. Therefore if $N_f+N_a<2(N-1)$, then again the only non-vanishing contribution is $Z_0^{({\rm fund})}$ and 
\begin{equation}
    Z^{({\rm fund})} \underset{w_{N-1} \rightarrow \infty}{\longrightarrow } 1, \quad N_f+N_a<2(N-1).
\end{equation}
In the case of the critical number of matters $N_f+N_a=2(N-1)$ the limit of $Z^{({\rm fund})}$ is a constant not depending on ${\bf a}$. Let us find this constant.
\par Only the leading term of $Z_k$ matters in this case, so 
\begin{eqnarray} \label{Zkdunflim}
     Z_k^{({\rm fund})} \underset{w_{N-1} \rightarrow \infty}{\longrightarrow }&&\sum_{\underset{|\vec{Y}|=k}{\vec{Y}}}\frac{\varepsilon^{(2N-2)\sum_{u=1}^{N}u |Y_u|}}{\prod_{u=1}^N(\prod_{v \neq u}(\varepsilon^v-\varepsilon^u)^{|Y_u|})\cdot\prod_{v=1}^N(\prod_{u \neq v}(\varepsilon^v-\varepsilon^u)^{|Y_v|})} \frac{1}{\prod_{u=1}^N  \prod_{(i,j)\in Y_u} \epsilon_1\epsilon_2 f_{(i,j)}} \nonumber \\
     &=&\sum_{\underset{|\vec{Y}|=k}{\vec{Y}}}\prod_{u=1}^N\frac{1}{\prod_{v \neq N}(-(1-\varepsilon^v))^{2|Y_u|}} \frac{1}{ \prod_{(i,j)\in Y_u} \epsilon_1\epsilon_2 f_{(i,j)}}.
\end{eqnarray}
Note that
\begin{equation}
    \prod_{v \neq N}(1-\varepsilon^v)=\underset{x \rightarrow 1}{\rm lim}\frac{x^N-1}{x-1}=\frac{{\rm d} x^N}{{\rm d}x}\bigg|_{x=1}=N.
\end{equation}
Therefore
\begin{equation} \label{Zkfactorised}
      Z_k^{({\rm fund})}
     \underset{w_{N-1} \rightarrow \infty}{\longrightarrow }
      \sum_{\underset{|\vec{Y}|=k}{\vec{Y}}}\prod_{u=1}^N \left(\frac{(-1)^{N-1}}{N^2\epsilon_1 \epsilon_2}\right)^{|Y_u|}\prod_{(i,j)\in Y_u}\frac{1}{f_{(i,j)}}.
\end{equation}
We see again that there is no interaction between the Young diagrams in (\ref{Zkfactorised}), and hence we can write
\begin{equation}
    Z^{({\rm fund})}
   \underset{w_{N-1} \rightarrow \infty}{\longrightarrow } \sum_{k} q^k \sum_{\underset{|\vec{Y}|=k}{\vec{Y}}}\prod_{u=1}^N\left(\frac{(-1)^{N-1}}{N^2\epsilon_1 \epsilon_2}\right)^{|Y_u|}\prod_{(i,j)\in Y_u}\frac{1}{f_{(i,j)}} =\left( \sum_{Y} q^{|Y|}\left(\frac{(-1)^{N-1}}{N^2\epsilon_1 \epsilon_2}\right)^{|Y|}\prod_{(i,j)\in Y_u}\frac{1}{f_{(i,j)}}\right)^N.
\end{equation}
Using a relation provided\footnote{To see (\ref{knownrelation}) from (4.5) of \cite{NakajimaYoshiokaZ} one has to replace $t_1 \rightarrow e^{\epsilon_1\delta}$, $t_2 \rightarrow e^{\epsilon_2\delta}$, $q \rightarrow \delta^2 q$ and send $\delta$ to zero.} in \cite{NakajimaYoshiokaZ}
\begin{equation} \label{knownrelation}
    \sum_{Y}x^{|Y|}\prod_{I \in Y}\frac{1}{f_{(i,j)}}=e^x
\end{equation}
we immediately get 
\begin{equation} \label{Zfundasympt}
      Z^{({\rm fund})}\underset{w_{N-1} \rightarrow \infty}{\longrightarrow }{\rm exp}\left( (-1)^{N-1}\frac{q}{N \epsilon_1 \epsilon_2}\right), \quad N_f+N_a=2(N-1).
\end{equation}
\par Finally if the number of fundamental and anti-fundamental hypermultiplets is above critical $N_f+N_a>2(N-1)$ the asymptotic behaviour of $Z^{({\rm fund})}({\bf a})$ can be a nontrivial function of ${\bf a}$ and finding it goes beyond this paper. We refer an interested reader to \cite{Poghossian2}, \cite{Poghossian}, where this behaviour was studied with the AGT approach in $SU(2)$ and $SU(3)$ theories. Although our residue formula (\ref{zinstrelation}) is valid also in this case, the recurrence relation which we will find below does not describe it.
\end{itemize}
\par The recurrence relations will be written for a partition function $\bar{Z}({\bf a})$ normalised to the constants discussed above, so the asymptotic behaviour of the normalised partition function is
\begin{equation}
    \bar{Z}^{({\rm R})} \underset{w_{N-1} \rightarrow \infty}{\longrightarrow } 1.
\end{equation}

\paragraph{The recurrent relation via symmetric variables.} Putting all together and taking into account the behaviour at infinity we get the recurrent Zamolodchikov-like relations in terms of the symmetric variables.
\begin{equation}
    \bar{Z}^{\rm (R)}({\bm \omega},w_{N-1})=1+\sum_{k=1}^{N-1}\sum_{m,n=1}^{\infty} \frac{q^{mn} J^{(m,n)}({\bm \omega},\bar{w}_{N-1}^{(k|m,n)}) }{(w_{N-1}-\bar{w}^{(k|m,n)}_{N-1})}\frac{\mathcal{P}^{(m,n)}_{N, {\rm R}}({\bm \omega},\bar{w}_{N-1}^{(k|m,n)})}{\mathcal{P}^{(m,n)}_{N}({\bm \omega},\bar{w}_{N-1}^{(k|m,n)})}\bar{Z}^{\rm (R)}({\bf \hat{w}}^{(k|m,n)}).
\end{equation}

\paragraph{Explicit examples.} 
Let us now explicitly write the recurrence relations in several simplest cases.
\begin{itemize}
    \item  SU(2) case.
\par For completeness let us formulate the $SU(2)$ case in terms of symmetric variables, although the Weyl symmetry is evident even in terms of the variables ${\bf a}$ in this case.
\par The only symmetric variable in this case is
\begin{equation*}
    w_1=a_1 a_2 = - \frac{1}{4} a_{12}^2.
\end{equation*}
The polynomial $Q^{({\bf w})}(x)$ is quadratic
\begin{equation*}
    Q^{({\bf w})}(x)=x^2+w_1
\end{equation*}
and the resultant gives us the denominator
\begin{equation*}
    \Delta^{(m,n)}({\bf w})=-4 w_1 - \epsilon_{m,n}^2,
\end{equation*}
so the pole is located at
\begin{equation*}
    \bar{w}_1=-\frac{1}{4}\epsilon_{m,n}^2
\end{equation*}
as it should.
\par The dual point is
\begin{equation*}
    \hat{w}_1^{(m,n)}=\bar{w}_1-\Delta w_1^{(m,n)} =-\frac{1}{4}\epsilon_{m,n}^2+m n \, \epsilon_1 \epsilon_2=-\frac{1}{4} \epsilon_{m,-n}^2,
\end{equation*}
which obviously corresponds to the partial Weyl permutation in $a_1$, $a_2$.
\par The polynomial $\Delta Q(x)$ in this case is just a constant  $\Delta Q(x)=1$, so the most part of   $\mathcal{P}^{(m,n)}$ and $J^{(m,n)}(\bar{w}_1)$ disappears, and the recurrent relation for the pure theory is just
\begin{equation}
    \bar{Z}^{({\rm R})}(w_1)=1+\sum_{m,n=1}^{\infty}\frac{ q^{mn} 2 \epsilon_{m,n} }{  \left(-4w_1-\epsilon_{m,n}^2 \right)}\frac{\mathcal{P}^{(m,n)}_{2,{\rm R}}}{\mathcal{P}^{(m,n)}_{2}}\bar{Z}^{({\rm R})}\left(-\frac{1}{4}\epsilon_{m,-n}^2\right),
\end{equation}
where 
\begin{equation}
    \mathcal{P}^{(m,n)}_{2}=  \prod_{i=-m}^{m-1}\sideset{}{'} \prod_{j=-n}^{n-1} \epsilon_{i,j},
\end{equation}
\begin{equation}
    \mathcal{P}^{(m,n)}_{2,{\rm adj}}= \prod_{i=-m}^{m-1} \prod_{j=-n}^{n-1} (\epsilon_{i,j}-M),
\end{equation}
\begin{equation}
    \mathcal{P}^{(m,n)}_{2,{\rm fund}}=\prod_{i=1}^{m}\prod_{j=1}^{n} \Bigg[ \prod_{t=1}^{N_f}   \left(-\frac{1}{2}\epsilon_{m,n}+\epsilon_{i,j}-m_t  \right)\cdot  \prod_{t=1}^{N_a}  \left(-\frac{1}{2}\epsilon_{m,n}+\epsilon_{i,j}+m_t \right) \Bigg].
\end{equation}

\item SU(3) case.
\par Let us now compare our result in the $SU(3)$ case with the one obtained in \cite{Poghossian}.
\par The variables $(u,v)$ used in  \cite{Poghossian} differ from ours by a numerical factor
\begin{eqnarray}
    w_1&=&\sum_{i<j} a_i a_j=-\frac{1}{3} (a_{12}^2+a_{12}a_{23}+a_{23}^2)= 
  -\frac{1}{3} u, \nonumber \\
    w_2&=& a_1 a_2 a_3=-\frac{1}{27} (a_{12}-a_{23})(2a_{12}+a_{23})(a_{12}+2a_{23})=-\frac{1}{27} v. \nonumber
\end{eqnarray}
The polynomial $Q(x)$ in this case is
\begin{equation*}
    Q^{({\bf w})}(x)=x^3+x w_1 - w_2,
\end{equation*}
which gives us a square equation for the poles
\begin{equation}
    27 {w}_2^2+4 w_1^3+9w_1^2\epsilon_{m,n}^2+6w_1\epsilon_{m,n}^4+\epsilon_{m,n}^6=0
\end{equation}
and hence the positions of the poles are
\begin{equation} \label{wmn}
   \bar{w}^{(k|m,n)}_2 = \pm \frac{1}{3}(w_1+\epsilon_{m,n}^2)\sqrt{-\frac{1}{3}(4 w_1+\epsilon_{m,n}^2)}\triangleq\pm w_{(m,n)}(w_1).
\end{equation}
System of recurrent equations (\ref{recsys}) boils down to only one equation
\begin{equation} \label{recn3}
    w_1+\frac{3}{4}\frac{(\Delta w_2^{(k|m,n)} )^2}{(\Delta w_1^{(m,n)})^2}+\frac{1}{4}\epsilon_{m,n}^2=0.
\end{equation}
It has two roots and we have to pick one for each of $\bar{w}_2^{(k|m,n)}$ using the Euclidean algorithm to divide $\bar{Q}(x+\epsilon_{m,n})$ by $\bar{Q}(x)$. After two steps of the division we get a linear remainder
\begin{equation}
    r_2(x)=x\left(\frac{2 w_1}{3}+\frac{2 \epsilon_{m,n}^2}{3} \right)-\bar{w}_2^{(k|m,n)}+\frac{w_1 \epsilon_{m,n}}{3}+\frac{\epsilon_{m,n}^3}{3}
\end{equation}
and demanding that $r_2(a_2(\Delta w_2^{(k|m,n)} ))=0$ we find the shift $\Delta w_2^{(k|m,n)}$
\begin{equation} \label{dw2}
    \Delta w_2^{(k|m,n)} =-\frac{3 \bar{w}_2^{(k|m,n)} \Delta w_1^{(m,n)} }{w_1+\epsilon_{m,n}}.
\end{equation}
Substituting two roots $w_{2}^{(k|m,n)}$ given by (\ref{wmn}) we see that (\ref{dw2}) is indeed the two roots of (\ref{recn3}).
\par Therefore the dual points $(\hat{w}_1^{(m,n)},\hat{w}_2^{(k|m,n)})$ are
\begin{eqnarray}
    \hat{w}_1^{(m,n)} &=&w_1+ m n \,\epsilon_1 \epsilon_2, \nonumber \\
    \hat{w}_2^{(k|m,n)}&=& \pm w_{(m,n)}(w_1) \left(1+\frac{3\Delta w_1^{(m,n)} }{w_1+\epsilon_{m,n}} \right)=\pm  w_{(m,-n)}(\hat{w}_1^{(m,n)} )
\end{eqnarray}
in agreement with (\ref{wshiftedroots}).
\par The polynomial $\Delta Q(x)$ is linear
\begin{equation}
    \Delta Q(x)=x-\frac{\Delta w_2^{(k|m,n)} }{\Delta w_1^{(m,n)} }.
\end{equation}
\par The recurrence relation in $SU(3)$ theory is the following
\begin{eqnarray} \label{recsymn3}
    \bar{Z}^{({\rm R})}(w_1,w_2)=1+\sum_{k=1}^{2}\sum_{m,n=1}^{\infty}\bigg( &&\frac{q^{mn} (3w_1+\epsilon_{m,n}^2)(-w_1-\epsilon_{m,n}^2)\epsilon_{mn}}{9\bar{w}_2^{(k|m,n)} (w_2-\hat{w}_2^{(k|m,n)})} \nonumber \\ && \cdot\frac{\mathcal{P}^{(m,n)}_{3,{\rm R}}(w_1,\bar{w}_2^{(k|m,n)})}{\mathcal{P}^{(m,n)}_{3}(w_1,\bar{w}_2^{(k|m,n)})}\bar{Z}^{({\rm R})}(\hat{w}_1^{(m,n)} ,\hat{w}_2^{(k|m,n)}) \bigg),
\end{eqnarray}
where
\begin{equation} 
    \mathcal{P}^{(m,n)}_{3}(w_1,\bar{w}_2^{(k|m,n)})=\prod_{i=-m}^{m-1}\sideset{}{'} \prod_{j=-n}^{n-1} \epsilon_{i,j} \cdot \prod_{i=1}^{m} \prod_{j=1}^{n}(3 w_1+\epsilon_{m,n}^2-\epsilon_{i,j}\epsilon_{m-i,n-j}),
\end{equation}
\begin{equation}
    \mathcal{P}^{(m,n)}_{3,{\rm adj}}(w_1,\bar{w}_2^{(k|m,n)})= \prod_{i=-m}^{m-1} \prod_{j=-n}^{n-1} (\epsilon_{i,j}-M) \cdot \prod_{i=1}^{m} \prod_{j=1}^{n}(3 w_1+\epsilon_{m,n}^2-(\epsilon_{i,j}+M)(\epsilon_{m-i,n-j}-M)),
\end{equation}
and $\mathcal{P}_{3, {\rm fund}}^{(m,n)}({\bm \omega}, \bar{w}_{N-1}^{(k|m,n)})$ is always the same and given by (\ref{Psymfund}).
\par One can recurrently see that for a pure theory and a theory with the adjoint matter the partition function actually depends only on $(\Delta w_2^{(k|m,n)})^2$, \textit{i.e.} only on  $w_{(m,n)}^2$, so we can write (\ref{recsymn3}) as
\begin{equation}
    \bar{Z}^{({\rm R})}(w_1,w_2^2)=1+\sum_{m,n=1}^{\infty}\frac{q^{mn}(3w_1+\epsilon_{m,n}^2)(-w_1-\epsilon_{m,n}^2)2 \epsilon_{m,n}}{9(w_2^2-w_{(m,n)}^2)}\frac{\mathcal{P}^{(m,n)}_{3,{\rm R}}(w_1)}{\mathcal{P}^{(m,n)}_{3}(w_1)}\bar{Z}^{({\rm R})}(\hat{w}_1^{(m,n)} ,w_{(m,-n)}^2(\hat{w}_1^{(m,n)} )).
\end{equation}
The results for pure theory and theory with adjoint hypermultiplet coincide with the ones found in \cite{Poghossian} up to the sign of the mass of adjoint multiplet $M$. To compare also the case with fundamental and anti-fundamental matter we should consider a particular case of $N_f=N_a=N$ and redefine the masses as 
\begin{eqnarray*}
    m_t \rightarrow \epsilon-m_{t} \quad {\rm fundamental} \\
    m_{t} \rightarrow -m_{t}  \quad {\rm anti-fundamental}\\
\end{eqnarray*}
In this case we do not know the behaviour of the partition function at infinity since $2N$ is above the critical number $2(N-1)$, but if we embrace the asymptotic behaviour provided in \cite{Poghossian}, we will recover exactly the recurrence relation found in there.

\item SU(4) case. 
\par Finally we are to write the recurrence relations for a non considered before case of $SU(4)$ theory.
\par The polynomial $Q^{({\bf w})}(x)$ in this case is
\begin{equation}
    Q^{({\bf w})}(x)=x^4+w_1 x^2-w_2 x+w_3,
\end{equation}
and the equation for the poles
\begin{equation} \label{polesSU4}
    {\rm \bf res}(Q^{({\bm \omega},{w}_{3})}(x), Q^{({\bm \omega},{w}_{3})}(x+\epsilon_{m,n}))=0
\end{equation}
is a cubic equation on ${w}_{3}$ of general form with three roots $\bar{w}^{(k|m,n)}_{3}({\bm \omega})$.
\par The system of recurrence equations (\ref{recsys}) consists of two equations
\begin{equation}
    \begin{cases}
    w_2+\frac{\Delta w_3^{(k|m,n)} \Delta w_2^{(k|m,n)} }{(\Delta w_1^{(m,n)} )^2}+\frac{1}{4}\frac{\Delta w_2^{(k|m,n)} }{\Delta w_1^{(m,n)} } \left( \epsilon_{m,n}^2-\frac{(\Delta w_2^{(k|m,n)} )^2}{(\Delta w_1^{(m,n)} )^2}\right)=0 \\
    w_1-\frac{\Delta w_3^{(k|m,n)} }{\Delta w_1^{(m,n)} }+\frac{3}{4}\frac{(\Delta w_2^{(k|m,n)} )^2}{(\Delta w_1^{(m,n)} )^2}+\frac{1}{4}\epsilon_{m,n}^2=0
    \end{cases}
\end{equation}
For the shift $\Delta w_3\textbf{}$ we get
\begin{equation}
    \Delta w_{3}\textbf{}=\frac{4 \bar{w}_{3}^{(k|m,n)}}{\frac{(\Delta w_2^{(k|m,n)} )^2}{(\Delta w_1^{(m,n)} )^2}-\epsilon_{m,n}^2}.
\end{equation}
After three steps of the Euclidean algorithm of the division of $\bar{Q}(x+\epsilon_{m,n})$ by $\bar{Q}(x)$ we find the remainder
\begin{eqnarray}
    r_3(x)=\frac{2\epsilon_{m,n}}{(2w_1+5\epsilon_{m,n}^2)^2}&\cdot&\bigg( \big( 2 x+\epsilon_{m,n})(-8w_1 \bar{w}_3^{(k|m,n)}-20 \bar{w}_3^{(k|m,n)}\epsilon_{m,n}^2 \nonumber \\ &+& 2 w_1^3+9w_2^2+9w_1^2\epsilon_{m,n}^2+12w_1\epsilon_{m,n}^4+5\epsilon_{m,n}^6 \big) \\ \nonumber
    &-&2 w_2 (12\bar{w}_3^{(k|m,n)}+w_1^2+8w_1 \epsilon_{m,n}^2+7\epsilon_{m,n}^4 ) \bigg)
\end{eqnarray}
and since $r_3(a_2(\Delta w_2^{(k|m,n)} ))=0$ we get
\begin{equation}
    \frac{\Delta w_2^{(k|m,n)} }{\Delta w_1^{(m,n)} }=\frac{-2 w_2 (12\bar{w}_3^{(k|m,n)}+w_1^2+8w_1 \epsilon_{m,n}^2+7\epsilon_{m,n}^4 )}{(-8w_1 \bar{w}_3^{(k|m,n)}-20 \bar{w}_3^{(k|m,n)}\epsilon_{m,n}^2+2 w_1^3+9w_2^2+9w_1^2\epsilon_{m,n}^2+12w_1\epsilon_{m,n}^4+5\epsilon_{m,n}^6) }.
\end{equation}
The polynomial $\Delta Q(x)$ in this case is
\begin{equation}
    \Delta Q(x)=x^2-\frac{\Delta w_2^{(k|m,n)}}{\Delta w_1^{(m,n)} }x+\frac{\Delta w_3^{(k|m,n)}}{\Delta w_1^{(m,n)} }.
\end{equation}
The recurrence relation is
\begin{equation}
    \bar{Z}^{({\rm R})}({\bm \omega},w_3)=1+\sum_{k=1}^{3}\sum_{m,n=1}^{\infty} \frac{q^{mn}J^{(m,n)}({\bm \omega},\bar{w}_{3}^{(k|m,n)})}{(w_3-\bar{w}^{(k|m,n)}_3)}\frac{ \mathcal{P}^{(m,n)}_{4,{\rm R}} ({\bm \omega},\bar{w}_3^{(k|m,n)})}{\mathcal{P}^{(m,n)}_{4}({\bm \omega},\bar{w}_3^{(k|m,n)})}\bar{Z}^{({\rm R})}({\hat{\bm \omega}^{(k|m,n)} ,\hat{w}_3^{(k|m,n)})}),
\end{equation} 
where $\bar{w}^{(k|m,n)}_3$ are the roots of equation (\ref{polesSU4}),
\begin{eqnarray}
     J^{(m,n)}({\bm \omega},\bar{w}_{3}^{(k|m,n)})&=&(-1)\frac{1}{3}\epsilon_{m,n}(w_1+2\frac{\Delta w_3^{(k|m,n)} }{\Delta w_1^{(m,n)} }+\epsilon^2) \nonumber \\ &\cdot& (w_1^2-4w_1\frac{\Delta w_3^{(k|m,n)} }{\Delta w_1^{(m,n)} }+4\frac{(\Delta w_3^{(k|m,n)})^2}{(\Delta w_1^{(m,n)})^2}+8w_1 \epsilon_{m,n}^2-4\frac{\Delta w_3^{(k|m,n)}}{\Delta w_1^{(m,n)}}\epsilon_{m,n}^2+7\epsilon_{m,n}^4) \nonumber \\&\cdot&(w_1^2-4w_1\frac{\Delta w_3^{(k|m,n)} }{\Delta w_1^{(m,n)} }+4\frac{(\Delta w_3^{(k|m,n)})^2}{(\Delta w_1^{(m,n)})^2}+\frac{4}{3}w_1 \epsilon_{m,n}^2-\frac{4}{3}\frac{\Delta w_3^{(k|m,n)} }{\Delta w_1^{(m,n)} }\epsilon_{m,n}^2+\frac{1}{3}\epsilon_{m,n}^4) \nonumber\\
     &\cdot&\bigg( 2 w_1^4+18 w_1 w_2^2-32 w_1^2 \bar{w}^{(k|m,n)}_3+96 (\bar{w}^{(k|m,n)}_3)^2+  4w_1^3\epsilon_{m,n}^2+27w_2^2\epsilon_{m,n}^2 \nonumber \\ &-&48w_1\bar{w}^{(k|m,n)}_3\epsilon_{m,n}^2+3 w_1^2 \epsilon_{m,n}^4-28\bar{w}^{(k|m,n)}_3 \epsilon_{m,n}^4+2w_1\epsilon_{m,n}^6+\epsilon_{m,n}^8 \bigg)^{-1}
\end{eqnarray}
and
\begin{equation}
    \mathcal{P}^{(m,n)}_{4}({\bm \omega},\bar{w}_3^{(k|m,n)})=\prod_{i=-m}^{m-1} \sideset{}{'} \prod_{j=-n}^{n-1} \epsilon_{i,j} \cdot \prod_{i=1}^{m} \prod_{j=1}^{n}\left( (w_1-2\frac{\Delta w_3^{(k|m,n)} }{\Delta w_1^{(m,n)} }+\epsilon_{i,j}\epsilon_{m-i,n-j})^2 -\frac{(\Delta w_2^{(k|m,n)} )^2}{(\Delta w_1^{(m,n)} )^2}\epsilon_{m-2i,n-2j}^2 \right),
\end{equation}
\begin{eqnarray}
    \mathcal{P}^{(m,n)}_{4,{\rm adj}}({\bm \omega},\bar{w}_3^{(k|m,n)})&=& \prod_{i=-m}^{m-1}\prod_{j=-n}^{n-1} (\epsilon_{i,j}-M)  \\
    &\cdot& \prod_{i=1}^{m} \prod_{j=1}^{n}\bigg( (w_1-2\frac{\Delta w_3^{(k|m,n)} }{\Delta w_1^{(m,n)} }+(\epsilon_{i,j}+M)(\epsilon_{m-i,n-j}-M))^2 \nonumber \\ &-& \frac{(\Delta w_2^{(k|m,n)} )^2}{(\Delta w_1^{(m,n)} )^2}(\epsilon_{m-2i,n-2j}-2M)^2 \bigg), \nonumber
\end{eqnarray}
and $\mathcal{P}_{4, {\rm fund}}^{(m,n)}({\bm \omega}, \bar{w}_{N-1}^{(k|m,n)})$ is given by (\ref{Psymfund}).

\end{itemize}

\section{Summary of results} \label{summary}
In this Section we collect the main relations obtained throughout the paper.
\par We showed that the instanton partition function has poles only at the points $a_{uv}=\epsilon_{m,n}$ with $m\cdot n>0$, the poles are simple and for the positive $m$, $n$ the residue of the instanton partition function with respect to the variable $a_{uv}$ can be expressed via its value at the point distinguished by the partial Weyl permutation between $a_u$ and $a_v$.
\begin{equation}
    {\rm Res}_{a_{uv}=\epsilon_{m,n}} Z^{({\rm R})}( {\bf a})=q^{m n}\frac{\mathcal{P}^{(uv)}_{N,{\rm R}}(m,n|{\bf a})}{\mathcal{P}^{(uv)}_{N}(m,n|{\bf a}) }Z^{({\rm R})}(\hat{{\bf a}}^{(uv)}),
\end{equation}
where
\begin{eqnarray}
   \mathcal{P}^{(uv)}_{N}(m,n|{\bf a})=
   \prod_{i=-m}^{m-1}\sideset{}{'} \prod_{j=-n}^{n-1} \epsilon_{i,j} \cdot\underset{w \neq u, \, v}{\prod_{w=1}^{N}} \prod_{i=1}^m\prod_{j=1}^n\left[  (a_{vw}+\epsilon_{i,j})(-a_{uw}+\epsilon_{i,j}) \right],
\end{eqnarray}
for the pure theory we found a trivial numerator
\begin{equation}
    \mathcal{P}^{(uv)}_{N,{\rm 0}}=1,
\end{equation}
for a theory with the adjoint matter there is a polynomial
\begin{eqnarray}
 \mathcal{P}^{(uv)}_{N,{\rm adj}}(m,n|{\bf a})=    \prod_{i=-m}^{m-1}\prod_{j=-n}^{n-1}  \left(\epsilon_{i,j}-M \right)  \cdot\underset{w \neq u, \, v}{\prod_{w=1}^{N}} \prod_{i=1}^m\prod_{j=1}^n\left[  (a_{vw}+\epsilon_{i,j}+M)(-a_{uw}+\epsilon_{i,j}+M) \right],
\end{eqnarray}
and finally for the theory with fundamental and anti-fundamental multiplets the polynomial is
\begin{eqnarray}
 \mathcal{P}^{(uv)}_{N,{\rm fund}}(m,n|{\bf a})=\prod_{i=1}^{m}\prod_{j=1}^{n} \Bigg[ \prod_{t=1}^{N_f}   \left(-\frac{1}{2}\epsilon_{m,n}+\epsilon_{i,j}-m_t-\frac{1}{2}\sum_{\underset{w \neq u,v}{w=1}}^N a_w \right) \cdot \\
 \prod_{t=1}^{N_a}  \left(-\frac{1}{2}\epsilon_{m,n}+\epsilon_{i,j}+m_t+\frac{1}{2}\sum_{\underset{w \neq u,v}{w=1}}^N a_w \right) \Bigg]. \nonumber
\end{eqnarray}
One can easily see from the proof of these relations that these hypermultiplets can be considered together, and the polynomial in the numerator will be simply the product of the polynomials above\footnote{It is actually not difficult to see that the residue formula holds for the matter in any representation, but in the general case the polynomial in the numerator is too long to write in a paper.}.
\par For the residue at the point $a_{uv}=\epsilon_{-m,-n}$ there is an additional minus sign.
\par There is an equivalent form in terms of the full partition function
\begin{equation}
    \lim_{a_{uv} \rightarrow \epsilon_{m,n}}\frac{\mathcal{Z}^{({\rm R})}({\bf a})}{\mathcal{Z}^{({\rm R})}({\bf \hat{a}}^{(uv)})}=-{\rm Sign}(\epsilon_1), \quad m,n \in \mathbb{Z}\setminus \{0\}
\end{equation}
for the permutation of $\epsilon_2$-coefficients (and $-{\rm Sign}(\epsilon_2)$ for the permutation of $\epsilon_1$-coefficients).
\par For the pure theory we found a recurrence relation for the instanton partition function in terms of the variables $a_{uv}$.
\begin{eqnarray}
{Z}^{({\rm 0})}({\bf a})=  1+
 \sum_{w=2}^{N}\sum_{m,\, n=1}^{\infty} \frac{q^{mn} {Z}^{({\rm 0})}(\hat{\bf a}^{(1w)})}{(a_{1N}-a_{wN}+\epsilon_{m,n})(a_{1N}-a_{wN}-\epsilon_{m,n})} \frac{2 \epsilon_{m,n}\mathcal{P}^{(1w)}_{N,{\rm 0}}(m,n|{\bf a})}{\mathcal{P}^{(1w)}_{N}(m,n|{\bf a})}.
\end{eqnarray}
To write the recurrence relation for theories with matter hypermultiplets we switched to the symmetrical variables defined as 
\begin{equation}
 \omega_{l}=\sum_{i_1<\ldots<i_k} a_{i_{1}}\cdot \ldots \cdot a_{i_{l+1}} \, \quad l=1, \ldots, N-1.
\end{equation}
The recurrence relation for the normalised instanton partition function $\bar{Z}^{\rm (R)}$ is 
\begin{equation}
    \bar{Z}^{\rm (R)}({\bm \omega},w_{N-1})=1+\sum_{k=1}^{N-1}\sum_{m,n=1}^{\infty} \frac{q^{mn} J^{(m,n)}({\bm \omega},\bar{w}_{N-1}^{(k|m,n)}) }{(w_{N-1}-\bar{w}^{(k|m,n)}_{N-1})}\frac{\mathcal{P}^{(m,n)}_{N, {\rm R}}({\bm \omega},\bar{w}_{N-1}^{(k|m,n)})}{\mathcal{P}^{(m,n)}_{N}({\bm \omega},\bar{w}_{N-1}^{(k|m,n)})}\bar{Z}^{\rm (R)}({\bf \hat{w}}^{(k|m,n)}),
\end{equation}
where $\bar{w}^{(k|m,n)}_{N-1}$ are the roots of equation (\ref{wroots}), the polynomials $\mathcal{P}^{(m,n)}_{N}$, $\mathcal{P}^{(m,n)}_{N,{\rm R}}$ and Jacobian $J^{(m,n)}$ are defined in (\ref{Psym}), (\ref{Psymadj}), (\ref{Psymfund}) and (\ref{jacobian}).
\par The normalisation constants are defined by the behaviour at infinity, which is the following
 \begin{equation}
    Z^{({\rm 0})} \underset{w_{N-1} \rightarrow \infty}{\longrightarrow } 1.
\end{equation}
\begin{equation}
    Z^{({\rm fund})} \underset{w_{N-1} \rightarrow \infty}{\longrightarrow } 1, \quad N_f+N_a<2(N-1).
\end{equation}
\begin{equation}
      Z^{({\rm fund})}\underset{w_{N-1} \rightarrow \infty}{\longrightarrow }{\rm exp}\left( (-1)^{N-1}\frac{q}{N \epsilon_1 \epsilon_2}\right), \quad N_f+N_a=2(N-1).
\end{equation}
\begin{eqnarray}
Z^{({\rm adj})}\underset{w_{N-1} \rightarrow \infty}{\longrightarrow }\left( q^{-\frac{1}{24}}\eta(q)\right)^{-N\left(1+\frac{M}{\epsilon_1} \right)\left(1+\frac{M}{\epsilon_2} \right)}.
\end{eqnarray}

\end{document}